\documentclass[11pt]{article}

\usepackage[margin=1in]{geometry}
\usepackage{amsmath,amssymb,amsthm,mathtools}
\usepackage{indentfirst}
\usepackage{times}
\usepackage{amsmath}
\usepackage{amsthm}
\usepackage{color}
\usepackage{verbatim}
\usepackage[makeroom]{cancel}
\usepackage{rotating}
\usepackage{float}
\usepackage{amssymb}
\usepackage{amsthm}
\usepackage{mathrsfs}
\usepackage{epsfig}
\usepackage{graphicx}
\usepackage{indentfirst}
\usepackage{framed}
\usepackage[numbers]{natbib}
\usepackage{color}
\usepackage{hyperref}
\usepackage{graphicx}
\usepackage{float}
\usepackage{bm}
\usepackage{hyperref}
\usepackage[capitalize,noabbrev,nameinlink]{cleveref}
\usepackage{enumitem}
\usepackage{mathtools}
\usepackage{xfrac}
\usepackage{mleftright}
\usepackage{thmtools} 
\usepackage{relsize}\usepackage{scalerel}
\usepackage{tikz}
\usepackage{multicol}
\usetikzlibrary{positioning,shapes,shadows,arrows}
\usepackage[labelfont=bf]{caption}
\usepackage{algorithm}
\usepackage[noend]{algpseudocode}
\usepackage{tablefootnote}
\usepackage{marginnote}   %
\usepackage{mdframed}

\definecolor{corlinks}{RGB}{0,0,200}
\definecolor{cormenu}{RGB}{0,0,200}
\definecolor{corurl}{RGB}{0,0,200}

\hypersetup{
	colorlinks=true,
	urlcolor=corlinks,
	linkcolor=corlinks,
	menucolor=cormenu,
	citecolor=corlinks,
	pdfborder= 0 0 0
}

\definecolor{blue-violet}{rgb}{0.54, 0.17, 0.89}
\definecolor{deep-green}{rgb}{0,0.5,0}

\newcommand{\inote}[1]{\textcolor{blue-violet}{(Igor: #1)}}
\newcommand{\jiatu}[1]{\textcolor{blue}{(Jiatu: #1)}}

\definecolor{todollmcolor}{RGB}{0,128,0}
\definecolor{todoauthorcolor}{RGB}{0,128,128}

\definecolor{claudcolor}{RGB}{204,85,0}

\definecolor{gptcolor}{RGB}{0,96,128}

\newif\ifAIfootnotes
\AIfootnotesfalse
\newcommand{\AIprefix}{AI assistance disclosure:}
\newcommand{\AIsymbol}{\ensuremath{\dagger}}
\newif\ifAIpending
\newcommand{\AIdetails}{}
\newcommand{\AI}[1]{\ifAIfootnotes\gdef\AIdetails{#1}\global\AIpendingtrue\fi}
\newcommand{\AIfootnote}[1]{%
  \textsuperscript{\AIsymbol}%
  \begingroup
    \renewcommand{\thefootnote}{\AIsymbol}%
    \footnotetext[\value{footnote}]{#1}%
  \endgroup
}
\expandafter\let\expandafter\AIproofhead\csname\string\proof\endcsname
\let\AIendproof\endproof
\renewenvironment{proof}[1][\proofname]{%
  \AIproofhead[#1]%
  \ifAIpending\global\AIpendingfalse\AIfootnote{\AIprefix\ \AIdetails}\fi
  \ignorespaces
}{\AIendproof}

\newtheorem{theorem}{Theorem}[section]
\newtheorem{lemma}[theorem]{Lemma}

\newtheorem{observation}[theorem]{Observation}

\newtheorem{corollary}[theorem]{Corollary}
\newtheorem{proposition}[theorem]{Proposition}

\theoremstyle{definition}
\newtheorem{definition}[theorem]{Definition}
\newtheorem{example}[theorem]{Example}

\theoremstyle{remark}
\newtheorem{preremark}[theorem]{Remark}
\newenvironment{remark}%
  {\vspace{0.5em}\begin{mdframed}[innertopmargin=-0em,skipabove=0em]\small
	\begin{preremark}\upshape}
{\end{preremark}\end{mdframed}}

\newenvironment{remarkn}[1]%
  {\vspace{0.5em}\begin{mdframed}[innertopmargin=0em,skipabove=0em]\small
	\begin{preremark}[#1]\upshape}
{\end{preremark}\end{mdframed}}

\theoremstyle{plain}
\newtheorem*{theorem*}{Theorem}
\newtheorem*{lemma*}{Lemma}
\newtheorem*{fact*}{Fact}
\newtheorem*{claim*}{Claim}
\newtheorem*{corollary*}{Corollary}
\newtheorem*{proposition*}{Proposition}
\newtheorem*{hypothesis*}{Hypothesis}
\newtheorem*{conjecture*}{Conjecture}
\theoremstyle{definition}
\newtheorem*{definition*}{Definition}
\newtheorem*{example*}{Example}
\newtheorem*{problem*}{Problem}

\numberwithin{equation}{section}

\SetEnumitemKey{compact}{itemsep=0pt,parsep=\parskip,topsep=5pt}
\SetEnumitemKey{lean}{itemsep=2pt,parsep=2pt,topsep=7pt}

\newenvironment{axioms}[1]{%
  \par\addvspace{\medskipamount}%
  \begin{framed}%
  {\centering\scshape #1\par}%
  \begin{description}[compact,topsep=6pt,parsep=3pt,style=unboxed,%
                      leftmargin=1.2em,labelindent=0pt]%
}{%
  \end{description}%
  \end{framed}%
  \addvspace{\medskipamount}%
}
\makeatletter
\newcommand{\axiomlabel}[2]{%
  \phantomsection\protected@edef\@currentlabelname{\textsc{#1}}\label{#2}%
  \ignorespaces}
\makeatother

\newcommand{\eqdef}{\triangleq}
\newcommand{\zo}{\{0,1\}}
\newcommand{\bits}{\{0,1\}}
\newcommand{\eps}{\varepsilon}
\newcommand{\lsim}{\lesssim}
\newcommand{\mlsim}{\lsim^*}
\newcommand{\mapprox}{\approx^*}

\newcommand{\Jerabek}{\text{Je\v{r}\'abek}}

\newcommand{\bbM}{\mathbb{M}}
\newcommand{\N}{\mathbb{N}}

\newcommand{\Q}{\mathbb{Q}}
\newcommand{\R}{\mathbb{R}}

\newcommand{\cI}{\mathcal{I}}
\newcommand{\cL}{\mathcal{L}}
\newcommand{\cM}{\mathcal{M}}
\newcommand{\cU}{\mathcal{U}}
\newcommand{\calt}{\mathcal{T}}
\newcommand{\disD}{\mathcal{D}}
\newcommand{\disU}{\mathcal{U}}
\newcommand{\cT}{\mathcal{T}}

\newcommand{\APX}{\mathsf{APX}_1}
\newcommand{\mAPX}{\APX^*}
\newcommand{\APC}{\mathsf{APC}_1}
\newcommand{\PV}{\mathsf{PV}_1}

\newcommand{\Log}{\mathsf{Log}}

\renewcommand{\P}{\mathsf{P}}
\newcommand{\FP}{\mathsf{FP}}
\newcommand{\BPP}{\mathsf{BPP}}
\newcommand{\prSBP}{\mathsf{prSBP}}
\newcommand{\TFNP}{\mathsf{TFNP}}
\newcommand{\TFZPP}{\mathsf{TFZPP}}
\newcommand{\TFSigma}[1]{\mathsf{TF\Sigma}_{#1}}
\newcommand{\PWPP}{\mathsf{PWPP}}

\newcommand{\AC}{\operatorname{AC}}
\newcommand{\NP}{\mathsf{NP}}

\newcommand{\wPHP}{\mathsf{wPHP}}
\newcommand{\dWPHP}{\mathsf{dwPHP}}
\newcommand{\dwPHP}{\mathsf{dwPHP}}
\newcommand{\rwPHP}{\mathsf{rwPHP}}

\newcommand{\RefYao}{\mathsf{Refuter}(\mathsf{Yao})}
\newcommand{\Local}{\mathsf{Local}}
\newcommand{\LossyCode}{\mathsf{LossyCode}}

\newcommand{\WeakPigeon}{\mathsf{WeakPigeon}}
\newcommand{\Avoid}{\mathsf{Avoid}}

\newcommand{\poly}{\operatorname{poly}}
\newcommand{\Fix}{\operatorname{Fix}}
\newcommand{\Val}{\operatorname{Val}}

\newcommand{\TT}{\mathsf{TT}}

\newcommand{\capp}{\mathsf{capp}}
\newcommand{\sT}{\mathsf{T}}

\newcommand{\emp}{\mathrm{emp}}
\newcommand{\true}{\mathrm{true}}
\newcommand{\final}{\mathrm{final}}
\newcommand{\range}{\mathrm{range}}
\newcommand{\out}{\operatorname{out}}
\newcommand{\cS}{\mathcal{S}}

\newcommand{\trueAPX}{\mathsf{True}_{\forall} \APX}

\newgeometry{left=0.86in,right=0.86in,top=1in,bottom=1in}

\title{Approximate Counting and the Provability of Combinatorial Principles\vspace{0.4cm}}
\author{Oliver Korten\thanks{Institute for Advanced Study. E-mail:~\texttt{korten@ias.edu}} \and Jiatu Li\thanks{Massachusetts Institute of Technology. E-mail:~\texttt{jiatuli@mit.edu}} \and Igor C. Oliveira\thanks{University of Warwick. E-mail:~\texttt{igor.oliveira@warwick.ac.uk}} \and Hanlin Ren\thanks{Institute for Advanced Study and Max Planck Institute for Informatics. E-mail:~\texttt{h4n1in.r3n@gmail.com}}\vspace{0.6cm}}
\date{\today}

\begin{document}
\pagenumbering{gobble}

\maketitle

\vspace{-0.3cm}

\begin{abstract}
Understanding which axioms suffice for the provability of known results in algorithms and complexity and whether they can settle longstanding conjectures in these areas is a fundamental direction at the intersection of logic and computer science. Motivated by this research program and its  applications, we investigate bounded arithmetic theories designed to serve as a logical foundation for \emph{probabilistic reasoning}, the relation between these theories, and whether they can prove more theorems than their underlying base theory for \emph{deterministic reasoning}. 

In more detail, for nearly twenty years, \Jerabek's theory $\APC$ \citep{Jerabek07} has provided a standard framework for the formalization of results from theoretical computer science relying on probabilistic techniques. In a  recent work, \citep{CLOW26} introduced a (possibly weaker) theory called $\APX$. This theory is able to capture a large set of probabilistic arguments using four simple axioms for \emph{approximate counting}. Both  theories extend Cook's theory $\PV$ \citep{Coo75} for deterministic polynomial-time reasoning.

Our main result establishes that, in the relativized setting, $\APX$ does not prove the retraction weak pigeonhole principle ($\rwPHP$). Consequently, $\APC$ is not conservative over $\APX$ with respect to relativized $\forall\Sigma^b_1$ sentences. Moreover, we establish that the unprovability of $\rwPHP$ is part of a more general phenomenon: for a large class of finite combinatorial principles, the existence of an \emph{infinite model} of the negation of the principle \emph{characterizes} its unprovability in $\APX$. This result, which is analogous to a theorem of Riis \citep{Riis94}, implies that $\PV$ and $\APX$ prove exactly the same finite combinatorial principles of this form.
\end{abstract}

\newpage 

\newgeometry{left=1in,right=1in,top=1in,bottom=1in}

\tableofcontents

\newpage 
\pagenumbering{arabic}

\section{Introduction}
\subsection{Context and Motivation}\label{sec:intro_motivation}

Randomness and probability theory have shaped many central developments in
theoretical computer science. Probabilistic techniques underpin the
design and analysis of efficient algorithms~\citep{MotwaniRaghavan95},
establish the existence of combinatorial objects such as expanders
and error-correcting codes~\citep{Shannon48, HooryLinialWigderson06, AlonSpencer},
and enable advances in cryptography~\citep{GoldwasserMicali84} and
proof verification~\citep{AroraSafra98}. They also provide powerful
tools for establishing computational limitations, as illustrated 
by the use of random restrictions to prove lower bounds on the size of Boolean circuits~\citep{DBLP:journals/mst/FurstSS84, DBLP:conf/stoc/Hastad86}.

These applications raise a natural foundational question: which axioms
suffice to formalize probabilistic reasoning? In a seminal work, \Jerabek~\citep{Jerabek07} introduced the bounded arithmetic theory $\APC$ and developed a robust framework for formalizing results from theoretical computer science that rely on probabilistic techniques. For nearly two decades, $\APC$ and its extensions have provided a standard setting for such formalizations. Its applications range from reasoning about randomized algorithms~\citep{Jerabek07, LeCook12}, cryptography \citep{DBLP:conf/stoc/HsiehJLM26}, and learning \citep{DBLP:journals/jml/PichS26},  to proofs of a variety of circuit size lower bounds~\citep{DBLP:journals/apal/MullerP20}. 

While the axioms of $\APC$ \emph{suffice} to prove these and many other results, it is unclear whether they are \emph{necessary}. Understanding this question is particularly important in light of recent developments in logic, complexity theory, and cryptography. On the one hand, there is a growing body of work on the unprovability of conjectures of interest in bounded arithmetic theories. Here, meaningful results arise when a theory $\mathsf{T}$ formalizes a broad range of proofs yet remains amenable to techniques for establishing unprovability results (see, e.g.,~\citep{LO23, Oliveira25, DBLP:journals/jml/AtseriasBM25} and references therein). On the other hand, recent work has revealed unexpected connections between unprovability results and the construction of advanced cryptographic primitives \citep{DBLP:conf/focs/Ilango25, DBLP:conf/stoc/HsiehJLM26, HJLM26}. Weakening the theory underlying the unprovability assumption strengthens these results and makes the security of the resulting cryptographic primitives more plausible.

Recently, Chen, Li, Oliveira, and Williams~\citep{CLOW26} investigated
whether probabilistic arguments from theoretical computer science
could be formalized using weaker axioms. They introduced the theory
$\APX$ as a base theory for feasible probabilistic reasoning and
showed that it supports formalizations of several influential results from algorithms and complexity theory,
including the lower bound
$\mathsf{PARITY}\notin\AC^0$~\citep{DBLP:journals/mst/FurstSS84},
the correctness of the BLR linearity test~\citep{BLR93},
and the Schwartz-Zippel lemma~\citep{Schwartz80,Zippel79}. We refer to their paper for further applications of $\APX$ to logic and complexity theory.

Both $\APX$ and $\APC$ extend Cook's theory $\PV$ \citep{Coo75}, a first-order theory of polynomial-time reasoning whose language contains function symbols for polynomial-time algorithms. While $\APC$ relies on the \emph{dual weak pigeonhole principle} ($\dwPHP$), which asserts that a function from a finite set to a set twice as large cannot be surjective, $\APX$ builds its probabilistic toolkit from four natural axioms for a fresh \emph{approximate counting} function symbol $\P$.  \citep{CLOW26} showed that $\APC$ simulates $\APX$ for statements in the language of $\PV$, and that $\APX$ does not prove $\dwPHP$ under  cryptographic and complexity-theoretic assumptions. Thus, under plausible computational assumptions, $\APX$ is strictly weaker than $\APC$, at least for statements of higher quantifier complexity such as $\dwPHP$.

\medskip 

\begin{remarkn}{Theory $\APX$ \citep{CLOW26}}
This theory extends $\PV(\P)$, which allows polynomial-time computations to query an approximate counting function $\P(C,\Delta)$. Its intended output is an estimate of the acceptance probability of a Boolean circuit $C$ under a random input, with additive error $\leq 1/\|\Delta\|$, where $\|\Delta\|$ denotes bit length. Its four axioms postulate that estimates lie in $[0,1]$, agree with direct evaluation for circuits that do not read their inputs, approximately agree across precision parameters, and approximately respect averaging over the two restrictions obtained by fixing one input bit of $C$. See \Cref{sec:theories} for the formal definition.
\end{remarkn}

\medskip

In this work, we investigate the gap between $\PV$, $\APX$, and $\APC$, and the strength of $\APX$ and its extensions with respect to the provability of combinatorial principles of lower quantifier complexity. Specifically, we are interested in the provability of $\forall \Sigma^b_1(\PV)$ sentences, i.e., statements of the form\footnote{To avoid confusion with the absolute value $|\cdot|$, in this paper we use $\|x\|$ to denote the bit-length of a binary string $x$.}
\begin{equation}\label{eq:total}
\forall x\, \exists y \leq \poly(\|x\|)\; \phi(x,y),
\end{equation}
where $\phi(x,y)$ is a quantifier-free formula in the language of $\PV$. Such statements  express the totality of search problems in $\TFNP$ (see, e.g., \citep[Chapter~7]{Krajicek-book} and \citep[Section~10.3]{Papa}), an active research area with several connections to algorithms and complexity theory (see, e.g., \citep{DBLP:journals/sigact/RezendeGR22, Thapen25}). We ask which such problems are provably total in $\APX$, and how they compare with those of $\PV$ and $\APC$.

This investigation is also motivated by the desire to understand the power of randomness in proofs. While the computational power of randomness is central to the $\P$ versus $\BPP$ question, asking whether $\APX$ proves more combinatorial principles than $\PV$ provides an analogous question in proof complexity. In other words, it asks whether access to a framework for probabilistic reasoning yields existence proofs beyond those available in the underlying theory of deterministic computation.

In the context of $\APX$ versus $\APC$, a result of Thapen~\citep{thapen2002weak} (see \citep[Proposition~1.14]{Jerabek04}) implies that $\APX$ and $\APC$ have the same $\forall\Sigma^b_1(\PV)$ consequences if and only if $\APX$ proves the \emph{retraction weak pigeonhole principle} ($\rwPHP$). This combinatorial   principle asserts the totality of the search problem $\LossyCode$, which has attracted recent attention in complexity theory~\citep{Korten22,LPT24, CLO24, DBLP:conf/stoc/LiLR26, FlemingG00RS026}. Given Boolean circuits $\operatorname{Enc} \colon \zo^n\to\zo^{n-1}$ and $\operatorname{Dec} \colon \zo^{n-1}\to\zo^n$, the task is to find $x$ such that $\operatorname{Dec}(\operatorname{Enc}(x))\neq x$. Thus a solution witnesses the failure of a proposed compression scheme to preserve every input. 

Whether $\APX$ proves $\rwPHP$ was one of the main questions left open in~\citep{CLOW26}. In this direction, they showed that every $\TFNP$ problem provably total in $\APX$ reduces in polynomial time to a problem called $\RefYao$. Although a reduction from $\RefYao$ to $\LossyCode$ was known, the converse remained open. Moreover, $\APX$ was not shown to prove the totality of $\RefYao$ (i.e.,~\Cref{eq:total} for the relevant formula $\phi$), leaving open whether this computational problem is harder than the provably total search problems of $\APX$.

\subsection{Results}\label{sec:intro_results}

Our contributions can be summarised as follows:
\begin{enumerate}
    \item First, we show that $\RefYao$ and $\LossyCode$ (corresponding to the principle $\rwPHP$) are equivalent under polynomial-time reductions, and that their totality statements are equivalent over $\PV$. We also describe a complete  problem for the $\TFNP$ consequences of $\APX$ by applying the framework of Thapen~\citep{Thapen25}. This problem, called $\Local(\APX)$, reduces to $\RefYao$ and $\LossyCode$.
    \item Then, we investigate the provability of $\rwPHP$ and other pigeonhole principles in $\APX$ and its extensions:
    \begin{itemize}
        \item We obtain unconditional results in the relativized setting, showing that $\dwPHP$ and the weak pigeonhole principle $\wPHP$ are not provable in $\APX$. These unprovability results extend to the true universal theory $\trueAPX$. In contrast, $\rwPHP$ is provable in $\trueAPX$.
        \item Through a substantially more involved argument, we show that $\rwPHP$ is not provable in relativized $\APX$. Thus, in the relativized setting, $\APC$ is not $\forall\Sigma^b_1$-conservative over $\APX$. 
        \item Complementing these results, we show that $\wPHP$ and $\dwPHP$ are provable in a multiplicative extension $\mAPX$ of $\APX$, which allows approximate counting with relative rather than additive precision. Together with the preceding lower bounds, this separates $\APX$ from $\mAPX$ in the relativized setting.
    \end{itemize}
    \item Finally, we establish that the unprovability of $\rwPHP$ in relativized $\APX$ is part of a more general phenomenon: for the class of finite combinatorial principles specified in \Cref{sec:intro_results_riis}, the existence of an \emph{infinite model} of the negation of a principle \emph{characterizes} its unprovability in $\APX$. This result, analogous to a theorem of Riis \citep{Riis94}, applies to $\rwPHP$ and to many other natural combinatorial principles. Furthermore, it allows us to show that, in the relativized setting, $\PV$ and $\APX$ prove exactly the same finite combinatorial principles of this form.
\end{enumerate}

In the following sections, we explain our results in detail.

\subsubsection{LossyCode, Refuter(Yao), and  Total Search Problems of \texorpdfstring{$\APX$}{APX1}}

We first recall the search problem $\RefYao$ introduced
in~\citep{CLOW26}. An instance consists of a circuit $G \colon (\{0,1\}^n)^m \to [n] \times \{0,1\}^s$, called a
\emph{predictor generator}, that takes a list $\mathcal D$ of $m$
strings in $\zo^n$ and outputs an index $i\in\{1,\ldots,n\}$
and a circuit $P\colon\zo^{i-1}\to\zo$ of description length
at most $s$. For an advantage parameter $\delta>0$, the task is
to find a list $\mathcal D$ such that,
writing $(i,P)=G(\mathcal D)$,
$$
\Pr_{x\gets\mathcal D}\bigl[P(x_{<i})=x_i\bigr]
<\frac12+\delta,
$$
where $x$ is a uniformly chosen entry of the list and $x_{<i}$
denotes its first $i-1$ bits. Thus a solution refutes the claim
that $G$ always produces a predictor with advantage at least
$\delta$ for the distribution it receives. The parameters satisfy
$(\delta^2/10)m\geq s+\lceil\log n\rceil+1$, which guarantees
totality. Our first result identifies the complexity of this
problem with that of $\LossyCode$.

\begin{theorem}[Equivalence of $\RefYao$ and $\LossyCode$]
\label{thm:intro_lossy_refyao}
The problems $\RefYao$ and $\LossyCode$ are equivalent under
deterministic polynomial-time mapping reductions. Moreover, their totality
statements are equivalent over $\PV$.
\end{theorem}

This equivalence does not settle whether $\APX$ proves the
totality of $\RefYao$. As recalled above, every $\TFNP$ problem
provably total in $\APX$ reduces in polynomial time to $\RefYao$ \citep{CLOW26},
but this witnessing result does not establish that $\RefYao$
itself is provably total in $\APX$.

To obtain a problem that captures precisely the $\TFNP$
consequences of $\APX$, we apply Thapen's framework~\citep{Thapen25}
and define $\Local(\APX)$. An instance is a sequence of circuits
describing an inspector's strategy for a finite interaction with
a purported approximate counting oracle. In each round, the
inspector uses the previous answers to choose a query
$(C,\Delta)$, and the oracle responds with a rational number of
the prescribed bit-length. A solution is a transcript of answers
that agree on repeated queries and violate no counting axiom
whose oracle values are all present in the transcript.
The inspector therefore fails to expose an inconsistency.
Both the length of a transcript and the time needed to verify
it are polynomial in the size of the instance. We refer to \Cref{sec: witnessing} for the formal definition  of $\Local(\APX)$.

\begin{theorem}[A complete problem for the $\TFNP$ consequences of $\APX$]
\label{thm:intro_local_apx}
The theory $\APX$ proves the totality of $\Local(\APX)$.
Moreover, every $\TFNP$ problem provably total in $\APX$
reduces to $\Local(\APX)$ by a deterministic  polynomial-time mapping reduction.
\end{theorem}

Writing $\TFNP(\APX)$ for the $\TFNP$ problems provably total in $\APX$,
we can summarize these results as follows:
$$
\TFNP(\APX)
\;\overset{
  \text{\Cref{thm:intro_local_apx}}\rule[-0.6ex]{0pt}{0.6ex}
}{\equiv_p}\;
\Local(\APX)
\;\;\;\leq_p\;\;\;
\bigl(
\RefYao
\;\overset{
  \text{\Cref{thm:intro_lossy_refyao}}\rule[-0.6ex]{0pt}{0.6ex}
}{\equiv_p}\;
\LossyCode
\bigr).
$$
Here the relations $\leq_p$ and $\equiv_p$ refer to deterministic polynomial-time mapping reductions. The reduction from $\Local(\APX)$ to $\RefYao$ follows
by applying the witnessing theorem of~\citep{CLOW26} to the totality
of $\Local(\APX)$.\footnote{The result stated in \citep{CLOW26} provides a Turing reduction, but one can check that a mapping reduction suffices. In fact, it is not hard to show that every $\TFNP$ problem
that admits a deterministic polynomial-time Turing reduction
to $\RefYao$ also admits a deterministic polynomial-time
mapping reduction to $\RefYao$.} We prove the two theorems above
in~\Cref{sec:TFNP-APX1}.

These results leave open whether $\APX \vdash^?\rwPHP$ and
whether the reverse reduction from $\LossyCode$ to $\Local(\APX)$
exists, making all three problems equivalent under polynomial-time reductions. A proof of $\rwPHP$ in $\APX$ would
imply such a reduction by the aforementioned witnessing theorem.

Establishing the unprovability of $\rwPHP$ in $\APX$ under
computational assumptions appears difficult. 
A standard approach is to use witnessing theorems to extract an
algorithm from a hypothetical proof and then contradict a
hardness assumption. For example, randomized witnessing rules
out the provability of $\wPHP$ in $\APX$ assuming the existence
of collision resistant hash functions. A more sophisticated witnessing
argument from \citep{ILW23} rules out $\dwPHP$ under
sub-exponentially secure indistinguishability obfuscation together
with the assumption that $\mathsf{coNP}$ is not contained
infinitely often in $\NP/\mathsf{poly}$ \citep{CLOW26}. However, randomized
witnessing applied to a hypothetical proof of $\rwPHP$ would
yield a randomized algorithm for $\LossyCode$ running in expected
polynomial time, and such an algorithm already exists. Indeed,
at most $2^{n-1}$ strings $x\in\zo^n$ can satisfy
$\operatorname{Dec}(\operatorname{Enc}(x))=x$, since
$\operatorname{Enc}$ is injective on the set of correctly decoded
strings. A uniformly random $x$ is therefore a solution with
probability at least $1/2$, and repeated sampling with verification
finds and confirms a solution in expected polynomial time. Consequently,
the existence of the extracted algorithm provides no
contradiction.

This limitation motivates our investigation of unconditional
unprovability results in the relativized setting, to which we
turn next.

\subsubsection{Approximate Counting and Weak Pigeonhole Principles}\label{sec:weak-PHP-in-APX}

Before stating our next results, we explain how $\rwPHP$ and
$\APX$ can be relativized in a natural way.

\medskip

\begin{remarkn}{Relativizations of $\APX$ and $\rwPHP$}
    Relativization replaces the explicit compression and decompression
circuits in $\rwPHP$ by fresh function symbols $f$ and $g$. The relativized retraction principle asserts that a map $f$
from $[2N]$ to $[N]$ cannot have a left inverse $g$, where
$[N]=\{0,\ldots,N-1\}$, and we typically think of $N = 2^n$. This is captured by the sentence
$$
\rwPHP(f,g) \;\eqdef\; \forall N\;\bigl(
N>1\;\rightarrow\;
\exists x<2N\;
\bigl(f(x)\geq N\;\lor\;g(f(x))\neq x\bigr)
\bigr).
$$
It is also possible to formulate the principle using binary strings, in which case the oracle $f$ maps $n$ bits to $n-1$ bits,
the oracle $g$ maps $n-1$ bits to $n$ bits, and the principle
asserts that some $x\in\zo^n$ satisfies $g(f(x))\neq x$. The chosen formulation is inessential in our results.

The theory $\APX(f,g)$ allows these same oracles both in the
underlying polynomial-time computations and in the Boolean
circuits submitted to $\P$. Thus $\P(C,\Delta)$ is intended
to approximate $\Pr_z[C^{f,g}(z)=1]$ with additive error at most
$1/\|\Delta\|$. Formally, the theory extends $\PV(\P,f,g)$
by the four approximate counting axioms, now applied to circuits
with oracle gates for $f$ and $g$. We refer to \Cref{sec:theories} for the details.
\end{remarkn}

\medskip

We obtain the following unconditional result.\footnote{In a concurrent work, Chen, Jain, Li, and Roy \cite{CJLR26} showed that under $\mathsf{prBPP}=\mathsf{prP}$, LWE, and an assumption related to the unprovability of (unrelativized) $\rwPHP$ from $\APX$ augmented with non-logical axioms ($\mathsf{prBPP}=\mathsf{prP}$ and LWE), there is a new construction of SNARGs for $\NP$. \Cref{thm:intro_retraction} provides evidence for $\APX\nvdash\rwPHP$ and shows that the assumption from \cite{CJLR26} is unlikely to be trivially false.}

\begin{theorem}[Unprovability of the retraction weak pigeonhole principle]
\label{thm:intro_retraction}
$$
\APX(f,g)\nvdash\rwPHP(f,g).
$$
\end{theorem}

This theorem identifies a limitation of the approximate
counting axioms. They support an internally consistent notion
of size, but do not suffice to establish the expected lower
bound on the size of the set of inputs on which decoding
fails. Indeed, although this set contains at least half of
all inputs in every standard interpretation, the theory cannot
prove that it is non-empty for every input length. At the same
time, the formalizations in~\citep{CLOW26} show that these axioms support
many substantial probabilistic arguments from algorithms and
complexity theory.

We derive \Cref{thm:intro_retraction} from a more general
result about finite combinatorial principles, discussed in
the next subsection.

Both $\wPHP$ and $\dwPHP$ imply $\rwPHP$ already over $\PV$,
and the implications hold with oracle functions as well. This is because a collision in $f$ forces decoding to fail on at least one of
the two colliding inputs, while a point outside the range of
$g$ cannot be decoded correctly. It is therefore useful to
study the (un)provability of these stronger principles in $\APX$
and its extensions. Their unprovability proofs provide a
starting point for our argument, while also revealing why
the retraction principle requires additional ideas.

For this comparison, we introduce two extensions of $\APX$
and their natural relativizations. The first one is the true universal theory of $\APX$, denoted $\trueAPX$.

\medskip

\begin{remarkn}{True relativized universal theories $\trueAPX$}
The theory $\trueAPX(f,g)$ extends $\PV(\P,f,g)$ by every
universal sentence that holds in every standard interpretation
of these symbols. Here a universal sentence is the universal
closure of a quantifier-free formula. In a standard
interpretation, $f$ and $g$ are arbitrary functions of the
prescribed lengths, and $\P$ is any correct additive approximate
counting oracle for circuits with access to $f$ and $g$. The remaining function
symbols are interpreted by their defining oracle algorithms.
The four axioms of $\APX(f,g)$
are among these universal truths, so $\trueAPX(f,g)$ extends
$\APX(f,g)$. We define $\trueAPX(f)$ and $\trueAPX(g)$
similarly. See \Cref{def:standard_model} and \Cref{def:true_universal}
for the precise definitions.
\end{remarkn}

\medskip

We also introduce a natural variant of $\APX$, denoted by $\mAPX$, designed to support a stronger form of approximate counting.

\medskip

\begin{remarkn}{Multiplicative approximate counting and $\mAPX$}
The theory $\mAPX$ strengthens the precision and local
consistency axioms so that the intended counting estimates
have relative error. More precisely, if $p(C)$ is the true
acceptance probability of $C$, the intended guarantee is
$$
(1-\delta)\,p(C)
\leq \P(C,\Delta)
\leq (1+\delta)\,p(C),
~~ \text{where} ~~ \delta=\frac{1}{\|\Delta\|}.
$$
This gives useful precision even when $p(C)$ is very small.
The resulting axioms imply those of $\APX$. 
\end{remarkn}

\medskip

The relativized theories
$\mAPX(f)$, $\mAPX(g)$, and $\mAPX(f,g)$ are defined by
allowing the corresponding oracle gates in counting queries
and oracle calls in the underlying computations, just as
for $\APX$.

We obtain the following positive and negative results.

\begin{theorem}[(Un)provability of pigeonhole principles in extensions of $\APX$]
\label{thm:intro_pigeon}
The following statements hold:
\begin{enumerate}
    \item $\trueAPX(f,g)\vdash\rwPHP(f,g)$.
    \item $\trueAPX(g)\nvdash\dwPHP(g)$.
    \item $\trueAPX(f)\nvdash\wPHP(f)$.
    \item $\mAPX(g)\vdash\dwPHP(g)$.
    In particular, $\mAPX\vdash\dwPHP$.
    \item $\mAPX(f)\vdash\wPHP(f)$.
    In particular, $\mAPX\vdash\wPHP$.
\end{enumerate}
\end{theorem}

The first item of \Cref{thm:intro_pigeon} highlights a distinction essential to
\Cref{thm:intro_retraction}: adding true universal axioms
already suffices to prove $\rwPHP$. To see why, consider
the oracle circuit $C^{f,g}$ that accepts precisely when $g(f(x))\neq x$.
Its acceptance probability is at least $1/2$, and the
corresponding lower bound on its approximate count is a
true universal statement. With this estimate available, $\APX(f,g)$ can prove the existence of an input $x$ on which
decoding fails. In contrast, the unprovability results for $\wPHP$
and $\dwPHP$ in the second and third items of \Cref{thm:intro_pigeon} remain valid even after all true universal
axioms are added.

A consequence is that, in the case of $\wPHP$ and $\dwPHP$, the corresponding lower bound argument can provide genuine approximate counting answers
as the adversary constructs the oracle showing unprovability. On the other hand, our adversary for $\rwPHP$
must preserve the axiom instances exposed during
the interaction without requiring the counting answers
to be accurate. Exploiting the distinction between the approximate counting axioms and their intended interpretation is a central difficulty
in establishing \Cref{thm:intro_retraction}.

The remaining items of \Cref{thm:intro_pigeon} separate additive and multiplicative
approximate counting in the relativized setting.
The theory $\mAPX(f)$ proves $\wPHP(f)$, which is unprovable
even in $\trueAPX(f)$, and $\mAPX(g)$ proves $\dwPHP(g)$,
which is unprovable even in $\trueAPX(g)$. Since the
multiplicative theory extends the  additive
theory, it follows that in the relativized setting, $\mAPX$ is strictly stronger than $\APX$. We refer to \Cref{sec:upper_bounds_pigeons} for some computational consequences of \Cref{thm:intro_pigeon}   that might be of independent interest.

The more delicate argument for $\rwPHP$ is part of a general
connection between unprovability in $\APX$ and the existence
of infinite models. We turn to this connection next.

\subsubsection{Finite Combinatorial Principles, Infinite Models, and the Limits of \texorpdfstring{$\APX$}{APX1}}\label{sec:intro_results_riis}

Our next results connect the provability of finite combinatorial
principles with their behavior on infinite structures. For a natural
class of principles that hold in every non-empty finite structure,
we show that their finite formulations are unprovable in relativized
$\APX$ exactly when they admit an infinite counterexample. When no
such counterexample exists, the finite formulation is already
provable in the corresponding relativization of $\PV$. Thus, for
this class of principles, the approximate counting axioms yield
no additional proving power.

For example, an infinite counterexample to $\rwPHP$ is obtained
on $\mathbb N$ by defining
$$
f_0(x)=2x,\qquad
f_1(x)=2x+1,\qquad
h(y)=\lfloor y/2\rfloor.
$$
The functions $f_0$ and $f_1$ encode two disjoint copies of
$\mathbb N$ into a single copy: one uses the even numbers,
and the other uses the odd numbers. Decoding recovers the
original copy from the parity of the encoded value and the
original element by applying $h$. Thus every input is recovered
correctly. This gives an infinite counterexample to the
retraction principle.

To formally state the results of this section, we first need to introduce the relevant terminology.

\paragraph{$\mathcal{L}$-Principles, $\widehat{\mathcal{L}}$-Translations, and $\APX(\widehat{\mathcal L})$.}
Let $\mathcal L$ be a non-empty finite language consisting of function
and constant symbols and disjoint from the language of $\APX$.
Let $\mathcal C$ be its set of constant symbols and $\mathcal F$
its set of function symbols. 

In order to discuss the provability of finite combinatorial principles expressed as $\mathcal{L}$-statements but restricted to a domain $[N]$, it will be convenient to replace each function
symbol $f\in\mathcal F$ of arity $d$ by a symbol
$\widehat f(s,x_1,\ldots,x_d)$, and each constant $c\in\mathcal C$
by a unary symbol $\widehat c(s)$. Write $\widehat{\mathcal L}$
for the resulting language. Now define the following formula over the language $\widehat{\mathcal L} \cup \{<\}$:
$$
\operatorname{Closed}_{\widehat{\mathcal L}}(N)
\eqdef
\bigwedge_{c\in\mathcal C}\widehat c(N)<N
\;\land\;
\bigwedge_{f\in\mathcal F}
\forall u_1,\ldots,u_{\operatorname{arity}(f)}<N\;
\widehat f(N,u_1,\ldots,u_{\operatorname{arity}(f)})<N,
$$
where $<$ is the $\PV$  symbol corresponding to the order relation $<$.
This formula will be useful when we consider
$\mathcal L$-structures on $[N]=\{0,\ldots,N-1\}$.

Let $\APX(\widehat{\mathcal L})$ denote the relativized theory
in which the symbols of $\widehat{\mathcal L}$ are available as
oracles both to polynomial-time computations and to the circuits
submitted to the approximate counting oracle $\P(C,\Delta)$, with the length axioms  
$$
\|\widehat f(s,\bar x)\|\leq\|s\| \quad \text{and} \quad
\|\widehat c(s)\|\leq\|s\|
$$
included before admitting derived
functions in the Cobham's style definition of relativized $\PV(\P, \widehat{\mathcal{L}})$. The notation $\PV(\widehat{\mathcal L})$ uses the
same parameterized oracles and length axioms, with the counting
oracle omitted. 

Finally, for an $\mathcal L$-formula $\psi$, write
$\psi^{[N]}$ for its \emph{finitary translation at size $N$} (where $N$ is a fresh variable), which is defined recursively by
\begin{align*}
& x^{[N]}\eqdef x,
\qquad
c^{[N]}\eqdef\widehat c(N),
\qquad
\bigl(f(t_1,\ldots,t_d)\bigr)^{[N]}
\eqdef
\widehat f(N,t_1^{[N]},\ldots,t_d^{[N]}), \\ 
& \bigl(\forall x~\phi\bigr)^{[N]}\eqdef \forall x < N~\phi^{[N]},
\qquad 
\bigl(\exists x~\phi\bigr)^{[N]}\eqdef \exists x < N~\phi^{[N]}.
\end{align*}
and commutes with equality and Boolean connectives.
All references below to infinite structures
concern the original language $\mathcal L$.\\

We are now ready to state the main result of this section. In the statement below, we think of a tuple
$(x_1,\ldots,x_k)$ as a potential solution to the search
problem encoded by $\phi$. 

\begin{restatable}[Riis-style dichotomy for relativized $\APX$]
{theorem}{RiisThm}\label{thm:intro-riis}
Let $\phi(x_1,\ldots,x_k)$ be a formula in
the language $\mathcal L$ with equality.
If \;$\forall x_1\cdots\forall x_k\;\neg\phi(x_1,\ldots,x_k)$
has an infinite model, then
$$
\APX(\widehat{\mathcal L})\nvdash
\forall N\geq1\;
\left(
\operatorname{Closed}_{\widehat{\mathcal L}}(N)
\;\rightarrow\;
\exists x_1,\ldots,x_k<N\;
\phi^{[N]}(x_1,\ldots,x_k)
\right).
$$
\end{restatable}

\Cref{thm:intro-riis} is an analogue of Riis' finitisation
theorems~\citep{Riis94} (see also \citep{DBLP:journals/apal/Muller21} for a more recent related work). For sentences expressed in an
additional relational language, Riis showed that 
provability of their finite restrictions in the relativized
theories $\mathsf{T}^1_2$ or $\mathsf{S}^2_2$ is equivalent
to validity in all structures for that relational language.
Crucially, the proof of the finite restriction may use all the
axioms of the ambient arithmetic theory, whereas an infinite
counterexample need only interpret the additional combinatorial
symbols, i.e., it need not carry any arithmetic structure or satisfy
the axioms of the theory. Thus a purely combinatorial infinite
counterexample can obstruct an arithmetic proof. \Cref{thm:intro-riis} establishes an obstruction of the same form in the context of relativized
$\APX$.
 Note that this phenomenon does not extend to relativized
$\APC$, which proves $\rwPHP$ despite the existence of an infinite
counterexample.

\begin{remarkn}{Example: The weak pigeonhole principle}
A map $f\colon[2N]\to[N]$ can be represented by two maps
$f_0,f_1\colon[N]\to[N]$, defined by
$f_0(x)=f(x)$ and $f_1(x)=f(N+x)$.
Conversely, any such pair determines a map $f\colon[2N]\to[N]$.
A collision in $f$ therefore corresponds either to a collision
within one of $f_0,f_1$, or to an intersection of their ranges.
To capture the relativized weak pigeonhole principle, let
$\mathcal L=\{f_0,f_1\}$ consist of two unary function symbols,
and define
$$
\phi(x,y)\eqdef
\left(
x\neq y
\;\land\;
\bigl(f_0(x)=f_0(y)\lor f_1(x)=f_1(y)\bigr)
\right)
\;\lor\;
f_0(x)=f_1(y).
$$
The sentence $\exists x,y\;\phi(x,y)$ asserts that $f_0$ and $f_1$
cannot both be injective with disjoint ranges. It holds in every
non-empty finite structure. However, it fails in the infinite
structure with universe $\mathbb N$ and interpretations
$f_0(x)=2x$ and $f_1(x)=2x+1$.
Thus \Cref{thm:intro-riis} gives
$$
\APX(\widehat{\mathcal L})\nvdash
\forall N\geq1\;
\left(
\operatorname{Closed}_{\widehat{\mathcal L}}(N)
\;\rightarrow\;
\exists x,y<N\;\phi^{[N]}(x,y)
\right).
$$
Under the encoding above, the slices
$\widehat f_0(N,\cdot)$ and $\widehat f_1(N,\cdot)$ represent
the two halves of a map $f\colon[2N]\to[N]$.
The displayed sentence is therefore the relativized weak
pigeonhole principle asserting that this map has a collision.
\end{remarkn}

In the next example, we explain how a principle that relies on a
relation symbol, in addition to function and constant symbols,
can be reformulated by replacing it with a function symbol.%

\medskip

\begin{remarkn}{Example: The Herbrandized ordering principle
\citep{DBLP:journals/tocl/AtseriasT14}}
The Herbrandized ordering principle ($\mathsf{HOP}$) asserts that
a finite strict linear order cannot have a total immediate
predecessor function: at a least element, such a function must fail.
In the vocabulary $\alpha=(\prec,h)$ of
\citep{DBLP:journals/tocl/AtseriasT14}, where $\prec$ is a binary
relation symbol and $h$ is a unary function symbol, its natural
formulation for $h\colon[N]\to[N]$ is
$$
\operatorname{LinOrd}_{\prec}(N)
\;\rightarrow\;
\exists x<N\;
\left(
\neg(h(x)\prec x)
\;\lor\;
\exists y<N\;(h(x)\prec y\land y\prec x)
\right),
$$
where $\operatorname{LinOrd}_{\prec}(N)$ asserts that $\prec$
is a strict linear order on $[N]$. Thus a solution witnesses
either a violation of the order axioms or a failure of $h(x)$
to be the immediate predecessor of $x$.

To express this principle in our framework, let
$\mathcal L=\{h,r,c_T\}$, where $h$ is a unary function,
$r$ is a binary function, and $c_T$ is a constant.
Abbreviate $r(x,y)=c_T$ by $x\prec y$. Now define
$$
\begin{aligned}
\phi(x,y,z)\eqdef {}&
x\prec x
\lor \bigl(x\neq y\land\neg(x\prec y)\land\neg(y\prec x)\bigr)\\
&\lor \bigl(x\prec y\land y\prec z\land\neg(x\prec z)\bigr)\\
&\lor \neg(h(x)\prec x)
\lor \bigl(h(x)\prec y\land y\prec x\bigr).
\end{aligned}
$$
The first three disjuncts witness a failure of $\prec$ to be a
strict linear order, while the last two express the remaining
conditions from $\mathsf{HOP}$. Hence
$\exists x,y,z\;\phi(x,y,z)$ holds in every non-empty finite
$\mathcal L$-structure. In the formula translation, the
abbreviation $x\prec y$ becomes
$\widehat r(N,x,y)=\widehat c_T(N)$.

An infinite counterexample is obtained on $\mathbb N$ by taking
$c_T=0$, $h(x)=x+1$, and $r(x,y)=0$ if $x>y$, and $1$ otherwise.
Then $\prec$ is the reverse of the usual order and $h(x)$ is
always the immediate predecessor of $x$. Consequently,
$$
\APX(\widehat{\mathcal L})\nvdash
\forall N\geq1\;
\left(
\operatorname{Closed}_{\widehat{\mathcal L}}(N)
\;\rightarrow\;
\exists x,y,z<N\;\phi^{[N]}(x,y,z)
\right).
$$
\end{remarkn}

\medskip

\Cref{thm:intro-riis} clarifies the weakness of the approximate counting
axioms of $\APX$. While they support a broad range of probabilistic
arguments \citep{CLOW26}, 
$\APX$ cannot distinguish the finite from the infinite. 

As a consequence of \Cref{thm:intro-riis}, for combinatorial
principles of the above form that hold in every non-empty finite
$\mathcal L$-structure, provability of their
translations in $\APX(\widehat{\mathcal L})$ is \emph{equivalent} to
the absence of an infinite counterexample. Indeed, one obtains
the following characterization.

\begin{restatable}[Finite relativized combinatorial principles
provable in $\APX$]{corollary}{CorFinite}\label{cor:intro-riis}
Let $\phi(x_1,\ldots,x_k)$ be any formula in
$\mathcal L$ with equality, and suppose that
$\exists\bar x\,\phi(\bar x)$ holds in every non-empty finite
$\mathcal L$-structure. Write
$$
\Theta_\phi\eqdef
\forall N\geq1\;
\left(
\operatorname{Closed}_{\widehat{\mathcal L}}(N)
\;\rightarrow\;
\exists x_1,\ldots,x_k<N\;
\phi^{[N]}(x_1,\ldots,x_k)
\right).
$$
Then the following are equivalent:
\begin{enumerate}
    \item $\APX(\widehat{\mathcal L})\vdash\Theta_\phi$.
    \item $\PV(\widehat{\mathcal L})\vdash\Theta_\phi$.
    \item $\exists\bar x\,\phi(\bar x)$ holds in every infinite
    $\mathcal L$-structure.
\end{enumerate}
\end{restatable}

\Cref{cor:intro-riis} indicates that in terms of the provability of a large class of  combinatorial principles, $\APX$ behaves more like its deterministic base theory $\PV$ than like  $\APC$.

\subsection{Techniques} \label{sec:intro_techniques}

We now explain the main ideas  underlying our most technical results (\Cref{thm:intro_retraction} and \Cref{thm:intro-riis}). As a warm-up, we provide an overview of the unprovability of $\dwPHP$ in $\trueAPX$ in the relativized setting (\Cref{thm:intro_pigeon} Item 2). 

\begin{paragraph}{Unprovability of $\bm{\dwPHP}$ (\Cref{thm:intro_pigeon} Item 2).}
  Assume towards a contradiction that $\trueAPX$ proves $\dwPHP$ in the relativized setting. Applying the KPT witnessing theorem~\cite{KrajicekPT91}, we obtain a winning strategy for the ``inspector'' in the following ``$\dwPHP$ game''. The game is defined in terms of functions $g \colon [N] \to [2N]$ and a finite number of true universal sentences $\Psi_1, \Psi_2, \dots, \Psi_\ell$ (i.e., universal sentences in the language of $\PV(g,\P)$ that hold in every interpretation of $g$ and any valid interpretation of the approximate counting oracle $\P$), and consists of $m = \poly\log(N)$ rounds. In each round, the inspector can either make a query to $g$, make a query to $\P$, or make an inversion query. Here, a query to $\P$ consists of an oracle circuit $C^{(-)}$ of size at most $s = \poly\log(N)$, and the adversary must return a value $p \in [0, 1]$ as the purported value of $\Pr_{z}[C^g(z) = 1]$ within additive error $\eps = 1/\poly\log(N)$. An inversion query consists of an element $y \in [2N]$, and the adversary must return a preimage $x\in [N]$ such that $g(x) = y$; if the adversary is unable to do so, the inspector wins. Finally, if the inspector is able to find a violation of one of the axioms $\Psi_i$, then the inspector also wins.

We will construct a winning strategy for the adversary as long as $m, s, \eps^{-1}$ are sufficiently small compared to $N$ (which leads to the contradiction). If there were no $\P$ queries, then the adversary could win by constructing a partial function $g$ lazily: For each query $g(x)$ such that $x$ is not already in the current domain of $g$, we assign $g(x) = 0 \in [2N]$; for each inversion query $y$, pick a random element $x\in [N]$ outside the current domain of $g$ and assign $g(x) = y$. As long as $m < N$, the domain size of $g$ will be less than $N$ and the adversary will not run out of preimages for inversion queries.

It remains to handle the $\P$ oracle queries. The idea is to fix a small pseudorandom set $\mathcal{S}$ such that for every small circuit $C$ and every ``valid'' $g$,
\[\Pr_{z}[C^g(z) = 1] \approx \Pr_{z\gets \mathcal{S}}[C^g(z) = 1],\]
and return $\Pr_{z\gets \mathcal{S}}[C^g(z) = 1]$ as the estimate. During this process, we need to evaluate $C^g(z)$ for every $z\in \mathcal{S}$. Again, whenever we make a query $g(x)$ such that $x$ is not already in the current domain of $g$, we assign $g(x) = 0$. At first glance, it may appear that this is self-defeating: during the game, the domain size of $g$ is upper bounded by $|\mathcal{S}|\cdot m\cdot s$, so we want a pseudorandom set of size $|\mathcal{S}|$ that is pseudorandom against every $g$ with this domain size; on the other hand, a standard union bound only gives a pseudorandom set of large size that is pseudorandom against functions with a smaller description. However, notice that we lazily assigned $0$ to all $g$-values except in the inversion queries, hence the number of preimages mapped to a non-zero element in $[2N]$ is at most $m$. We say $g$ is ``valid'' if there are at most $m$ preimages that map to a non-zero element. It is then easy to obtain a pseudorandom set of size $|\mathcal{S}| = \poly(m, s, 1/\eps)$ that $\eps$-fools every $C^g$ for small circuit $C$ and valid $g$. As long as the domain size of $g$ is $\poly(m, |\mathcal{S}|) < N$, the adversary wins the game.
\end{paragraph}

\medskip \medskip

We note that in the above proof, the adversary replies with \emph{honest} probabilities in the following sense. If we let $g^{\rm final}$ be the function $g$ at the end of the game, then every $\P(C)$ query is answered with an $\eps$-approximation of $\Pr_z[C^{g^{\rm final}}(z) = 1]$. Hence, these answers will satisfy any universal axiom $\Psi_i$ and the adversary wins the game. Looking ahead, to show the unprovability of $\rwPHP$ in $\APX$, the adversary cannot reply with honest probabilities, as $\trueAPX$ indeed proves $\rwPHP$ (\Cref{thm:intro_pigeon} Item 1).

\begin{paragraph}{Unprovability of $\bm{\rwPHP}$ (\Cref{thm:intro_retraction}).} We now explain at a high level the  unprovability of the retraction weak pigeonhole principle ($\rwPHP$), which asserts for $\PV$ function symbols $f \colon \{0,1\}^{n+1} \to \{0,1\}^n$ and $g \colon \{0,1\}^n \to \{0,1\}^{n+1}$ that there exists $x \in \{0, 1\}^{n+1}$ such that $g(f(x)) \neq x$. The general Riis-style dichotomy follows a very similar argument.

We will establish using witnessing methods that if $\APX$ proves $\rwPHP$ in the relativized setting, then there is a winning strategy for the ``inspector'' in the following ``$\rwPHP$ game'' (we will actually prove a broad generalization of this in \Cref{sec:riis}). The $\rwPHP$ game is defined in terms of functions $f \colon [2N] \to [N]$ and $g \colon [N] \to [2N]$. The game proceeds for $m = \poly\log(N)$ rounds and in each round, the inspector supplies a distribution ${\cal T}_i$ over decision trees $T$ of height $h \leq \poly\log(N)$ which query the functions $f,g$ on various inputs and output a value in $\{0,1\}$. The adversary must reply with a value $p_i \in [0,1]$. At the end of the game, the inspector outputs a value $x \in [2N]$, and wins the game if $g(f(x)) \neq x$. The values $p_i$ are supposed to represent $\Pr_{T \sim {\mathcal{T}}_i}[ T(f,g) = 1]$. However, the adversary is allowed to deviate from this ``honest answer,'' as long as they obey the following ``local consistency'' rules for some error parameter $\epsilon \geq (\poly\log(N))^{-1}$:
\begin{enumerate}
    \item {\it \emph{(}Deterministic Consistency\emph{)}}\; If the inspector supplies a distribution supported on a single decision tree $T_i$, then the adversary must reply with the true probability $\Pr[T_i(f,g)= 1] \in \{0,1\}$.
    \item {\it \emph{(}Mixture Consistency\emph{)}}\; For any three round indices $i_1,i_2,i_3$ such that $\calt_{i_1}=\tfrac12\calt_{i_2}+\tfrac12\calt_{i_3}$, if the adversary responded with values $p_{i_1}, p_{i_2}, p_{i_3}$, it must be that  $\left|p_{i_1}-\tfrac12p_{i_2}-\tfrac12p_{i_3}\right|\leq\eps$.
\end{enumerate}

Our goal is to construct an adversary that always wins when $m, h, \epsilon^{-1} \leq \poly\log(N)$. As discussed above, this lower bound cannot be proven by an adversary that always gives honest answers: In this case the game amounts to a standard randomized decision tree which aims to solve the search problem associated with $\rwPHP$, which is easy using binary search (using the fact that a random $x$ satisfies $g(f(x)) \neq x$ with probability $\geq \frac{1}{2}$). This highlights the subtlety of the lower bound: to construct our adversary, we must answer ``fake probabilities'' in such a way that (1) they preclude the determination of any particular solution to the $\rwPHP$ search problem, and (2) they are indistinguishable from honest probabilities according to the two local consistency rules.

Our strategy will be as follows. Over the course of the game, the adversary will maintain a partial matching $\rho_i$ after round $i$ has been completed. When the next tree distribution ${\cal T}_i$ is supplied, the adversary will take $s$ independent samples $T_{i,1},\ldots,T_{i,s} \sim {\cal T}_i$ (for some parameter $s = \poly(h,m,\epsilon^{-1})$ to be determined later). It then runs over each of the trees $T_{i,j}$ in order and does the following. Starting at the root of $T_{i,j}$, if the tree queries $f(x)$ and $x$ is matched in $\rho_i$, we traverse the tree as if the answer to the query was the mate of $x$ in $\rho_i$. If $x$ is unmatched in $\rho_i$, we choose a uniformly random unmatched element $y\in [N]$, extend $\rho_i$ to add the pair $(x,y)$, and traverse the tree according to $f(x)=y$. We perform the symmetric operation for queries of the form $g(y)$. We can always perform an appropriate move provided $|\rho_i| + sh \leq N$ which we will make sure to maintain throughout the game. When all trees $T_{i,1},\ldots,T_{i,s} $ have been traversed in order, we reach a leaf in each tree and hence determine the corresponding output bits $b_{i,1},\ldots,b_{i,s} \in \{0,1\}$ of these trees at those leaves. The adversary then responds with the value $p_i = \frac{1}{s}\sum_j b_{i,j}$, and sets $\rho_{i+1} \supseteq \rho_i$ to be the extended matching obtained from all the extensions made to $\rho_i$ during these traversals.

Now, when we reach the end of the game and $\rho_m$ is the final matching, we will guarantee by our choice of parameters that $|\rho_m| < N$. The inspector must decide on some $x_\final \in [2N]$ to output as their final answer; we extend $\rho_m$ to $\rho_\final$ such that $x_\final$ is matched in $\rho_\final$, which is possible since $|\rho_m|< N$. We then choose any pair of functions $f, g$ such that for all $(x,y) \in \rho_\final$, $f(x) = y$ and $g(y) = x$. We want to argue that with high probability over the randomness of the adversary (corresponding to the random matching choices and random trees sampled), this pair $(f,g)$ will witness the failure of the inspector. First observe that, since $g(f(x_\final)) = x_\final$, the inspector failed to find an $\rwPHP$ solution. Second, since $f(x) = y$ and $g(y) = x$ for all $(x,y) \in \rho_\final$, the deterministic consistency rule is obeyed with probability $1$. So it remains only to show that with high probability the mixture consistency rule is obeyed by the adversary's answers $p_i$.

To prove mixture consistency, we consider an alternate hypothetical sequence of answers $p_i^{\final}$ defined as the probability of outputting $b=1$ in the following experiment for each $i$. Sample a fresh $T_i \sim {\cal T}_i$ and randomly traverse through $T_i$ starting from $\rho_\final$: each query $f(x)$ is answered according to $\rho_\final$ if $x$ is matched in $\rho_\final$, and is otherwise set to a uniform element of $[N]$ which is unmatched in the current extension of $\rho_\final $ obtained thus far. Queries of the form $g(y)$ are handled similarly. We end up at a leaf of $T_i$ with bit $b$ and output $b$. We perform this process \emph{in parallel} for each $i \leq m$ to define the values $p_i^\final$, instead of iterating over the rounds in order and updating the matching each time (this is crucial).  Now, for these values $p_i^{\final}$, the mixture consistency rules are obeyed precisely: $p_i^{\final}$ is the probability of outputting $1$ in some experiment in which the first step is to take a uniform sample from ${\cal T}_i$, so its value is linear in the distribution of ${\cal T}_i$. So it would suffice to show that with nonzero probability over the randomness of the adversary, $|p_i - p_i^\final| \leq  \epsilon/2$ for every $i \leq m$. From this, we immediately get mixture consistency for each relevant triple $p_{i_1}, p_{i_2}, p_{i_3}$.

To prove that $|p_i - p_i^\final| \leq  \epsilon/2$, we will consider two more sets of intermediate $p$ values and proceed by a kind of hybrid argument. The key claim underlying each step of the hybrid argument is a general lemma (\Cref{lem:riis:swapping-orders}) which says the following. Consider two decision trees $T_1, T_2$, a partial matching $\rho$, and two experiments. In the first experiment, we randomly traverse $T_1$ followed by $T_2$, using the initial matching $\rho$ and randomly answering each query $f(x)$ (resp.~$g(y)$) by choosing a random element in $[N]$ (resp.~$[2N]$) which has so far been unmatched. After the traversal of $T_1$ we reach some leaf $\ell_1 \in T_1$, and then after the traversal of $T_2$ we reach some $\ell_2 \in T_2$. The second experiment is defined identically except we first traverse $T_2$ and then $T_1$. \Cref{lem:riis:swapping-orders} states that if the heights of the trees $T_1,T_2$ and the size of $\rho$ are all much smaller than $N$, then the distribution of the leaves  $(\ell_1,\ell_2)$ we reach in these two experiments will be close in statistical distance. Roughly speaking, this allows us to undo the sequence of matching extensions used to determine $p_i^\final$ and show that $p_i$ is close to $p_i^\final$.

To prove \Cref{lem:riis:swapping-orders} (and more generally, to make all of the probabilistic arguments easier during the analysis of the adversary), we define a global random variable $\omega$ which simultaneously determines the path we will reach in any tree $T$ by such a traversal starting with any initial matching $\rho$. This random variable is defined by choosing, for each $x \in [2N]$ a uniformly random ordering of the elements of $[N]$ (and symmetrically for each $y\in[N]$ an ordering of $[2N]$). Traversals of decision trees are then defined by answering queries according to the current partial assignment or else by matching $x/y$ to the least unpaired element in its associated ordering. In this terminology, \Cref{lem:riis:swapping-orders} will say that with high probability over the global randomness $\omega$, the pair of leaves $\ell_1,\ell_2$ obtained in the $T_1,T_2$ and $T_2,T_1$ experiments are the same (this is actually a stronger statement than the former statistical distance bound, and is what we will directly prove). The argument that $T_1 \circ T_2$ behaves as $T_2 \circ T_1$ with high probability regarding matching extensions and reached leaves proceeds by induction on the total depth $h_1 + h_2$ of these trees, ultimately relying on an elementary probabilistic analysis of the effect of a single swap of two basic matching operations. 
\end{paragraph}

We remark that  the probabilities $p_i$ (and $p_i^\final$) that our adversary provides are indeed different from honest probabilities. They are evaluated over $\rho_\final$ where the $f$- and $g$-queries \emph{always get matched}, which ensures that the decision tree queries never find any unmatched element. For example, if the inspector submits the randomized decision tree $\{T_z\}_z$ such that for some $\rwPHP$ instance $(f', g')$, $T_z = 1$ if and only if $g'(f'(z)) = z$, then our adversary outputs $1$ while the true probability is always at most $1/2$.

\medskip \medskip

Finally, we briefly mention how the aforementioned ideas can be extended to a substantially large class of principles.

\begin{paragraph}{Generalization to finite combinatorial principles (\Cref{thm:intro-riis}).} The above proof strategy extends to any finite combinatorial principle that can be captured using quantifier-free first-order formulas and  admits an infinite counter-model $\mathcal{M}$ (a weaker form of \Cref{thm:intro-riis}). In a bit more detail, instead of lazy-sampling a partial matching between $[N]$ and $[2N]$, we lazy-sample a partial matching between $[N]$ and $\mathcal{M}$. For this, we define a global random variable $\omega$ that chooses a uniformly random ordering of $[N]$ for each element in $\mathcal{M}$, and a uniformly random ordering of (a finite subset of) $\mathcal{M}$ for each element in $[N]$. The partial injection from $[N]$ to $\mathcal{M}$ together with a natural adaptation of the argument described above guarantees the existence of a winning strategy for the adversary. In turn, this implies that there is no proof of the finitisation of the principle in relativized $\APX$, since one can show that a winning inspector can be extracted from a candidate proof using witnessing and other ideas. We refer to \Cref{sec:riis} for more details. The argument can be extended to quantified formulas (as in the actual statement of \Cref{thm:intro-riis}) using Herbrandization, a standard technique from mathematical logic.  

Our approach  shares with Riis' forcing proof \citep{Riis94} the use of partial maps to transfer information from an infinite counter-model. In the forcing approach, this construction takes place over a nonstandard model of arithmetic, and a central task is to ensure that it yields a model of the required arithmetic axioms (see Atserias and Müller \citep{DBLP:journals/aml/AtseriasM15} and Müller \citep{DBLP:journals/apal/Muller21} for related results and techniques). Here, witnessing replaces the construction of a model with a finite adversary argument. The additional challenge is to maintain the local consistency of approximate counting answers as the partial structure grows. As explained above, our construction and its analysis supply this consistency without requiring the answers to approximate the true probabilities of a completed finite structure.
\end{paragraph}

\vspace{0.45cm}

\noindent\rule{\textwidth}{0.4pt}

\vspace{0.55cm}

\noindent \textbf{Acknowledgements.} We would like to thank Ondra Je\v{z}il for discussions about the provability of weak pigeonhole principles in bounded arithmetic. We also thank Moritz Müller for  conversations about Riis' result.

\vspace{0.65cm}

\noindent \textbf{Funding Information.} This work received support from the UKRI Frontier Research Guarantee Grant EP/Y007999/1 and the Centre for Discrete Mathematics and its Applications (DIMAP) at the University of Warwick. The authors would also like to thank the Isaac Newton Institute for Mathematical Sciences, Cambridge, for support and hospitality during the programme Logical Foundations of Complexity Theory (LFC), where part of the work on this paper was undertaken. The LFC programme was supported by EPSRC grant EP/Z000580/1. Jiatu Li is supported by the MIT Schwarzman College of Computing PhD Fellowship. Oliver Korten is supported by the National Science Foundation under Grant No. DMS-2424441.

\vspace{0.65cm}

\noindent \textbf{AI Methodology.} Following the recommendation of the ``AI and TCS: The Next Six Months'' white paper, we describe the usage of AI tools in this research project in detail.  

\begin{flushleft}
\renewcommand{\arraystretch}{1.3}
\begin{tabular}{@{}l@{\qquad}c@{}}
\textbf{Activity} & \textbf{AI used} \\
Asking the research question & No \\
Coming up with the approach & No \\
Proof development & Yes \\
Writing and exposition & Yes \\
Checking for bugs & Yes \\
Other supporting tasks & Yes \\
\end{tabular}
\end{flushleft}

\medskip

All AI-generated content was independently verified and edited by the authors, who take full responsibility for the correctness of all claims and proofs in this paper.

The main technical contribution of LLM tools is in the proof of \Cref{thm:intro_retraction}. The authors formulated a framework (similar to \Cref{sec:adversary_winning} but only for $\rwPHP$) that reduces unprovability of $\rwPHP$ to lower bounds for a query problem related to randomized decision trees, and worked out proofs of partial cases (e.g.~non-adaptive queries with ``dishonest'' adversaries or $n/\poly(\log n)$ adaptive queries with ``honest'' adversaries that provide genuine approximate count estimates). The authors then formulated a proof strategy for the general case (fully describing the behavior of the adversary) and reduced its correctness to a key ``robustification lemma'' which they could not prove; they asked GPT 5.6 (from OpenAI) to prove this robustification lemma, and it succeeded. The machine-generated proof involved a formulation and proof of a key order-swapping lemma (see \Cref{lem:riis:swapping-orders}). After confirming this original proof, the authors suggested a significant simplification to the adversary's strategy to GPT 5.6, and the machine subsequently proved that this simpler strategy worked (and gave an analysis of the strategy which was much simpler than the analysis in the original proof). This latter proof is the one given in the current manuscript; it still relies on the key order-swapping lemma devised by GPT 5.6 for the original proof. Finally, the authors worked out a mostly complete proof sketch of an extension of this $\rwPHP$ result to \Cref{thm:intro-riis} and presented it to GPT 5.6, which filled in some minor missing details in this argument.

LLM tools are also used for routine mathematics, including drafting proofs (with the ideas given as part of the prompt), adapting proof techniques to a similar setting, calculating parameters, and fixing errors. This affects \Cref{cor:intro-riis} (GPT 6),  \Cref{thm: RefYao Lossy} (by Claude Fable 5), \Cref{sec:unprov_rwPHP} (by GPT 6), \Cref{thm:wphp_unprov} (by GPT 5.6 and Claude Fable 5), \Cref{thm: UB dwPHP} (by Claude Opus 5.5), \Cref{thm: ub WPHP} (by Claude Opus 5.5), \Cref{lem:riis:winning_inspector} (by GPT 6), and results in the Appendix (by Claude Fable 5, Claude Fable 5.1, and Claude Opus 5.5). In these cases, LLM tools have not provided substantial conceptual or technical contributions.

Moreover, LLM tools (primarily GPT 6) are used to improve the presentation of the introduction through interaction with the authors.

\section{Preliminaries}\label{sec:preliminaries}

\subsection{Basic Notation}

For a string $x\in\zo^*$, we use $\|x\|$ to denote the bit length of $x$ (this is to distinguish from the absolute value $|\cdot|$). We use $\|v\|_1$ to denote the $\ell_1$-norm of a vector $v\in\R^n$, and in particular, $\|x\|_1$ to denote the Hamming weight of a string $x\in\zo^n$. 

We write $\mathsf{T} \vdash \phi$ to denote that $\phi$ is provable in first-order logic from $\mathsf{T}$. We write $\vdash_{\mathrm{FO}} \phi$ to emphasise provability of the sentence in first-order predicate logic, i.e., when the sentence is logically valid.

\subsection{Bounded Arithmetic and Theory \texorpdfstring{$\PV$}{PV}}

We recall the background in bounded arithmetic needed below.
For further details about bounded arithmetic, we refer to \citep{CLOW26}
and the survey \citep{Oliveira25}.

\paragraph{Theory $\PV$.} Our base theory $\PV$ is a first-order theory formalizing
deterministic polynomial-time reasoning. We use a standard
presentation over finite binary strings, with natural numbers,
rational numbers, circuits, and finite sequences represented
by suitable efficient encodings. The language of $\PV$ contains
function symbols for polynomial-time algorithms, introduced
through composition and limited recursion on notation, together
with axioms describing these constructions. In the standard
model, terms therefore compute polynomial-time functions, and
quantifier-free formulas define polynomial-time predicates.
Conversely, every polynomial-time predicate can be represented
by a quantifier-free formula using its characteristic function.

The theory $\PV$ supports induction on notation for quantifier-free
formulas. It also proves ordinary induction for such formulas, i.e., 
for every quantifier-free formula $\varphi(i,\bar a)$,
$$
\PV\vdash
\forall\bar a\,\forall b\;
\left(
\varphi(0,\bar a)\land
\forall i<b\,
\bigl(\varphi(i,\bar a)\rightarrow\varphi(i+1,\bar a)\bigr)
\;\rightarrow\;
\varphi(b,\bar a)
\right).
$$
A further property essential to our applications is that $\PV$
admits a universal axiomatization, meaning an equivalent
presentation whose axioms are universal closures of
quantifier-free formulas. We refer to \citep{DBLP:journals/eccc/Li25} for a detailed presentation of $\PV$.\\

A quantifier is \emph{bounded} if it has the form
$\exists y\le t(\bar x)$ or $\forall y\le t(\bar x)$, where the
term $t$ does not contain $y$. It is \emph{sharply bounded}
if its bound has the form $\|t(\bar x)\|$ (recall $\|\cdot\|$
denotes bit length). The statements expressing totality of
search problems that we consider typically have the form
$$
\forall x\,\exists y\le t(x)\;R(x,y),
$$
where $R$ is quantifier-free in the language of $\PV$.
Such statements belong to the class of sentences $\forall\Sigma^b_1(\PV)$, which 
 assert that every input has a witness of polynomially
bounded bit length whose validity can be checked in polynomial
time. 

\medskip

We write $n\in\Log$ to indicate that $n=\|u\|$ for an
auxiliary string $u$. Formally, $\forall n\in\Log\;\psi(n)$
abbreviates $\forall u\,\psi(\|u\|)$.
This convention supplies a parameter of length $n$, allowing
constructions of size polynomial in $n$, including $1^n$
and the binary representation of $2^n$.
Similarly, $n\in\Log\Log$ indicates
$n=\|\|u\|\|$. For precision parameters, the notation
$\delta^{-1}\in\Log$ means that
$\delta=1/\|\Delta\|$ for a string $\Delta$.

\medskip

We also use relativized theories $\PV(F)$, where $F$ is a
tuple of fresh constant and function symbols, typically with specified
polynomial bounds on output length. These symbols and their
length axioms are added before closing the language under
composition and limited recursion on notation, with induction
available for quantifier-free formulas in the resulting
language. Terms then describe polynomial-time computations
with oracle access to $F$. These theories likewise admit
universal axiomatizations. The particular oracle symbols
and length bounds used here are specified in
\Cref{sec:theories}.

\subsection{Witnessing Theorems}

We will need the following formulation of Herbrand's Theorem.

\begin{theorem}[Herbrand's Theorem; finite instance form (see, e.g., {\citep[Section~2.5.1]{Buss1998IntroProofTheory}})]
\label{thm:finite_herbrand}
Let $T$ be a universal theory in a first-order language
$\mathcal K$ with equality and at least one constant symbol.
Let $\bar c$ be a \emph{(}possibly empty\emph{)} tuple of fresh parameter
constants, and let $\chi(\bar c,\bar x)$ be a quantifier-free
formula in $\mathcal K(\bar c)$ whose free variables are among
$\bar x$. If
$$
T\vdash\exists\bar x\,\chi(\bar c,\bar x),
$$
then there are finitely many ground instances
$\theta_j(\bar s_j(\bar c))$, $1\leq j\leq m$, of axioms
$\forall\bar u_j\,\theta_j(\bar u_j)$ of $T$, and finitely many
tuples of ground terms $\bar t_i(\bar c)$, $1\leq i\leq r$,
of the same length as $\bar x$, such that
$$
\vdash_{\mathrm{FO}}\;
\bigwedge_{j=1}^{m}\theta_j(\bar s_j(\bar c))
\;\longrightarrow\;
\bigvee_{i=1}^{r}\chi(\bar c,\bar t_i(\bar c)).
$$
Here $\vdash_{\mathrm{FO}}$ denotes derivability in first-order
logic with equality, every formula $\theta_j$ is quantifier-free,
and all terms belong to $\mathcal K(\bar c)$.
\end{theorem}

We will also need to extract information from a proof of a sentence of higher quantifier complexity. The following result will be useful.

\begin{theorem}[KPT Witnessing Theorem {\citep{KrajicekPT91}}]
\label{thm:KPT}
Let $T$ be a universal first-order theory in a language
$\mathcal L$, and let $\varphi(x,y,z)$ be a quantifier-free
$\mathcal L$-formula. Suppose that
$$
T\vdash \forall x\,\exists y\,\forall z\;\varphi(x,y,z).
$$
Then there exist a constant $r\ge 1$ and $\mathcal L$-terms
$t_1(x), t_2(x,z_1), \dots,t_r(x,z_1,\dots,z_{r-1})
$
such that
$$
T\vdash
\forall x\,\forall z_1\cdots\forall z_r\;
\bigvee_{i=1}^{r}
\varphi\bigl(x,t_i(x,z_1,\dots,z_{i-1}),z_i\bigr).
$$
\emph{(}The variables $x,y,z$ may also denote tuples, in which case
each $t_i$ is a corresponding tuple of terms.\emph{)}
\end{theorem}

\subsection{Unrelativized Combinatorial Principles}

\subsubsection{Refuter(Yao)}

We review the definition of the $\TFZPP$ problem called $\RefYao$ \citep{CLOW26}. Recall that Yao's distinguisher-to-predictor transformation \cite{Yao82} shows that if a distribution $\disD$ over $\zo^n$ is not $\eps$-pseudorandom, i.e., there is a circuit $C:\zo^n\to\zo$ (called \emph{distinguisher}) such that 
\[
\Big |\Pr[C(\disD)] - \Pr[C(\disU_n)]\Big | > \eps, 
\]
then there exists $i\in[n]$ and a \emph{predictor} $P:\zo^{i-1}\to\zo$ such that 
\[
\Pr_{x\gets \disD}[P(x_{<i}) = x_i] \ge \frac{1}{2} + \frac{\eps}{4n},
\]
i.e., $P$ predicts the $i$-th bit of $\mathcal{D}$ with advantage at least $\varepsilon/4n$.

We say that a discrete probability distribution $\mathcal{D}$ is \emph{flat} if it is the uniform distribution over a multiset. The \emph{size} of the distribution is the size of this multiset. We will represent flat distributions explicitly as a list of strings. In the subsequent discussions, we might tacitly assume that the relevant distribution is flat and explicitly represented.

\begin{definition}
The search problem $\RefYao$ is defined as follows. 
\begin{itemize}[compact]
\item (\emph{Parameters}). Length of strings $n$, distribution size $m$, predictor size $s$, and advantage $\delta\in[0,1]$.
\item (\emph{Input}). A circuit $G:\zo^{nm}\to[n]\times \zo^s$ (called \emph{predictor generator}). 
\item (\emph{Solution}). Any explicit flat distribution $\disD\in(\zo^{n})^m$ of size $m$ such that the following holds:

\vspace{0.2cm}

Let $(i,P) \eqdef G(\disD)$, where $P:\zo^{i-1}\to\zo$ is parsed as a circuit of description length $\leq s$. Then 
\[
\Pr_{x\gets \disD}[P(x_{<i}) = x_i] < \frac{1}{2} + \delta. 
\]
In other words, $P$ is not a predictor of the $i$-th bit of $\disD$ with advantage $\delta$.  
\end{itemize}
\end{definition}

At a high level, $\RefYao$ asks to generate a distribution $\disD$ that is unpredictable against a \emph{given} predictor generator $G$ -- a deterministic algorithm that aims to output a predictor $P$ for $\disD$. The distribution $\disD$ is not necessarily an unpredictable (or equivalently, pseudorandom) distribution against \emph{small circuits}; it suffices to fool the given \emph{deterministic} predictor generator $G$. 

To ensure that $\RefYao$ is a total search problem, we assume $(\delta^2/10)\cdot m \ge s + \lceil \log n\rceil +1$ if it is not explicitly specified (see \cite{CLOW26} for more discussions). 

\subsubsection{Pigeonhole Principles and Total Search Problems}
\label{sec: principles}

\begin{definition}[Lossy Code]
We define $\LossyCode$ as the following total search problem: Given circuits $C:\zo^n\to\zo^m$ and $D:\zo^m\to\zo^n$, where $m<n$, the solution is a string $x\in\zo^n$ such that $D(C(x))\ne x$. 
\end{definition}

\begin{definition}[$\rwPHP$]
The retraction weak Pigeonhole Principle ($\rwPHP$) is defined as the $\forall\Sigma_1$-sentence in the language of $\PV$ stating that ``$\LossyCode$ is total''. Specifically: for every pair of circuits $C:\zo^n\to\zo^m$ and $D:\zo^m\to\zo^n$ satisfying $n,m\in\Log$, $m<n$, there exists $x\in\zo^n$ such that $D(C(x))\ne x$. 
\end{definition}

\section{\texorpdfstring{$\APX$}{APX},  \texorpdfstring{$\mAPX$}{APX*}, and Related Theories}\label{sec:theories}

\subsection{\texorpdfstring{$\APX$}{APX} and \texorpdfstring{$\mAPX$}{APX*}}
\label{sec: APX and mAPX}

We first recall the theory $\APX$ introduced in \cite{CLOW26}. The language of $\APX$ is the relativized theory $\PV(\P)$, where $\|\P(C,\Delta)\|\le \|C\|+2\|\Delta\|$ is intended to output a rational number $p\in\Pr_{x\in\cU}[C(x)=1]\pm 1/\|\Delta\|$. For simplicity, we use $\P_\delta(C)$ to denote $\P(C,1^{\delta^{-1}})$ for $\delta^{-1}\in\Log$. 

Besides the ordinary axioms for relativized $\PV$ (where functions are interpreted as polynomial-time oracle algorithms), four axioms are introduced to govern the behavior of the oracle $\P$: 
\begin{axioms}{$\P$-oracle axioms for $\APX$}\label{axiom: APX}
    \item[Basic Axiom]\axiomlabel{Basic Axiom for $\APX$}{axiom: basic APX} $\P_\delta(C)\in[0,1]\cap\mathbb{Q}$.
    \item[Boundary Axiom]\axiomlabel{Boundary Axiom for $\APX$}{axiom: boundary APX} If $C$ is a circuit that does not read its input bits, then $\P_\delta(C)=\Val(C)\in\zo$, where $\Val(C)$ denotes the output of $C$.
    \item[Precision Consistency Axiom]\axiomlabel{Precision Consistency Axiom for $\APX$}{axiom: precision consistency APX} For every circuit $C$ and every $\delta_1^{-1},\delta_2^{-1},\beta^{-1}\in\Log$, $|\P_{\delta_1}(C)-\P_{\delta_2}(C)| \le\delta_1+\delta_2+\beta$.
    \item[Local Consistency Axiom]\axiomlabel{Local Consistency Axiom for $\APX$}{axiom: local consistency APX} For every circuit $C$ and $\delta^{-1},\beta^{-1}\in\Log$, let $\Fix_b(C)$ be the circuit obtained by fixing the rightmost input bit of $C$ to $b$, then
    \[
    \left|\P_\delta(C)-\frac{\P_\delta(\Fix_0(C))+\P_\delta(\Fix_1(C))}{2}\right| \le 2\delta + \beta. 
    \]
\end{axioms}
Intuitively, the theory captures minimal mathematical reasoning capability corresponding to probabilistic polynomial time. It is shown that $\APX$ provides a rich toolkit for probability theory, where probability is formalized using the oracle $\P$. 

\paragraph{The Theory $\mAPX$.} We define the theory $\mAPX$ corresponding to minimal theory for multiplicative-error approximate counting (i.e.~the complexity class $\prSBP$) by modifying the last two axioms to satisfy the behavior of a multiplicative-error approximate counting oracle: $\P_\delta(C)\in(1\pm \delta)\cdot\Pr_{x\gets\cU}[C(x)=1]$. 

For simplicity, we introduce the notation $u\mlsim v$ for numbers $u,v$ as the abbreviation for the formula ``$\forall\beta^{-1}\in\Log~u \le (1+\beta)\cdot v$'', and $u\mapprox v$ as the abbreviation of the formula ``$u\mlsim v\land v\mlsim u$''; this hides the slack parameter $\beta$. We can then describe the axioms of $\mAPX$: 

\begin{axioms}{$\P$-oracle axioms for $\mAPX$}
\item[Basic and Boundary Axioms]\axiomlabel{Basic Axiom for $\mAPX$}{axiom: basic mAPX}\axiomlabel{Boundary Axiom for $\mAPX$}{axiom: boundary mAPX} Same as the corresponding axioms in $\APX$. 
\item[Precision Consistency Axiom]\axiomlabel{Precision Consistency Axiom for $\mAPX$}{axiom: precision consistency mAPX} For every circuit $C$ and every $\delta_1^{-1},\delta_2^{-1}\in\Log$, 
\[ 
(1-\delta_1) \P_{\delta_2}(C) \mlsim (1+\delta_2)\P_{\delta_1}(C)
\] 
\item[Local Consistency Axiom]\axiomlabel{Local Consistency Axiom for $\mAPX$}{axiom: local consistency mAPX} For every circuit $C$ and every $\delta^{-1}\in\Log$, then  
\begin{align*}
(1-\delta) p_M \mlsim (1+\delta) \P_\delta(C) \quad \text{and}\quad (1-\delta) \P_\delta(C) \mlsim (1+\delta) p_M
\end{align*}
where $p_M\eqdef \frac{1}{2}(\P_\delta(\Fix_0(C))+\P_\delta(\Fix_1(C)))$. 
\end{axioms}

Note that an additive-error approximate counting oracle $\P^+_\delta(\cdot)$ can be implemented by a multiplicative-error approximate counting oracle $\P^*_\delta(\cdot)$, as $\P^*_\delta(C)\in (1\pm\delta)\cdot p$ implies that $\P^*_\delta(C)\in p\pm\delta$. Accordingly, we can show that $\mAPX$ is indeed an extension of $\APX$.

\begin{proposition}\label{prop: mAPX extension}
    For every sentence $\phi$ in the language of $\PV(\P)$, $\APX\vdash\phi$ implies that $\mAPX\vdash\phi$. 
\end{proposition}
\begin{proof}
    It suffices to show that axioms of $\APX$ can be proved in $\mAPX$. As the two theories share basic and boundary axioms, it suffices to verify the other two axioms. Fix the slack parameter $\beta^{-1}\in\Log$ of the target $\APX$ axiom and let $\beta'\eqdef\beta/3$ (so that $\beta'^{-1}\in\Log$); below, the notation $\mlsim$ in the axioms of $\mAPX$ is instantiated at the slack $\beta'$.
    \begin{itemize}[compact]
        \item (\nameref{axiom: precision consistency APX}). We may assume that $\delta_1+\delta_2+\beta<1$, as otherwise the axiom is immediate from the \nameref{axiom: basic mAPX}; in particular $\delta_1,\delta_2<1$. The \nameref{axiom: precision consistency mAPX} (with the roles of $\delta_1,\delta_2$ exchanged) gives $(1-\delta_2)\P_{\delta_1}(C)\le(1+\beta')(1+\delta_1)\P_{\delta_2}(C)$, and therefore
        \[
        \P_{\delta_1}(C)-\P_{\delta_2}(C) \le \bigl((1+\beta')(1+\delta_1)-1\bigr)\cdot\P_{\delta_2}(C)+\delta_2\cdot\P_{\delta_1}(C) \le \delta_1+\delta_2+2\beta' \le \delta_1+\delta_2+\beta,
        \]
        where the second inequality uses $\P_{\delta_1}(C),\P_{\delta_2}(C)\in[0,1]$ (the \nameref{axiom: basic mAPX}) and $\delta_1<1$; the bound on $\P_{\delta_2}(C)-\P_{\delta_1}(C)$ is symmetric.
        \item (\nameref{axiom: local consistency APX}). Similarly, we may assume $2\delta+\beta<1$, so $\delta<1/2$; also $p_M\in[0,1]$ by the \nameref{axiom: basic mAPX}. From $(1-\delta)\P_\delta(C)\le(1+\beta')(1+\delta)p_M$ we obtain 
        \[ 
        \P_\delta(C)-p_M\le\bigl((1+\beta')(1+\delta)-1\bigr)\cdot p_M+\delta\cdot\P_\delta(C)\le 2\delta+2\beta'\le 2\delta+\beta,
        \]
        and the bound on $p_M-\P_\delta(C)$ follows from $(1-\delta)p_M\le(1+\beta')(1+\delta)\P_\delta(C)$ analogously. 
    \end{itemize}
    This completes the proof.
\end{proof}

\paragraph{Standard Models and Admissible Models.} Similar to \cite{CLOW26}, we now justify that the axioms for $\mAPX$ capture exactly the intended behavior of $\P$ by showing that standard models (i.e.~$\P$-oracle behaves as a multiplicative-error counting oracle) are exactly \emph{admissible models} (i.e.~$\P$-oracle satisfying axioms).

Let $\bbM$ denote the standard model of $\PV$: the universe is the set of all binary strings, and every function symbol of $\PV$ is interpreted by the polynomial-time algorithm defining it. For a function $\hat\P:\zo^*\times\zo^*\to\zo^*$, let $\bbM(\hat\P)$ denote the model of $\PV(\P)$ obtained from $\bbM$ in which $\P$ is interpreted by $\hat\P$.

\begin{definition}[Standard Models]\label{def: standard models mAPX}
    A function $\hat\P:\zo^*\times\zo^*\to\zo^*$ is a \emph{correct multiplicative-error approximate counting function} if the following conditions hold:
    \begin{enumerate}[compact]
        \item for every circuit $C:\zo^n\to\zo$ and every $\Delta$, $\hat\P(C,\Delta)$ is (the encoding of) a rational number $q\in[0,1]$ of length at most $\|C\|+2\|\Delta\|$ such that $q\in(1\pm 1/\|\Delta\|)\cdot p(C)$, where $p(C)\eqdef\Pr_{x\in\zo^n}[C(x)=1]$; and
        \item $\hat\P(C,\Delta)=\Val(C)\in\zo$ whenever $C$ does not read its input bits.
    \end{enumerate}
    We say that $\bbM(\hat\P)$ is a \emph{standard model} of $\mAPX$ if $\hat\P$ is a correct multiplicative-error approximate counting function.
\end{definition}

\begin{definition}[Admissible Models]\label{def: admissible models mAPX}
    Let $\hat\P:\zo^*\times\zo^*\to\zo^*$ be a function. We say that $\bbM(\hat\P)$ is an \emph{admissible model} of $\mAPX$ if it satisfies all axioms of $\mAPX$.
\end{definition}

The following proposition establishes the equivalence between standard and admissible models. Its proof is essentially an adaption of \cite[Theorem 2.5]{CLOW26}, and is deferred to \Cref{sec: missing standard admissible}. 

\begin{restatable}{proposition}{propStandardAdmissible}\label{prop: standard equiv admissible}
    Let $\hat\P:\zo^*\times\zo^*\to\zo^*$ be any function. Then $\bbM(\hat\P)$ is a standard model of $\mAPX$ if and only if it is an admissible model of $\mAPX$.
\end{restatable}

\subsection{Relativized Theories  and Relativized Principles}

To study the (un)provability of combinatorial principles such as $\wPHP$, $\dWPHP$ and $\rwPHP$ in bounded theories, we will also work with relativized theories and principles, where the compression and decompression circuits are replaced by fresh function symbols, see, e.g., \cite{Jerabek07-dWPHP}.

First, we discuss relativization in the context of pigeonhole principles.

\paragraph{Relativized $\APX$ and $\mAPX$ for Pigeonhole Principles.} Let $f,g$ be two new function symbols with the intended length bounds $\|f(u)\|=\|u\|-1$ and $\|g(v)\|=\|v\|+1$. We define $\PV(f,g)$ as the relativized theory of $\PV$, where $f$ and $g$ are added as base functions and the length bounds are added as axioms. Function symbols in $\PV(f,g)$ are therefore polynomial-time oracle
algorithms with oracle calls to $f$ and $g$. 

The relativized version of $\APX$ further extends $\PV(f,g)$ by containing an approximate-counting symbol
$\P(C,\Delta)$. The input $C$ is parsed as an $(f,g)$-oracle Boolean circuit, and $\Delta$ is the precision parameter. The intended semantics of $\P(C,\Delta)$ is that 
\begin{equation}
  \P(C,\Delta)\in \Pr_{z \in \bits^\ell}\bigl[C^{f,g}(z)=1\bigr]\pm \frac{1}{\|\Delta\|}.
\label{eq:rev-capp-intended}
\end{equation}
where $\ell$ is the input length of $C$. Similar to the unrelativized theory $\APX$, we do not allow the circuit $C$ to contain $\P$-oracles. 

The axioms of $\APX(f,g)$ consist of the axioms of $\PV(f,g)$ and the counting axioms for $\APX$ (see \Cref{sec: APX and mAPX}). The basic, precision and local consistency axioms are identical (with the change that the circuit can query $f$ and $g$). The boundary axiom will be: 
\begin{itemize}
    \item (\emph{Boundary Axiom}). If $C$ does not read its input bits and $\delta^{-1}\in\Log$, then $\P_\delta(C) = \Val^{f,g}(C) \in \{0,1\}$, where $\Val^{f,g}(C)$ denotes the output of $C$, which may involve oracle queries to $f$ and $g$. 
\end{itemize}
Similarly, we can define the relativized theory $\mAPX(f,g)$, where the local and precision consistency axioms are replaced by that of $\mAPX$ (relativized). Note that we may also define the relativized theory with only $f$ or $g$ accordingly. 

\paragraph{Relativized Pigeonhole Principles.} The purpose of introducing $f,g$ is to phrase combinatorial principles in a black-box fashion. The relativized retraction PHP, denoted by $\rwPHP(f,g)$, is defined as\footnote{Here we use $\{0,1\}^n$ instead of $N$ for convenience. This difference is inessential in our results.} 
\begin{equation}
\rwPHP(f,g) \eqdef \forall n\in\Log~\exists x\in\zo^{n} \big (\|f(x)\| \geq n \vee g(f(x))\ne x \big). 
\end{equation}
These sentences can be defined in the language of $\PV(f,g)$ (and thus also $\APX(f,g)$). Similarly, we define the relativized versions of $\dwPHP$ and $\wPHP$ as 
\begin{align}
\dwPHP(g) \eqdef&\,\forall n\in\Log~\exists y\in\zo^{n+1}~\forall x\in\zo^n~g(x)\ne y\\
\wPHP(f) \eqdef&\,\forall n\in\Log~\exists x_1,x_2\in\zo^n\nonumber\\
& ~~~~~~\big(x_1\ne x_2\land (f(x_1)=f(x_2) \lor \|f(x_1)\| \geq n \lor \|f(x_2)\| \geq n)\big ).
\end{align}
These sentences are in the language of $\PV(g)$ and $\PV(f)$, respectively.\\

We also consider a relativized extension of $\APX$ for
\Cref{thm:intro-riis}, where the additional function symbols represent
maps on a finite universe $[N]$. Unlike the functions used in the
pigeonhole principles above, these maps need not satisfy an output
length bound depending only on the lengths of their arguments,
uniformly in $N$. For example, $f(0)$ may equal $N-1$. To incorporate
such functions into the underlying theory of polynomial-time
computation, we introduce $N$ as an explicit argument of each new
symbol and impose the universal length bound
$$
\|f(N,\bar x)\|\leq\|N\|.
$$
These bounds are included before closing the language of the theory under
composition and limited recursion on notation. We give the precise definition next.

\paragraph{The Relativized Theory $\APX(\widehat{\mathcal{L}})$.} 
Let $\mathcal L$ be a finite language consisting of function and
constant symbols. For each $d$-ary function symbol $f\in\mathcal L$,
introduce a function symbol $\widehat f$ of arity $d+1$.
An original constant symbol $c$ is treated as the case $d=0$, giving
a unary symbol $\widehat c$. Write $\widehat{\mathcal L}$ for this
parameterized language.

The primitive symbols satisfy the universal length axioms
$$
\|\widehat f(s,x_1,\ldots,x_d)\|\leq\|s\|,
\qquad f\in\mathcal L,
$$
including $\|\widehat c(s)\|\leq\|s\|$ for constants.
The size argument $s$ is an ordinary input and may vary between
oracle calls. These are the only axioms imposed on the primitive
symbols of $\widehat{\mathcal L}$.

We first define the base theory
$B=\PV(\widehat{\mathcal L},\P)$.
Start with the ordinary $\PV$ functions and axioms, the primitive
symbols $\widehat{\mathcal L}\cup\{\P\}$, their length axioms above,
and
$$
\|\P(C,\Delta)\|\leq\|C\|+2\|\Delta\|.
$$
As usual, generate further functions in stages by composition and limited
recursion on notation. At each stage, include the defining axioms
for the functions admitted so far and quantifier-free induction
on notation, with parameters, in the current language.

The theory $\APX(\widehat{\mathcal{L}})$ extends $B$ by the four approximate
counting axioms, defined in the natural way. As before, $P(C,\Delta)$ can be applied to Boolean circuits with
$\widehat{\mathcal L}$-oracle gates and no $\P$ gates.\\

For applications to finite $\mathcal L$-structures on $[N]$, the
original symbols are represented by the slice
$\widehat f(N,\bar x)$ and $\widehat c(N)$. Closure of this slice
under the functions, and membership of its constants in $[N]$,
are hypotheses of the finite principles rather than axioms of
$\APX(\widehat{\mathcal{L}})$. Note that every $\mathcal L$-structure on $[N]$ can be
represented in this way: use its tables at the slice $s=N$ on
arguments from $[N]$, and return zero elsewhere. The
length axioms hold because every  value is below $N$.

\subsection{\texorpdfstring{$\trueAPX$}{True-APX}: True Universal Axioms of (Relativized) \texorpdfstring{$\APX$}{APX1}}

\begin{definition}[Standard models]\label{def:standard_model}
A \emph{standard model} of the language of $\APX(g)$ is a structure over the standard integers in which $g$ is interpreted by an arbitrary family of functions $\{g_m:\zo^{m-1}\to\zo^m\}$, the symbol $\P$ is interpreted by a correct approximate-counting oracle for $g$ (defined below), and every $\PV(g,\P)$-function symbol is interpreted by its defining polynomial-time machine with oracle access to $g$ and $\P$. Here a function $\P$ is a \emph{correct approximate-counting oracle} for $g$ if for every $g$-oracle circuit $C:\zo^m\to\zo$ and every $\Delta$:
\begin{enumerate}[compact]
  \item $\P(C,\Delta)\in\mathbb{Q}\cap[0,1]$;
  \item $\left|\P(C,\Delta)-\Pr_{z\gets\zo^m}[C^g(z)=1]\right|\le 1/\|\Delta\|$;
  \item $\P(C,\Delta)=\Val^g(C)$ whenever $C$ does not read its random input.
\end{enumerate}
\end{definition}

\begin{definition}[True universal axioms and $\trueAPX$]\label{def:true_universal}
A \emph{true universal axiom} of $\APX(g)$ is a universal sentence $\Psi=\forall\vec u\;\psi(\vec u)$, with $\psi$ quantifier-free in the language of $\APX(g)$, that holds in every standard model. The \emph{true universal theory} of $\APX(g)$ is the set of all true universal axioms. We use $\trueAPX(g)$ to denote the theory consisting of $\PV(P,g)$ extended with the set of all true universal axioms. We extend this definition to $\trueAPX(f,g)$ in the natural way.
\end{definition}

Specifically, the four $\APX(g)$ axioms are true universal axioms of $\APX(g)$. Moreover, the true universal theory includes combinatorial axioms that are unlikely to be provable in $\APX(g)$, as well as other mathematical statements; we provide two examples.

\begin{example}[$\#\rwPHP$]
The approximate counting version of retraction weak Pigeonhole Principle, denoted $\#\rwPHP$, is the following statement: 
\begin{itemize}
  \item For every $g$-oracle circuits $C:\zo^n\to\zo^{n-1}$, $D:\zo^{n-1}\to\zo^n$, and every $\delta^{-1},\beta^{-1}\in\Log$, let $T_{C,D}:\zo^n\to\zo$ be the circuit $T(x)\eqdef [D(C(x))=x]$, $\P_\delta(T_{C,D})\le 1/2 + \delta + \beta$. 
\end{itemize}
This is a true sentence for any correct approximate-counting oracle. By the Existence Lemma in $\APX(g)$ (see \cite{CLOW26}), if $\APX(g)\vdash\#\rwPHP$, then $\APX(g)$ proves the ordinary retraction weak Pigeonhole Principle.
\end{example}

We record the following result (\Cref{thm:intro_pigeon} Item 1), which follows from the discussion above using that $\#\rwPHP$ is a universal sentence that holds in every standard model.

\begin{proposition}\label{prop:rwPHP-in-TrueAPX}
$\trueAPX(f,g) \vdash \rwPHP(f,g)$.    
\end{proposition}

\begin{example}[$\mathsf{prBPP}=\mathsf{prP}$]
Assume that $\mathsf{prBPP}=\mathsf{prP}$ and let $\capp(C,\Delta)$ be the polynomial-time algorithm that outputs a number in $\Pr_{x}[C(x)=1]\pm 1/\|\Delta\|$ for circuits $C$ without oracle gates. The correctness of the algorithm can be formalized as the following universal statement:
\begin{itemize}
\item For every circuit $C:\zo^n\to\zo$ (without oracle gates), any $\delta,\beta\in\Log^{-1}$, it follows that $|\P_\delta(C)-\capp(C,1^{\delta^{-1}})|\le 2\delta + \beta$.
\end{itemize}
This is a true sentence (and thus a true universal axiom) if $\mathsf{prBPP}=\mathsf{prP}$ and $\capp$ is a correct algorithm for approximate counting. Note that this sentence does not involve the $g$-oracle.
\end{example}

\section{LossyCode, Refuter(Yao), and  Total Search Problems of \texorpdfstring{$\APX$}{APX1}}\label{sec:TFNP-APX1}

In this section, we prove \Cref{thm:intro_lossy_refyao} and \Cref{thm:intro_local_apx}.

\subsection{Equivalence Between LossyCode (rwPHP) and Refuter(Yao)}

Recall that \cite{CLOW26} proved a witnessing theorem for the $\forall\Sigma_1$-fragment of the theory $\APX$ via the search problem $\RefYao$, namely, $\TFNP$-problems that are provably total in $\APX$ can be reduced to $\RefYao$. Yet it is unclear whether $\RefYao$ is provably total in $\APX$. 

\begin{theorem}[Witnessing for $\APX$ \citep{CLOW26}]\label{thm:witness_APX}
Let $\varphi(x,y)$ be a quantifier-free formula in the language of $\PV$. If $\APX\vdash \forall x~\exists y~\varphi(x,y)$, there exists a deterministic polynomial-time Turing reduction from the search problem defined by $\varphi$ to $\RefYao$. 
\end{theorem}

We show that the answer is likely \emph{no} by proving $\PV$-\emph{equivalence} between the totality of $\RefYao$ and $\rwPHP$, which is unlikely provable in $\APX$ (as we will show later). In particular, we show that $\RefYao$ and the search problem $\LossyCode$ are equivalent with respect to polynomial-time reduction. 

\begin{theorem}\label{thm: RefYao Lossy}
    There are polynomial-time mapping reductions between $\RefYao$ and $\LossyCode$. Moreover, the totality of $\RefYao$ and $\rwPHP$ (i.e.~totality of $\LossyCode$), as $\forall\Sigma_1^b$-formulas in the language of $\PV$, are equivalent with respect to $\PV$. 
\end{theorem}

\AI{Drafted by Claude Fable 5 given a proof overview as prompt, verified and edited by authors.}
\begin{proof}
    We note that the reduction from $\RefYao$ to $\LossyCode$ is proved in \citep[Theorem 5.4]{CLOW26} and the idea originates from \citep{Korten22}; the formalization in $\PV$ is left as an exercise. In the rest of the proof, we construct the reduction from $\LossyCode$ to $\RefYao$. 

    Let $C:\zo^n\to\zo^{n-1}$ be the compressor and $D:\zo^{n-1}\to\zo^n$ be the decompressor of a $\LossyCode$ instance; we assume without loss of generality that the stretch is $1$. Let $\eps\in(0,1)$ be a parameter to be determined later. The reduction consists of the following two steps. 
    \begin{enumerate}[compact]
        \item Construct a new instance where the decompressor is the truth table generator $\mathsf{TT}_{s,\ell}:\zo^{s}\to\zo^{\ell}$ that takes a circuit of size $s$ and outputs the first $\ell$-bits of its truth-table, where $s=\ell^{\eps}$. More formally, we construct a polynomial-time mapping reduction from $(C,D)$ to a new instance $(E,\TT_{s,\ell})$ of $\LossyCode$. 
        \item We further reduce the instance $(E,\TT_{s,\ell})$ to $\RefYao$ with desired parameters. 
    \end{enumerate}
    
    We note that for any $\eps\in(0,1)$, the first step can be achieved by the standard ``tree-construction'' \cite{DBLP:conf/focs/Korten21} (see, e.g., \cite{Korten22,Korten-survey,LPT24}), and its formalization in $\PV$ is straightforward (see, e.g., \cite[Theorem 3.1 (iii)]{Jerabek07-dWPHP}). We then focus on the second step. 

    Now let $(E,\mathsf{TT}_{s,\ell})$ be the instance of $\LossyCode$ with $\eps\eqdef 1/20$ and $s=\ell^\eps$. We assume without loss of generality that $m$ is sufficiently large. Let $\ell=2^\nu$, we construct a $\RefYao$ instance with parameters
    \begin{equation}
    n'\eqdef\nu,\qquad m\eqdef 2^{\nu/10}=\ell^{1/10},\qquad s'\eqdef s+\nu^3,\qquad \delta\eqdef 1/4,\label{equ: refyao parameter}
    \end{equation}
    which satisfy $(\delta^2/10)\cdot m\ge s'+\lceil\log \ell\rceil+1$ since $s\le \ell^{1/20}$.

    The predictor generator $G:\zo^{\nu m}\to[\nu]\times\zo^{s'}$ parses its input as a list $\disD=(x^{(1)},\dots,x^{(m)})$ of $\nu$-bit strings. For $0\le j<\nu$, let $f_j:\zo^{j}\to\zo$ be the empirically optimal predictor of the $(j+1)$-st bit: $f_j(z)$ is the most common value of $(x^{(t)})_{j+1}$ among the samples with $(x^{(t)})_{1\dots j}=z$, counted with multiplicity ($0$ to break tie). We say that $f_j$ \emph{errs} on $x^{(t)}$ if $f_j\bigl((x^{(t)})_{1\dots j}\bigr)\ne(x^{(t)})_{j+1}$.

    \begin{claim*}\label{claim: halving}
        For every $t\in[m]$, the number of $j\in\{0,\dots,\nu-1\}$ such that $f_j$ errs on $x^{(t)}$ is at most $\log m$.
    \end{claim*}

    \begin{proof}
        Let $c_{t,j}$ be the number of indices $t'\in [m]$ such that $x^{(t')}_{\le j} = x^{(t)}_{\le j}$, so that $c_{t,0}=m$. The sequence $c_{t,j}$ is non-increasing in $j$, and $c_{t,\nu}\ge 1$. Suppose that $f_j$ errs on $x^{(t)}$ and let $b\eqdef(x^{(t)})_{j+1}$. Among the $c_{t,j}$ samples sharing the prefix $(x^{(t)})_{\le j}$, the number of those with $(j+1)$-st bit equal to $b$ is at most $c_{t,j}/2$: otherwise $b$ would be the strict majority value and $f_j$ would output $b$ instead. Hence $c_{t,j+1}\le c_{t,j}/2$, and there are at most $\log m$ erring indices.
    \end{proof}

    Summing the claim above over $t\in[m]$ and using $\log m=\nu/10$,
    \begin{equation}
    \sum_{j=0}^{\nu-1}\bigl|\{t\in[m]: f_j\text{ errs on }x^{(t)}\}\bigr|\;\le\; m\log m\;=\;\frac{m\nu}{10},\label{equ: summing err}
    \end{equation}
    so there exists $j<\nu$ such that $f_j$ errs on at most $m/10$ samples. The predictor generator $G$ then: 
    \begin{itemize}[compact]
        \item computes the smallest $j$ such that $f_j$ errs on at most $m/10$ samples;
        \item computes the truth table $F_j$ of the function $\tilde f_j:\zo^{\nu}\to\zo$ defined as $\tilde f_j(w)\eqdef f_j(w_{1\dots j})$; 
        \item computes $E(F_j)\in\zo^s$ and parses the output as a circuit $K:\zo^\nu\to\zo$; and 
        \item outputs $(j+1,P)$, where $P(z)\eqdef K(z\circ0^{\nu-j})$. 
    \end{itemize}
    It can be verified that the instance map outputs $G$ with the parameters in \Cref{equ: refyao parameter}. To see the correctness of the reduction, let $\disD$ be any solution of $G$, and suppose toward a contradiction that $\mathsf{TT}_{s,\ell}(E(F_j))=F_j$, i.e., $\mathrm{tt}(K)=F_j$. Then for every $z\in\zo^{j}$ we have $P(z)=f_j(z)$, so $P$ errs on at most $m/10$ samples, and subsequently
    \[
    \Pr_{x\gets\disD}\bigl[P(x_{<i})=x_i\bigr]\;\ge\;\frac{9}{10}\;>\;\frac12+\delta,
    \]
    contradicting that $\disD$ is a solution. 

    \smallskip\noindent\emph{Formalization in $\PV$}. As mentioned above, it suffices to verify that the step 2 (the reduction from the $\LossyCode$ instance $(E,\TT_{s,\ell})$ to $\RefYao$) can be formalized in $\PV$. Recall that $\rwPHP$ formalizes exactly $\LossyCode$ is total, and we formalize it with $n\in\Log$. It is worth noting that the construction of the truth table $F_j$ is feasible as $\nu\in\Log\Log$ (recall that $\nu=\log\ell$ and $\ell=\poly(n, |C|, |D|)$). The rest of the proof can be formalized straightforwardly, with some routine work that has been comprehensively described in literature (see, e.g., \cite{DBLP:journals/corr/Pich14,krajicek_2019,DBLP:journals/eccc/Li25}). 
\end{proof}

\subsection{Witnessing Theorem for the \texorpdfstring{$\forall\Sigma^b_1(\PV)$}{ForallSigmaPV}-Consequences of \texorpdfstring{$\APX$}{APX1}}
\label{sec: witnessing}

Both $\APX$ and $\mAPX$ are universal theories in the language of relativized $\PV$. Following a recent work of \cite{Thapen25}, we  derive witnessing theorems for the $\forall\Sigma_1^b$-fragments of these theories.

\begin{definition}[Witness of Axiom Violation]
Let $L=\{(C_i,\Delta_i,p_i)\}_{i\in[k]}$ be a list of query-answer pairs to the $\P$ oracle, where each $C_i$ is a circuit, $\Delta_i$ is a string, and $p_i\in[0,1]$ is a rational number with at most $\|C_i\|+2\|\Delta_i\|$ bits. A \emph{witness of axiom violation} for local consistency axiom of $\APX$ is a tuple $(u,v,w,B)$, $u,v,w\in[k]$, such that the followings hold. 
\begin{itemize}[compact]
\item $C_{v},C_w$ are exactly $C_u$ with last bit fixed to be $0$ and $1$, respectively. 
\item $|\Delta_u|=|\Delta_v|=|\Delta_w|$. Let $\delta\eqdef 1/|\Delta_u|$, and $\beta\eqdef 1/|B|$. 
\item $|p_u-(p_v+p_w)/2| > 2\delta + \beta$.  
\end{itemize}
Accordingly, we can define witness of violation for other axioms of $\APX$ and $\mAPX$. Note that given the list $L$ and $\tau$, it can be verified in polynomial time whether $\tau$ is indeed a witness of axiom violation. 
\end{definition}

\begin{definition}[Inspector-Adversary Game]\label{def: IAG}
    The inspector-adversary game for the theory $\APX$ (resp.~the theory $\mAPX$) is defined as follows.
    \begin{itemize}[compact]
        \item In each round, the inspector outputs a circuit $C_i$ and a string $\Delta_i$ (called inspector's \emph{challenge}), and the adversary outputs a rational $p_i \in [0,1]$ with at most $\|C_i\|+2\|\Delta_i\|$ bits (called adversary's \emph{response}). 
        \item Let $\{\P(C_i,\Delta_i)=p_i\}_{i\in[k]}$ be a transcript of $k$ rounds. The inspector wins after $k$ rounds, if it can output a valid witness of axiom violation for any axiom of $\APX$ (resp.~$\mAPX$). 
    \end{itemize} 
\end{definition}

\begin{definition}[$\Local$]\label{def: IAG APX}
    The search problem $\Local(\APX)$, associated with the Inspector-Adversary Game for $\APX$, is defined as follows: Given a query algorithm $Q=(M_1,M_2,\dots,M_k,M_{k+1})$, where $M_i$ is a circuit that takes the previous transcript as input and outputs the message in the $i$-th round, and in particular, $M_{k+1}$ is intended to output a witness of axiom violation for any $\APX$ axiom. The solution is a list of adversary's responses $p_1,\dots,p_k$ given which the inspector does not win after $k$ rounds. 

    Similarly, we can define the search problem $\Local(\mAPX)$ according to the Inspector-Adversary Game for $\mAPX$ (instead of $\APX$).
\end{definition}

We show that $\Local(\APX)$ (resp.~$\Local(\mAPX)$) captures $\forall\Sigma_1^b$-consequences of $\APX$ (resp.~$\mAPX$). First, Thapen \cite{Thapen25} shows that provably total $\TFNP$ problems reduce to their inspector-adversary game search problems; indeed, Thapen proves a more general result that for any theory $\PV(f)$ augmented with universal axioms about the $f$-oracle, provably total TFNP problems in the theory reduce to their corresponding Inspector-Adversary Game. In particular, we have:

\begin{lemma}[{\cite[Lemma 3.7]{Thapen25}}]\label{lmm: witnessing TFNP}
    Let $R(x,y)$ be a $\PV$-predicate. Then: 
    \begin{itemize}[compact]
        \item if $\APX\vdash \forall x~\exists y~R(x,y)$, then $R\le_p \Local(\APX)$;
        \item if $\mAPX\vdash \forall x~\exists y~R(x,y)$, then $R\le_p \Local(\mAPX)$;
    \end{itemize} 
    (both via single-query polynomial-time reducibility.)
\end{lemma}

The above lemma shows that each computational problem $\Local(\cdot)$ is strong enough for witnessing $\forall\Sigma_1^b$-consequences of the corresponding theory. The next lemma shows that each of them is also the weakest possible $\TFNP$ problem for witnessing. 

\begin{lemma}
    $\APX$ (resp.~$\mAPX$) proves that $\Local(\APX)$ is total (resp.~$\Local(\mAPX)$ is total).
\end{lemma}

\begin{proof}
    We work in $\APX$ (resp.~$\mAPX$). We need to prove that for any given query algorithm $Q=(M_1,M_2,\dots,M_k,M_{k+1})$, there exists a transcript given which the inspector does not win in $k$ rounds. Indeed, we compute iterative as follows: 
    \begin{itemize}[compact]
        \item Let $(C_1,\Delta_1)=M_1(\tau_1:=\varnothing)$, we define $p_1\eqdef\P(C_1,\Delta_1)$ 
        \item Let $\tau_2$ be the transcript before this step, and $(C_2,\Delta_2)=M_2(\tau_2)$, then $p_2\eqdef\P(C_2,\Delta_2)$.
        \item $\dots$
        \item Finally, $p_k\eqdef\P(C_k,\Delta_k)$. 
    \end{itemize}
    Note that $p_1,\dots,p_k$ can be computed via a $\P$-oracle polynomial time algorithm, and can thus be defined. Moreover, with the axioms of $\APX$ (resp.~$\mAPX$), we can conclude that the inspector must fail. 
\end{proof}

\section{Approximate Counting and Weak Pigeonhole Principles}

In this section, we study upper bounds and (relativized) lower bounds for weak pigeonhole principles  in bounded theories for approximate counting. In particular, we derive \Cref{thm:intro_retraction} and prove the remaining items of \Cref{thm:intro_pigeon}.

\subsection{Unprovability of Relativized rwPHP in \texorpdfstring{$\APX$}{APX1}}\label{sec:unprov_rwPHP}

In this section, we use \Cref{thm:intro-riis} to show that
$\APX(f,g) \nvdash \mathsf{rwPHP}(f,g)$
(\Cref{thm:intro_retraction}). Recall that
$$
\rwPHP(f,g) = \forall N \big (
N>1 \rightarrow
\exists x<2N \; (f(x)\geq N \vee g(f(x))\neq x)
\big).
$$

As in the example following \Cref{thm:intro-riis} in
\Cref{sec:intro_results_riis}, we represent the two halves of
the compression map by unary function symbols $f_0$ and $f_1$.
Introduce one further unary function symbol $h$, and let
$\mathcal L \eqdef \{f_0,f_1,h\}$. Define
$$
\phi(x)\eqdef
h(f_0(x))\neq x
\;\lor\;
h(f_1(x))\neq x
\;\lor\;
f_0(x)=f_1(x).
$$
Thus $\forall x\,\neg\phi(x)$ asserts that $h$ is a common left
inverse of $f_0$ and $f_1$, and that $f_0(x)\neq f_1(x)$ for every
$x$. This sentence has an infinite model with universe $\mathbb N$:
$$
f_0(x)=2x,\qquad
f_1(x)=2x+1,\qquad
h(y)=\lfloor y/2\rfloor.
$$
Consequently, \Cref{thm:intro-riis} gives
$\APX(\widehat{\mathcal L})\nvdash\Theta_\phi$, where
$$
\Theta_\phi\eqdef
\forall N\geq1\;
\left(
\operatorname{Closed}_{\widehat{\mathcal L}}(N)
\;\rightarrow\;
\exists x<N\;\phi^{[N]}(x)
\right).
$$

Finally, in order to obtain \Cref{thm:intro_retraction},
it is enough to show that a proof of $\rwPHP(f,g)$ in
$\APX(f,g)$ would yield a proof of $\Theta_\phi$ in
$\APX(\widehat{\mathcal L})$. For completeness, we describe
the argument below.\\

\noindent
\textbf{If $\bm{\APX(f,g) \vdash \rwPHP(f,g)}$ then
$\bm{\APX(\widehat{\mathcal L}) \vdash \Theta_\phi}$.}
For each $N>1$, define
$$
F_N(z)\eqdef
\begin{cases}
\widehat f_0(N,z), & z<N,\\
\widehat f_1(N,z-N), & N\leq z<2N,
\end{cases}
$$
and
$$
G_N(y)\eqdef
\begin{cases}
\widehat h(N,y),
& \widehat f_0(N,\widehat h(N,y))=y,\\
N+\widehat h(N,y),
& \widehat f_0(N,\widehat h(N,y))\neq y,
\end{cases}
\qquad(y<N).
$$
Extend $F_N$ and $G_N$ by $0$ outside $[2N]$ and $[N]$,
respectively. These are $\PV(\widehat{\mathcal L})$ functions,
defined uniformly in the parameter $N$.
Under $\operatorname{Closed}_{\widehat{\mathcal L}}(N)$,
they restrict to maps
$$
F_N\colon[2N]\to[N],
\qquad
G_N\colon[N]\to[2N].
$$

Given these definitions, it is not hard to check that the theory
$\PV(\widehat{\mathcal L})$ proves that, for every $N>1$,
the following implication holds:
$$
\begin{gathered}
\operatorname{Closed}_{\widehat{\mathcal L}}(N)
\;\land\;
\exists z<2N\;
\bigl(F_N(z)\geq N\vee G_N(F_N(z))\neq z\bigr)
\\
\longrightarrow\quad
\exists x<N\;\phi^{[N]}(x).
\end{gathered}
$$

Now suppose, towards a contradiction, that
$\APX(f,g)\vdash\rwPHP(f,g)$.
In this proof, replace calls to $f$ and $g$ by the
$\PV(\widehat{\mathcal L})$ functions $F_N$ and $G_N$,
treating $N>1$ as a parameter. Inside counting queries,
replace each oracle circuit $C$ by the oracle circuit $C_N$
over $\widehat{\mathcal L}$ obtained by the same substitutions
at its oracle gates. One can check by verifying axioms and induction (in the meta-language) on the complexity of proofs that this translation can
be carried out so that the resulting proof is valid in
$\APX(\widehat{\mathcal L})$. It therefore yields the $N$
instance of $\rwPHP(F_N,G_N)$.

By the preceding reduction, we get that
$$
\APX(\widehat{\mathcal L})\vdash
\forall N>1\;
\left(
\operatorname{Closed}_{\widehat{\mathcal L}}(N)
\;\rightarrow\;
\exists x<N\;\phi^{[N]}(x)
\right).
$$
Finally, for $N=1$, closure gives
$\widehat f_0(1,0)=\widehat f_1(1,0)=0$, so
$\phi^{[1]}(0)$ holds. Together with the displayed consequence,
this proves $\Theta_\phi$ in $\APX(\widehat{\mathcal L})$,
contradicting \Cref{thm:intro-riis}. Therefore
$\APX(f,g)\nvdash\rwPHP(f,g)$, as required.

\subsection{Unprovability of Relativized (d)wPHP in \texorpdfstring{$\trueAPX$}{True-APX1}}

\subsubsection{A Lower Bound for dwPHP}

We first prove the following lower bound on inspector algorithms for $\dwPHP$. Looking ahead, we will use this lower bound and the KPT witnessing theorem (\autoref{thm:KPT}) to show the unprovability of $\dwPHP(g)$ in the true universal theory of $\APX(g)$.

\begin{lemma}\label{lem:dwphp_lb}
  Let $k\in\N$ be a constant and let $\Psi_1,\dots,\Psi_k$ be true universal axioms of $\APX(g)$, where $\Psi_j=\forall\vec u\;\psi_j(\vec u)$ with $\psi_j$ quantifier-free. Let $t(n)\ge n$ be a non-decreasing function and $\cI(1^n)$ be a (non-uniform) query algorithm of size $t=t(n)$, $g=\{g_n:\zo^{n-1}\to\zo^n\}$, $\P:\zo^*\times\zo^*\to\zo^*$ such that $\|\P(C,\Delta)\|\le \|C\| + 2\|\Delta\|$. The algorithm (called inspector) may take one of the following three types of queries: 
  \begin{itemize}
  \item ($g$ queries): Given $x\in\zo^{m-1}$, the oracle outputs $g(x)\in\zo^{m}$. 
  \item ($\P$ queries): Given an oracle circuit $C$ and a string $\Delta$, the oracle outputs $\P(C,\Delta)$. 
  \item (inversion queries): Given $y\in\zo^n$, the oracle either outputs some string $x\in\zo^{n-1}$ such that $g(x)=y$, or $\bot$. 
  \end{itemize}
  Note that the algorithm $\cI(1^n)$ can query the $g$ oracle on input of any length $\le t$, while for inversion queries it is only allowed to query $y$ of length \emph{exactly} $n$. The inspector is said to succeed (resp.~fail) if at least one (resp.~neither) of the following two conditions holds: 
  \begin{itemize}
  \item (range avoidance): the oracle outputs $\bot$ in an inversion query $y\in\zo^n$. 
  \item (axiom violation): the inspector finds and outputs some $j\in[k]$ and a tuple $\vec u$ of strings such that the instance $\psi_j(\vec u)$ evaluates to false by the oracle answers to $g$ and $P$.
  \end{itemize}
  Then for some constant $\eps>0$, if $t(n)\le 2^{\eps n}$ for all sufficiently large $n$, there are $\{g_n\},\P$, and an inversion oracle such that for infinitely many $n$, $\cI(1^n)$ fails. %
\end{lemma}

\begin{proof}
Fix $k\in\N$ and axioms $\Psi_1,\dots,\Psi_k$, and let $d\ge 1$ be a constant such that for every $i\in[k]$, $\Psi_i=\forall\vec u~\psi(\vec u)$, $\psi(\vec u)$ will make $g$-oracle queries and $\P$-oracle queries on input length at most $n^{d}$ when $n=\|\vec u\|$ is sufficiently large. Note that the constant $d$ always exists as $\psi(\vec u)$ are quantifier-free formulas in the language of $\APX(g)$ and are thus polynomial-time algorithms with $(g,\P)$-oracles.  

Let $\eps>0$ be a parameter to be determined, and let $N=\{n_1,n_2,\dots\}$ be a set of input lengths on which we will ensure that $\cI$ does not succeed. Let $n\in N$, $t=t(n)$, $\tilde{t}=t^d$. Let $\ell=\ell(n)$ be a parameter to be determined. For $\vec x=(x_1,\dots,x_\ell),\vec y=(y_1,\dots,y_\ell)$, where $x_i$'s are distinct and $\|y_i\|=\|x_i\|+1\le {\tilde t}$ for every $i\in[\ell]$, we define $g_{\vec x,\vec y}$ as  
\[
g_{\vec x,\vec y}(x) \eqdef \begin{cases}
  y_i & x = x_i \\ 
  0^{\|x\|+1} & x\notin\{x_1,\dots,x_\ell\}
\end{cases}
\]
Let $D_n\subseteq\zo^{\tilde t}$ and let $D^{\le m}_n$ be its projection to the first $m$ bits. It is said to be pseudorandom if for every such $\vec x,\vec y$ and any oracle circuit $C:\zo^m\to\zo$ of size at most ${\tilde t}$, $D^{\le m}$ $(1/{\tilde t})$-fools the circuit $C^{g_{\vec x,\vec y}}$. Specifically,
\begin{equation}
\left|\Pr_{z\gets D^{\le m}_n}[C^{g_{\vec x,\vec y}}(z)=1] - \Pr_{z\gets\zo^m}[C^{g_{\vec x,\vec y}}(z)=1]\right| \le \frac{1}{{2\tilde t}}. \label{equ: PRS}
\end{equation}
By the Hoeffding bound, we know that for any fixed $\vec x,\vec y$ and circuit $C$, if we sample a size-$s$ multiset $D\subseteq\zo^{\tilde t}$ uniformly at random, $D^{\le m}_n$ fools $C$ with probability $1-2\exp(-s/(2{\tilde t}^2))$. There are at most $2^{2({\tilde t}+1)\ell}$ choices of $(\vec x,\vec y)$ and at most $2^{O({\tilde t}\log {\tilde t})}$ oracle circuits of size at most ${\tilde t}$. Therefore, by union bound over $\vec x,\vec y,C$, there exists a pseudorandom set $D_n\subseteq\zo^{{\tilde t}}$ of size at most $s(n)\le 100\cdot t^{3d}\ell^2\sqrt n$ ($\ast$).

\paragraph{Construction of Adversary.} Let $n_0\in\N$ be sufficiently large. We will construct the oracle $g$ and $\P$ iteratively, starting with $g(z)$ and $\P(C,\Delta)$ being undetermined for any input $z$ and $C,\Delta$. In the $i$-th iteration, we will fix entries of $g,\P$ to force $\cI(1^{n_i})$ fails; in addition, we will maintain the invariance that after the $i$-th iteration:
\begin{enumerate}[compact]
  \item We fix $g$ on all inputs of length at most $\tilde t(n_i)$ and $\P$ whenever $\|C\|,\|\Delta\|\le \tilde t(n_i)$.
  \item $g$ agrees with $g_{\vec x,\vec y}(x)$ on fixed strings for a list $\vec x,\vec y$ of length at most $\ell(n_i)$. 
  \item For any $\|C\|,\|\Delta\|\le \tilde t(n_i)$, $|\P(C,\Delta)-\Pr_{x}[C^g(x)=1]|\le 1/\|\Delta\|$.
  \item The procedure will not stall; see ($\diamond$) below. 
\end{enumerate}
For $i=0$, we fix $g$ to be all-zero on input length at most $\tilde t(n_0)$ and fix $\P$ to be the correct acceptance probability with respect to $g$. We now describe the $i$-th iteration of the procedure.

\smallskip\noindent (\emph{Iteration $i$}). Fix a pseudorandom set $D_n\subseteq\zo^{\tilde t}$. We run the algorithm $\cI(1^{n_i})$ and answer queries if the corresponding entry for $g$ and $\P$ are fixed. Otherwise we work as follows: 
\begin{itemize}[compact]
  \item For any $g$ query on $x$, we fix $g(x) = 0^{\|x\|+1}$ and answer it.
  \item For any $\P$ query on circuit $C:\zo^m\to\zo$ and parameter $\Delta$, we evaluate $C^{g}(z)$ for any $z\in D^{\le m}_n$, during which any unfixed $g$-query will be fixed to output the all zero string. Let
  \begin{equation}
  p \eqdef \Pr_{z\gets D^{\le m}_n}[C^{g}(z)=1], \label{equ: P by PRS}
  \end{equation}
  and fix and answer $\P(C,\Delta)$ to be any dyadic rational number that is $1/(2\|\Delta\|)$-close to $p$.

  \item For any inversion query $y\in\zo^{n_i}$, if we have set $x$ such that $g(x)=y$, we answer $x$. Otherwise, we find any $x\in\zo^{n_i-1}$ such that $g(x)$ has not been fixed, fix $g(x)=y$, and then answer $x$. We will ensure that such $x$ always exists later ($\diamond$).
\end{itemize}
After that, we fix $g(x)=0^{\|x\|+1}$ for any unfixed $x$ with $\|x\|\le \tilde t(n_i)$. Then for every circuit $C:\zo^m\to\zo$ of size at most $\tilde t(n_i)$ and any $\Delta$ of length at most $\tilde t(n_i)$, we fix $\P(C,\Delta)$ according to \eqref{equ: P by PRS}.

Let $g,\P$ be the union of all fixed $g,\P$ in the procedure for $i=1,2,\dots$. We argue that for the functions $g,\P$, it is always possible to answer inversion queries such that $\cI(1^{n_i})$ fails for any $i$.

\paragraph{Proof of Correctness.} We first show that the adversary will succeed if the four invariance properties hold. Notice that if $\cI(1^{n_i})$ succeeds instead, it either outputs an instance $\vec u$ such that $\psi_j(\vec u)$ is false for some $j\in[k]$, or finds a string $y$ on which $g^{-1}(y)=\varnothing$. It can be verified that the latter is ruled out by the fourth property and by the construction procedure. 

To rule out the former case, we first observe that after the $i$-th iteration, the functions $g$ and $\P$ are fixed on input length $\le \tilde t(n_i)$. Fix any $j\in[k]$. For any instance $\vec u$ provided by the inspector that intended to falsify $\Psi_j$, let $m=\|\vec u\|$, we know that $m\le t(n_i)$ as the inspector runs in at most $t(n_i)$ time. Recall that we set $d\ge 1$ so that to evaluate $\psi_j(\vec u)$, all $g$-oracle and $\P$-oracle queries are of length at most $m^d \le \tilde t(n_i)$, and thus are already fixed. Moreover, by the third invariant, we know that the $\P$-oracle is a correct approximate-counting oracle. (Note that conditions (1) and (3) of \Cref{def:standard_model} follow immediately from the implementation of $\P(\cdot,\cdot)$.) Therefore, the instance $\vec u$ cannot falsify $\Psi_j$ as it is true for any fixed $g$ and correct approximate-counting oracle $\P$.

\medskip 

It remains to verify the invariance properties.

\smallskip\noindent(\emph{Invariance 1}): This is straightforward from the construction procedure. 

\smallskip\noindent(\emph{Invariance 2}): Note that $g(x)$ is set to be non-zero in the $i$-th iteration only if $x$ was chosen to answer an inversion query made by $\cI(1^{n_i})$. In the $i$-th iteration, we will answer at most $t(n_i)$ inversion queries from $\cI(1^{n_i})$. Subsequently, the statement holds provided that $\ell(n_i) \ge t(n_i) + \ell(n_{i-1})$. Set $\ell(n_i)\eqdef n_i\cdot t(n_i)$ ($\star$).  

\smallskip\noindent(\emph{Invariance 3}): Fix any $C\in\zo^m\to\zo$ and $\Delta$ of length at most $\tilde t(n_i)$. By invariance 2, we know that $g$ agrees with $g_{\vec x,\vec y}$ in the $i$-th iteration for a list $\vec x,\vec y$ of length $\ell(n_i)$. Recall that $D_{n_i}$ is a pseudorandom set, and in particular $D_{n_i}^{\le m}$ $(1/(2\tilde t(n_i)))$-fools any size-$\tilde{t}(n_i)$ oracle circuit with $g_{\vec x,\vec y}$ to answer the oracle queries. Moreover, we know by the construction that whenever we fix $\P$ in the $i$-th iteration, it is fixed to be the empirical estimation over the set $D_{n_i}$. Subsequently, it agrees with the actual acceptance probability with error at most $1/(2\tilde{t}(n_i)) + 1/(2\|\Delta\|)\le 1/\|\Delta\|$.

\smallskip\noindent(\emph{Invariance 4}): We will show that the number of fixed $g(x)$ for $x\in\zo^{n_i-1}$ during the run of $\cI(1^{n_i})$ is at most $2^{n_i-2}$; as a result, there is always an unfixed $x\in\zo^{n_i-1}$ when we answer inversion oracle queries. We set $n_i \ge 2\tilde{t}(n_{i-1})$ so $g(x)$ is unfixed for any $x\in\zo^{n_i-1}$ before the $i$-th iteration. In the $i$-th iteration, each $g$-oracle query or inversion query will fix at most $1$ entry, and each $\P$-oracle query will fix at most $s(n_i)\cdot t(n_i)$ entries of $g$. Therefore in total, the number of entries in $g$ on input length $n_i-1$ fixed by these three types of queries is at most
\[
2\cdot(t(n_i))^2\cdot s(n_i) \le 100\cdot (t(n_i))^{5d}\cdot \ell(n_i)^2\cdot \sqrt{n_i} \le 100\cdot (t(n_i))^{7d}\cdot n_i^{2.5},
\]
where the first inequality follows from ($\ast$) and the second inequality follows from ($\star$). By setting $\eps := 1/(10d)$ and the assumption $t(n_i)\le 2^{\eps n_i}$, this is at most $100\cdot 2^{0.7n_i}\cdot n_i^{2.5}\le 2^{n_i-2}$ for sufficiently large $n_i$.
\end{proof}

The following is then an immediate corollary of the KPT witnessing theorem (\Cref{thm:KPT}).

\begin{theorem}\label{thm:dwphp_unprov}
$\trueAPX(g) \nvdash\dWPHP(g)$. In particular, $\APX(g)\nvdash\dWPHP(g)$. 
\end{theorem}

\begin{proof} Let $\sT$ be the set of true universal axioms of $\APX(g)$, and
suppose that $\PV(\P,g)+\sT\vdash \dWPHP(g)$. A proof invokes only finitely many axioms $\Psi_1,\dots,\Psi_k\in \sT$, so $\PV(\P,g)+\{\Psi_1,\dots,\Psi_k\}\vdash\dWPHP(g)$. For simplicity, we assume that all axioms $\Psi_i=\forall u_i~\psi(u_i)$ have only one variable. Then we know that $\PV(\P,g)\vdash \dWPHP(g)\lor\bigvee_{i\in[k]}\exists u_i~\lnot \psi_i(u_i)$. Subsequently we can see that: 
\[
\PV(\P,g)\vdash \forall n\in\Log~\exists y\in\zo^n~\exists (u_1,\dots,u_k)~\forall x\in\zo^{n-1}~g(x)\ne y\lor \bigvee_{i\in[k]}\lnot\psi_i(u_i). 
\]

By the KPT Witnessing Theorem (\Cref{thm:KPT}) applied to the (universal) theory $\PV(\P,g)$, there are a constant $r\in\N$ and a polynomial-size Inspector (or Student) that, on input $1^n$, makes $g$ and $\P$ oracle queries, that wins against any Adversary (or Teacher) in the following sense: 
\begin{itemize}
  \item In each round, the Inspector proposes (from previous transcript) a string $y\in\zo^n$ and $u_1,\dots,u_k$. If one of $u_i$ witnesses $\lnot\psi(u_i)$ the Inspector wins. 
  \item The Adversary outputs a string $x\in\zo^{n-1}$ such that $g(x)=y$; if impossible, the Inspector wins. 
\end{itemize}
This matches the game in \Cref{lem:dwphp_lb}, and thus leads to a contradiction as the Inspector has complexity $\poly(n)$ and makes at most $r=O(1)$ inversion queries for $g$. 
\end{proof}

\vspace{-0.9em}%
\begin{remark}
The query lower bound and unprovability result supplement the results of \cite{ILW23} (see also the follow up papers \cite{CL24,Ilango25,RWZ26,LRZ26}), which shows the hardness of the range avoidance problem (see \Cref{def: avoid}) and the unprovability of $\dwPHP$ in $\PV$ under cryptographic assumptions (see also \cite{CLOW26} for an easy generalization to $\APX$). 

It is worth noting that since the function $g$ is modelled as an oracle, the lower bound in \Cref{lem:dwphp_lb} allows exponentially many inversion queries, whereas the hardness result of \cite{ILW23,RWZ26} allows only $O(1)$ queries. In fact, assuming circuit lower bounds, there is an algorithm for solving range avoidance that makes only polynomially many queries (see \cite[Appendix A]{ILW23}). 
\end{remark}

\subsubsection{A Lower Bound for wPHP}

We now prove a similar lower bound for Weak PHP against $\APX$. Standard models and true universal axioms of $\APX(f)$ are defined as in \Cref{def:standard_model,def:true_universal}, with $g$ replaced by the length-decreasing oracle $f=\{f_n:\zo^{n}\to\zo^{n-1}\}$.

\begin{lemma}\label{lem:wphp_lb}
  Let $k\in\N$ be a constant and let $\Psi_1,\dots,\Psi_k$ be true universal axioms of $\APX(f)$, where $\Psi_j=\forall\vec u\;\psi_j(\vec u)$ with $\psi_j$ quantifier-free. Let $t(n)\ge n$ be a non-decreasing function and $\cI(1^n)$ be a (non-uniform) query algorithm in size $t=t(n)$, $f=\{f_n:\zo^{n}\to\zo^{n-1}\}$, $\P:\zo^*\times\zo^*\to\zo^*$ such that $\|\P(C,\Delta)\|\le \|C\| + 2 \|\Delta\|$. The algorithm (called inspector) may take one of the following two types of queries:
  \begin{itemize}
  \item ($f$ queries): Given $x\in\zo^{m}$, the oracle outputs $f(x)\in\zo^{m-1}$.
  \item ($\P$ queries): Given an oracle circuit $C$ and a string $\Delta$, the oracle outputs $\P(C,\Delta)$.
  \end{itemize}
  The algorithm $\cI(1^n)$ can query the $f$ oracle on inputs of any length $\le t$. It is said to succeed (resp.~fail) if at least one (resp.~neither) of the following two conditions holds:
  \begin{itemize}
  \item (collision): the inspector outputs distinct $x,x'\in\zo^{n}$ such that $f(x)=f(x')$.
  \item (axiom violation): the inspector outputs some $j\in[k]$ and a tuple $\vec u$ of strings such that the instance $\psi_j(\vec u)$ evaluates to false by the oracle answers.
  \end{itemize}
  Then for some constant $\eps>0$, if $t(n)\le 2^{\eps n}$ for all sufficiently large $n$, there are $\{f_n\}$ and $\P$ such that for infinitely many $n$, $\cI(1^n)$ fails.
\end{lemma}

\AI{Initial draft obtained via interaction with GPT \jiatu{to igor: please confirm this} \inote{yes, done with the help of GPT}; proof drafted by Claude Fable 5 given the initial draft, edited and verified by authors.}
\begin{proof}
Fix $k\in\N$ and axioms $\Psi_1,\dots,\Psi_k$, and let $d\ge 1$ be a constant such that for every $j\in[k]$, the evaluation of $\psi_j(\vec u)$ makes $f$-oracle and $\P$-oracle queries on inputs of length at most $n^d$ when $n=\|\vec u\|$ is sufficiently large; such $d$ exists as $\psi_j(\vec u)$ are quantifier-free formulas in the language of $\APX(f)$ and are thus polynomial-time algorithms with $(f,\P)$-oracles. Let $\tilde t(n)\eqdef (t(n))^d$. Let $N=\{n_1,n_2,\dots\}$ be a set of input lengths with $n_i\ge 2\tilde t(n_{i-1})$ and $n_0$ sufficiently large. We fix $f$ and $\P$ iteratively, maintaining that: 
\begin{enumerate}[compact]
\item After the $i$-th iteration, $f$ is fixed on all inputs of length $\le \tilde t(n_i)$, $\P$ is fixed when $\|C\|,\|\Delta\|\le \tilde t(n_i)$.
\item Every value of $\P$ fixed in the $i$-th iteration is within $1/\tilde t(n_i)\le 1/\|\Delta\|$ of the acceptance probability $\Pr_z[C^f(z)=1]$ with respect to the final $f$.
\end{enumerate}
For $i=0$, we fix $f$ on all input lengths at most $\tilde t(n_0)$ to be the truncation function $x\mapsto x_1\cdots x_{\|x\|-1}$ and fix $\P$ to be the correct acceptance probabilities with respect to $f$.

\smallskip\noindent(\emph{Iteration $i$}). Write $t=t(n_i)$ and $\tilde t=\tilde t(n_i)$. On every new input length $m\le \tilde t$ other than $n_i$, fix $f_m$ to be the truncation function. On input length $n_i$, choose a uniformly random function $f_{n_i}:\zo^{n_i}\to\zo^{n_i-1}$. Independently, sample a multiset $D\subseteq\zo^{\tilde t}$ of size $s\eqdef \tilde t^4$ uniformly at random. 

Now $f$ is fixed on all lengths at most $\tilde t$; for every pair $(C,\Delta)$ with $\|C\|,\|\Delta\|\le \tilde t$ that is not fixed yet, where $C:\zo^m\to\zo$, we fix $\P(C,\Delta)$ to be any dyadic rational that is $1/(2\|\Delta\|)$-close to $\Pr_{z\gets D^{\le m}}[C^f(z)=1]$ as in \eqref{equ: P by PRS}. Note that a circuit of size at most $\tilde t$ queries $f$ only on inputs of length at most $\tilde t$, so these values are determined at this point and are not affected by later iterations. We claim that with positive probability over the choice of $(f_{n_i},D)$, the following two events hold simultaneously.

\smallskip\noindent(\emph{Honesty}): every value of $\P$ fixed in this iteration is within $1/\tilde t$ of $\Pr_z[C^f(z)=1]$. Fix any choice of the function $f$ as mentioned above. For every circuit $C$ of size at most $\tilde t$, the empirical estimate over $s=\tilde t^4$ uniform samples deviates from the acceptance probability by more than $1/(2\tilde t)$ with probability at most $2\exp(-2s/\tilde t^2)=2\exp(-2\tilde t^2)$ by the Hoeffding bound. If this does not happen, then $P(C, \Delta)$ deviates from the acceptance probability by at most $1/(2\tilde t) + 1/(2\|\Delta\|) \le 1/\|\Delta\|$. Taking a union bound over the at most $2^{O(\tilde t\log \tilde t)}$ oracle circuits of size at most $\tilde t$, the event (\emph{Honesty}) fails with probability at most $1/6$.

\smallskip\noindent(\emph{No collision}): $\cI(1^{n_i})$ does not output distinct $x,x'\in\zo^{n_i}$ with $f(x)=f(x')$. Fix any $D$. We may assume that $\cI$ queries both $x$ and $x'$ before it halts, as this only adds two $f$-queries; it then suffices to show that with probability at least $5/6$, no two queried entries of length $n_i$ receive the same value. The answers received by the inspector are determined by the values of $f$ at its at most $t$ direct $f$-queries, together with the at most $s\cdot \tilde t$ $f$-queries when evaluating each of its at most $t$ queried circuits over the samples of $D$. Hence the run of $\cI(1^{n_i})$ queries at most
\[
B\eqdef 2+t+t\cdot s\cdot \tilde t\le 2\tilde t^{6}
\]
entries of $f$; moreover, $f$ is deterministic except on entries of length $n_i$.

For the analysis, we implement the answers to the entries of length $n_i$ by \emph{lazy sampling}: upon a query $x\in\zo^{n_i}$ that was not queried before, the oracle answers with an independent uniformly random string $v(x)\in\zo^{n_i-1}$; repeated queries are answered according to $v$. The joint distribution of the answers is identical to that of the function $f_{n_i}$ sampled upfront. By the union bound, the probability that any two queried entries receive the same value is at most $2B^2/2^{n_i}$. By setting $\eps := 1/(24d)$ and the assumption $t(n_i)\le 2^{\eps n_i}$, we have $B\le 2t^{6d}\le 2^{n_i/3}$ for sufficiently large $n_i$, and thus the event (\emph{No collision}) fails with probability at most $2B^2/2^{n_i}\le 1/6$.

\smallskip Since (\emph{Honesty}) fails with probability at most $1/6$ for every fixed $f_{n_i}$, and (\emph{No collision}) fails with probability at most $1/6$ for every fixed $D$, with positive probability both events hold; fix such a pair $(f_{n_i},D)$ and proceed to the next iteration.

Finally, suppose that $\cI(1^{n_i})$ succeeds. It cannot output a collision by (\emph{No collision}). It cannot output an instance $(j,\vec u)$ falsifying $\Psi_j$ either: since $\|\vec u\|\le t(n_i)$ and by the choice of $d$, all $f$-oracle and $\P$-oracle queries made to evaluate $\psi_j(\vec u)$ concern lengths at most $\|\vec u\|^d\le \tilde t(n_i)$ and thus are already fixed, and the pair $(f,\P)$ is a standard model of $\APX(f)$: every fixed value of $\P$ is within $1/\|\Delta\|$ of the acceptance probability with respect to the final $f$, is a rational in $[0,1]$, and equals $\Val^f(C)$ exactly whenever $C$ does not read its random input (in which case all evaluations over $D^{\le m}$ agree). Therefore $\cI(1^{n_i})$ fails for every $i$.
\end{proof}

The following is then an immediate corollary of the Herbrand witnessing theorem.

\begin{theorem}\label{thm:wphp_unprov} $\trueAPX(f)\nvdash\wPHP(f)$. In particular, $\APX(f)\nvdash \wPHP(f)$. 
\end{theorem}

\begin{proof} Let $\sT$ be the set of true universal axioms of $\APX(f)$. 
Suppose that $\PV(\P,f)+\sT\vdash \wPHP(f)$. A proof invokes only finitely many axioms $\Psi_1,\dots,\Psi_k\in \sT$, and for simplicity we assume that all axioms $\Psi_i=\forall u_i~\psi_i(u_i)$ have only one variable. As in the proof of \Cref{thm:dwphp_unprov}, we can see that:
\[
\PV(\P,f)\vdash \forall n\in\Log~\exists x,x'\in\zo^n~\exists (u_1,\dots,u_k)~\bigl(x\ne x'\land f(x)=f(x')\bigr)\lor \bigvee_{i\in[k]}\lnot\psi_i(u_i).
\]
The Herbrand witnessing theorem applied to the universal theory $\PV(\P,f)$ yields a \emph{single} polynomial-size inspector that, for every $f$ and $\P$, outputs a tuple $(x,x',u_1,\dots,u_k)$ satisfying the displayed disjunction: the inspector evaluates the finitely many Herbrand candidates extracted from the proof and outputs the first one that verifies. This inspector succeeds on every input length in the sense of \Cref{lem:wphp_lb}, which contradicts \Cref{lem:wphp_lb} as $\poly(n)\le 2^{\eps n}$ for all sufficiently large $n$.
\end{proof}

\begin{remark}
This unprovability result is closely related to an early result of Thapen \cite{Thapen05} that $S^1_2+\dwPHP\nvdash \wPHP$ (and therefore $\APC \nvdash\wPHP$) in the relativized setting. Thapen's result relies on the witnessing theorem for $S^1_2+\dwPHP$ (due to Wilkie \cite{Krajicek-book}) that if $S^1_2+\dwPHP\vdash \forall x~\exists y~\phi(x,y)$ for a quantifier-free $\phi$, then there is a \emph{zero-error} algorithm that finds such $y$ given $x$. It then suffices to use the standard fact that a random function is likely collision-resistant. Therefore, the unprovability result naturally generalizes to the theory that contains all true $\forall\Sigma_0^b(\PV)$ consequences of $S^1_2+\dwPHP$, for which the witnessing theorem still applies. 

\Cref{thm:wphp_unprov} is technically incomparable as the true universal theory $\APX(f)$ could contain sentences that are not in $S^1_2+\dwPHP$ (or even its extension with all true $\forall\Sigma_0^{b}(\PV)$ sentences). Nevertheless, the proof follows exactly the same route: For each fixed $f$, any $\forall\Sigma_1^b(\PV(f))$-consequence, including $\wPHP(f)$ admits a zero-error witnessing algorithm (i.e.~answering $\P$-oracle queries by empirical estimation). Then take $f$ to be a random function, which is likely collision-resistant. 
\end{remark}

\subsection{Provability of (d)wPHP in \texorpdfstring{$\mAPX$}{APX1*}}\label{sec:upper_bounds_pigeons}

We now show both $\wPHP$ and $\dwPHP$ are provable in $\mAPX$, the bounded theory for multiplicative-error approximate counting. We note that a computational version of these upper bounds, specifically 
\[
\WeakPigeon \in \FP^{\prSBP} \qquad\text{and}\qquad \Avoid \in \FP^{\prSBP},
\]
is probably folklore (see \cite{KP24,HV26} for discussions of the latter inclusion). Our main contribution is the proof complexity upper bounds of the principles in the (seemingly weak) theory $\mAPX$. Moreover, by Thapen's witnessing theorem (see \Cref{lmm: witnessing TFNP}), we obtain (possibly) better upper bounds for $\PWPP$, which might be of independent interest (see \Cref{thm: ub WPHP}). 

\subsubsection{Basic Counting Functionalities in \texorpdfstring{$\mAPX$}{APX*}}
\label{sec: counting in mAPX}

We first develop basic counting functionalities in the theory $\mAPX$, analogous to \cite[Section 3]{CLOW26} for $\APX$. These results follow from essentially the same idea in \cite{CLOW26}; we defer the proofs of \Cref{lmm: global consistency mAPX,lmm: nonexistence mAPX,lmm: averaging mAPX} to \Cref{sec: missing counting mAPX}. 

\begin{restatable}[Global Consistency for $\mAPX$]{lemma}{lmmGlobalConsistency}\label{lmm: global consistency mAPX}
  $\mAPX$ proves the following statement. Let $n,\delta^{-1}\in\Log$ and $C_1,C_2:\zo^n\to\zo$ be circuits such that $C_1(x)\le C_2(x)$ for every $x\in\zo^n$. Then
  \[
  (1-\delta)\cdot\P_\delta(C_1)\mlsim(1+\delta)\cdot\P_\delta(C_2).
  \]
  In particular, if $C_1(x)=C_2(x)$ for every $x\in\zo^n$, then both $(1-\delta)\cdot\P_\delta(C_1)\mlsim(1+\delta)\cdot\P_\delta(C_2)$ and $(1-\delta)\cdot\P_\delta(C_2)\mlsim(1+\delta)\cdot\P_\delta(C_1)$.
\end{restatable}

We will need the fact that if the acceptance probability of a circuit is noticeably smaller than $(1-\delta)\cdot 2^{-n}$, then it is unsatisfiable. This can be viewed as a multiplicative-error counterpart of \cite[Lemma 3.7]{CLOW26} on the acceptance probability of the ``less-than-$t$'' circuit.

\begin{restatable}[Non-Existence Lemma for $\mAPX$]{lemma}{lmmNonExistence}\label{lmm: nonexistence mAPX}
  $\mAPX$ proves the following statement. Let $n,\delta^{-1}\in\Log$ satisfying $\delta<1$ and $C:\zo^n\to\zo$ be a circuit. If there is $x\in\zo^n$ such that $C(x)=1$, then $(1-\delta)\cdot2^{-n}\mlsim\P_\delta(C)$. (Or equivalently, if $(1+\beta)\cdot\P_\delta(C)<(1-\delta)\cdot2^{-n}$ for some $\beta^{-1}\in\Log$, then $C(x)=0$ for every $x\in\zo^n$.)
\end{restatable}

We will need the following generic averaging argument, which implies several useful facts. The counterpart for $\APX$ is in \cite[Section 3.2.3]{CLOW26}. 

\begin{restatable}[Averaging on Linear Combination for $\mAPX$]{lemma}{lmmAveraging}\label{lmm: averaging mAPX}
  $\mAPX$ proves the following statement. Let $n,m,\delta^{-1}\in\Log$, $C_1,\dots,C_m:\zo^n\to\zo$ be circuits, and $\lambda^+_1,\dots,\lambda^+_m,\lambda^-_1,\dots,\lambda^-_m\in\Q$ be non-negative. Let
  \[
  \mu^+\eqdef\sum_{i\in[m]}\lambda_i^+\cdot\P_\delta(C_i)
  \qquad\text{and}\qquad
  \mu^-\eqdef\sum_{i\in[m]}\lambda_i^-\cdot\P_\delta(C_i).
  \]
  Then for every $\beta^{-1}\in\Log$, there is $z\in\zo^n$ such that
  \[
  \frac{(1+\beta)(1+\delta)\cdot\sum_{i\in[m]}\lambda_i^+\cdot C_i(z)}{(1-\delta)\cdot\sum_{i\in[m]}\lambda_i^-\cdot C_i(z)}\ \ge\ \frac{\mu^+}{\mu^-},
  \]
  where $a/b\ge c/d$ is an abbreviation of $ad \ge bc$. Moreover, if $\mu^+>0$ and $\mu^->0$, then $z$ can be chosen such that the inequality is strict, i.e., $ad>bc$.
\end{restatable}

\begin{corollary}[Existence Lemma for $\mAPX$]\label{cor: existence mAPX}
  $\mAPX$ proves the following statement. Let $n,\delta^{-1}\in\Log$, $\delta<1$, and $C:\zo^n\to\zo$ be a circuit. If $\P_\delta(C)>0$, then there is $x\in\zo^n$ such that $C(x)=1$.
\end{corollary}

\begin{proof}
  We work in $\mAPX$. Fix $n,\delta^{-1}\in\Log$. Let $\mu\eqdef \P_\delta(C)>0$ and $\beta=0.01$. By \nameref{lmm: averaging mAPX} with circuits $C$ and the constant-$1$ circuit $O:\zo^n\to\zo$ (that does not read the input and outputs $1$) and coefficients $\lambda_C^+=\lambda_O^-=1$, $\lambda_C^-=\lambda_O^+=0$, there exists $x\in\zo^n$ such that 
  \[ 
  \frac{(1+\beta)(1+\delta) C(x)}{(1-\delta) O(x)} \ge \frac{\mu}{\P_\delta(O)} = \frac{\mu}{1}.
  \] 
  Note that $O(x)=1$, and the equality follows from the \nameref{axiom: boundary mAPX}. This further implies that $C(x)\ge \frac{(1-\delta)\mu}{(1+\beta)(1+\delta)}>0$; subsequently, $C(x)=1$. 
\end{proof}

Note that the existence lemma for $\APX$ in \cite{CLOW26} requires $\P_\delta(C)>\delta+\beta$ for some $\beta^{-1}\in\Log$, whereas in $\mAPX$ it suffices that $\P_\delta(C)$ is positive.

\begin{corollary}[Inclusion-Exclusion for $\mAPX$]\label{cor: inclusion exclusion mAPX}
  $\mAPX$ proves the following statement. Let $n,\delta^{-1}\in\Log$ and $C_1,C_2,C_\lor,C_\land:\zo^n\to\zo$ be circuits such that $C_\lor(x)=C_1(x)\lor C_2(x)$ and $C_\land(x)=C_1(x)\land C_2(x)$ for every $x\in\zo^n$. Then
  \begin{align}
  (1-\delta)\cdot\big(\P_\delta(C_\lor)+\P_\delta(C_\land)\big)&\mlsim(1+\delta)\cdot\big(\P_\delta(C_1)+\P_\delta(C_2)\big)\quad\text{and}\label{equ: Inc Exc 1}\\
  (1-\delta)\cdot\big(\P_\delta(C_1)+\P_\delta(C_2)\big)&\mlsim(1+\delta)\cdot\big(\P_\delta(C_\lor)+\P_\delta(C_\land)\big)\label{equ: Inc Exc 2}.
  \end{align}
  In particular, if $C_1(x)\land C_2(x)=0$ for every $x\in\zo^n$, then $(1-\delta)\cdot\P_\delta(C_\lor)\mlsim(1+\delta)\cdot\big(\P_\delta(C_1)+\P_\delta(C_2)\big)$ and $(1-\delta)\cdot\big(\P_\delta(C_1)+\P_\delta(C_2)\big)\mlsim(1+\delta)\cdot\P_\delta(C_\lor)$.
\end{corollary}

\begin{proof}
  We work in $\mAPX$. Fix $n,\delta^{-1}\in\Log$ and circuits $C_1,C_2,C_\lor,C_\land:\zo^n\to\zo$. We only prove \eqref{equ: Inc Exc 2}; other parts are left as an exercise.\footnote{For \eqref{equ: Inc Exc 1}, the case $\P_\delta(C_1)=\P_\delta(C_2)=0$ additionally needs the \nameref{lmm: nonexistence mAPX}, which implies that $C_1$ and $C_2$, and hence $C_\lor$, are identically $0$.} Let $\lambda_1^+=\lambda_2^+=\lambda_\lor^-=\lambda_\land^-=1$ and $\lambda_\lor^+=\lambda_\land^+=\lambda_1^-=\lambda_2^-=0$. We apply \nameref{lmm: averaging mAPX} to circuits $C_1,C_2,C_\lor,C_\land$ and coefficients $\lambda_i^+,\lambda_i^-$. Let 
  \[
  \mu^+\eqdef \P_\delta(C_1) + \P_\delta(C_2), \qquad \mu^-\eqdef \P_\delta(C_\lor)+\P_\delta(C_\land).
  \] 
  
  We may assume that $\mu^+>0$, as otherwise \eqref{equ: Inc Exc 2} is trivial. We may also assume that $\mu^->0$. This is because if $\mu^-=0$, then $\P_\delta(C_\lor)=\P_\delta(C_\land)=0$ by the \nameref{axiom: basic mAPX}. Since $C_i(x)\le C_\lor(x)$ for every $x\in\zo^n$ and $i\in\{1,2\}$, \nameref{lmm: global consistency mAPX} gives $(1-\delta)\cdot\P_\delta(C_i)\mlsim(1+\delta)\cdot\P_\delta(C_\lor)=0$, i.e., $(1-\delta)\cdot\P_\delta(C_i)\le 0$; therefore $(1-\delta)\cdot\mu^+\le 0$, and \eqref{equ: Inc Exc 2} holds.

  Assume that $\mu^+,\mu^->0$. Then (the strict form of) \nameref{lmm: averaging mAPX} implies that for every $\beta^{-1}\in\Log$, there exists $z\in\zo^n$ such that 
  \[
  \frac{\mu^+}{\mu^-} < \frac{(1+\beta)(1+\delta)(C_1(z)+C_2(z))}{(1-\delta)(C_\lor(z)+C_\land(z))} . 
  \]
  The inequality means that $(1-\delta)(C_\lor(z)+C_\land(z))\cdot\mu^+<(1+\beta)(1+\delta)(C_1(z)+C_2(z))\cdot\mu^-$. Since the LHS is non-negative, $C_1(z)+C_2(z)>0$. We can then cancel the common positive factor $C_1(z)+C_2(z)=C_\lor(z)+C_\land(z)$, which gives \eqref{equ: Inc Exc 2}. 
\end{proof}

\subsubsection{An Upper Bound for dwPHP}
\label{sec: upper bound dwPHP}

\begin{restatable}{theorem}{UBdwPHP}\label{thm: UB dwPHP}
$\mAPX(g)\vdash \dwPHP(g)$. Consequently, $\mAPX\vdash\dwPHP$.  
\end{restatable}

Indeed, we will prove a slightly stronger version: there exists a $(g,\P)$-oracle algorithm $A(1^n)$ that finds $y\in\zo^n$ that is not in the range of $g$. This immediately implies \Cref{thm: UB dwPHP} by fixing $y\eqdef A(1^n)$. 

\begin{theorem}\label{thm: UB dwPHP stronger}
Let $g$ be a function symbol with intended output length $\|g(u)\|=\|u\|+1$. There is a $\PV(g,\P)$ function $A$ such that the following is provable in $\mAPX(g)$: for every $n\in\Log$, $A(1^n)$ outputs $y\in\zo^n$ such that for every $x\in\zo^{n-1}$, $g(x)\ne y$. 

Consequently, there is a $\PV(\P)$ function $B$ such that $\mAPX$ proves the following: for every $n\in\Log$ and circuit $C:\zo^{n-1}\to\zo^n$, $B(C)$ outputs $y\in\zo^n$ such that for any $x\in\zo^{n-1}$, $C(x)\ne y$. 
\end{theorem}

\AI{The proof overview is by the authors; the proof was drafted by Claude Opus 5.5 following the overview and the proof of \Cref{thm: ub WPHP}. The authors then reviewed and edited the proof for clarity.}
\begin{proof}
We argue in $\mAPX(g)$. Let $n\in\Log$ (we may assume $n\ge 1$) and $\eta^{-1}\in\Log$ be a parameter to be determined (where $\eta\le 1/100$), and $\kappa\eqdef(1-\eta)/(1+\eta)^2$. For $y\in\zo^{\le n}$, let $T_y:\zo^{n-1}\to\zo$ be the $g$-oracle circuit such that $T_y(x)=1$ iff $y$ is a prefix of $g(x)$, and $p(y)\eqdef \P_{\eta}(T_y)$. In particular, for $y=\eps$ being the empty string, $T_y$ is a circuit that is functionally equivalent to the constant-$1$ circuit. By the \nameref{axiom: boundary mAPX} and \nameref{lmm: global consistency mAPX}, we can show that $p(\eps)\le (1+\eta)^2/(1-\eta)=\kappa^{-1}$.

Consider the $(g,\P)$-oracle algorithm $A(1^n)$ that will find a string $y\in\zo^{n}$ that is not in the range of $g$. The algorithm starts with $y\gets \eps$, and iteratively constructs $y$ such that $p(y)$ is sufficiently small.
\begin{itemize}
  \item For $\|y\|<n$, it can be proved in $\PV$ that $T_{y0}(x)\land T_{y1}(x)=0$ for every $x\in\zo^{n-1}$, and $T_y(x)=T_{y0}(x)\lor T_{y1}(x)$ (using the length axiom $\|g(x)\|=n$). By the \nameref{cor: inclusion exclusion mAPX} with $C_1\eqdef T_{y0},C_2\eqdef T_{y1}$, and $C_{\lor}\eqdef T_y$, we can prove that
  \[
  p(y0)+p(y1)=\P_\eta(T_{y0}) + \P_{\eta}(T_{y1}) \le \frac{(1+\eta)^2}{1-\eta}\cdot \P_{\eta}(T_y)=\kappa^{-1}\cdot p(y).
  \]
  \item When $\|y\|<n$, the algorithm chooses $b\in\zo$ such that $p(yb)\le \frac{1}{2\kappa}\cdot p(y)$, and sets $y\gets yb$.
\end{itemize}
By induction on a quantifier-free formula, we can prove that the algorithm $A(1^n)$ outputs $y\in\zo^{n}$ such that $p(y)\le \kappa^{-1}\cdot(2\kappa)^{-n}=2^{-n}\cdot \kappa^{-(n+1)}$.

We now set the parameters. Let $\eta\eqdef 1/(100n)\le 1/100$, so we have $\kappa\ge 1-5\eta$ and $\kappa^{n+1}>0.8$. This means that
\[
(1+\eta)\cdot p(y)\le (1+\eta)\cdot\kappa^{-(n+1)}\cdot 2^{-n}< 1.3\cdot 2^{-n}<(1-\eta)\cdot 2^{-(n-1)}.
\]
Since $T_y$ has $n-1$ input bits, by the \nameref{lmm: nonexistence mAPX} (with $\delta\eqdef\eta$ and $\beta\eqdef\eta$), we know that $T_y(x)=0$ for every $x\in\zo^{n-1}$. This completes the proof, as $T_y(x)=0$ means that $y$ is not a prefix of $g(x)$, and in particular $g(x)\ne y$.
\end{proof}

\subsubsection{An Upper Bound for wPHP}
\label{sec: upper bound wPHP}

\begin{restatable}{theorem}{UBwPHP}\label{thm: ub WPHP}
$\mAPX(f)\vdash \wPHP(f)$. Consequently, $\mAPX\vdash\wPHP$.  
\end{restatable}

\AI{The proof is discovered and written by authors. Claude Opus 5.5 is used to fix bugs.}
\begin{proof}
We argue in $\mAPX(f)$. Let $n\in\Log$ and $\eta^{-1}\in\Log$ be a parameter to be determined (where $\eta\le 1/100$), and $\kappa\eqdef(1-\eta)/(1+\eta)^2$. For $y\in\zo^{\le n-1}$, let $T_y:\zo^n\to\zo$ be the $f$-oracle circuit such that $T_y(x)=1$ iff $y$ is a prefix of $f(x)$, and $p(y)\eqdef \P_{\eta}(T_y)$. In particular, for $y=\eps$ being the empty string, $T_y$ is a circuit that is functionally equivalent to the constant-$1$ circuit. By the \nameref{axiom: boundary mAPX} and \nameref{lmm: global consistency mAPX}, we can show that $p(\eps)\ge (1-\eta)/(1+\eta)^2=\kappa$. 

Consider the $(f,\P)$-oracle algorithm $A(1^n)$ that will find a string $y\in\zo^{n-1}$ such that $|f^{-1}(y)|\ge 2$. The algorithm starts with $y\gets \eps$, and iteratively constructs $y$ such that $p(y)$ is sufficiently large.  
\begin{itemize}
  \item It can be proved that $T_{y0}(x)\land T_{y1}(x)=0$ for every $x\in\zo^n$, and $T_y(x)=T_{y0}(x)\lor T_{y1}(x)$. By the \nameref{cor: inclusion exclusion mAPX} with $C_1\eqdef T_{y0},C_2\eqdef T_{y1}$, and $C_{\lor}\eqdef T_y$, we can prove that 
  \[
  p(y0)+p(y1)=\P_\eta(T_{y0}) + \P_{\eta}(T_{y1}) \ge \frac{1-\eta}{(1+\eta)^2}\cdot \P_{\eta}(T_y)=\kappa\cdot p(y). 
  \]
  \item When $\|y\|<n-1$, the algorithm chooses $b\in\zo$ such that $p(yb)\ge \frac{\kappa}{2}\cdot p(y)$, and sets $y\gets yb$.
\end{itemize}
By induction on a quantifier-free formula, we can prove that the algorithm $A(1^n)$ outputs $y\in\zo^{n-1}$ such that $p(y)\ge 2^{-(n-1)}\cdot \kappa^{n}$.  

For $x\in\zo^{\le n}$, let $T_{y,x}:\zo^{n-\|x\|}\to\zo$ be the circuit such that $T_{y,x}(z)=1$ iff $f(z\circ x)=y$. In particular, $T_{y,\eps}$ is functionally equivalent to $T_y$. Let $q_y(x)\eqdef \P_\eta(T_{y,x})$; by \nameref{lmm: global consistency mAPX}, we know that $q_y(\eps)\ge \kappa \cdot p(y)$. Next, we consider the following $(f,\P)$-oracle algorithm $E(1^n,y)$ that outputs $x\in\zo^{<n}$ such that $q_y(0x)>0$ and $q_y(1x)>0$. The algorithm starts with $x\gets \eps$ and constructs $x$ iteratively as follows. 
\begin{itemize}
  \item Note that for $b\in\zo$, $T_{y,bx}\eqdef\Fix_b(T_{y,x})$. Therefore by the \nameref{axiom: local consistency mAPX}, we know that 
  \[
  \frac{q_y(0x)+q_y(1x)}{2}=\frac{\P_\eta(T_{y,0x})+\P_\eta(T_{y,1x})}{2} \ge \frac{1-\eta}{(1+\eta)^2}\cdot \P_\eta(T_{y,x})=\kappa\cdot q_y(x). 
  \]
  \item In each iteration, the algorithm computes $q_y(0x)$ and $q_{y}(1x)$. If both are positive, it succeeds and outputs $x$. Otherwise, it sets $x\gets bx$ for the $b\in\zo$ maximizing $q_y(bx)$, so $q_y(bx)\ge 2\kappa\cdot q_y(x)$.
\end{itemize} 
By induction on a quantifier-free formula, we can prove that in the $i$-th iteration, the algorithm maintains $x\in\zo^i$ such that $q_{y}(x)\ge (2\kappa)^i\cdot q_y(\eps)\ge  (2\kappa)^i\cdot 2^{-(n-1)}\cdot \kappa^{n+1}$. If the algorithm does not succeed after $n$ iterations, we will construct $x\in\zo^n$ such that $q_{y}(x)\ge (2\kappa)^n\cdot 2^{-(n-1)}\cdot \kappa^{n+1}=2\kappa^{2n+1}$.

We now set the parameters. Let $\eta\eqdef 1/(100n)\le 1/100$, so we have $\kappa\ge 1-5\eta$ and $\kappa^{2n+1}>0.8$. This means if the algorithm does not succeed, we will have $q_y(x)>1$, which violates the \nameref{axiom: basic mAPX}. Subsequently, the algorithm $E(1^n,y)$ must succeed in finding $x\in\zo^{<n}$ such that $q_{y}(0x),q_y(1x)>0$. Then by the \nameref{cor: existence mAPX}, there exist $z_0,z_1$ such that
\[
T_{y,0x}(z_0) = 1,~\text{i.e.,}~ f(z_0\circ 0\circ x) = y, \qquad T_{y,1x}(z_1) = 1,~\text{i.e.,}~ f(z_1\circ 1\circ x) = y.
\]
This completes the proof, as $x_0\eqdef z_0\circ 0\circ x$ and $x_1\eqdef z_1\circ 1\circ x$ are distinct strings of length $n$ such that $f(x_0)=f(x_1)$. 
\end{proof}

\paragraph{Complexity-theoretic implication.} It is worth mentioning that \Cref{thm: ub WPHP} has an immediate consequence in complexity theory by the witnessing theorems of Thapen \cite{Thapen25} (see \Cref{lmm: witnessing TFNP}). Recall that $\Local(\APX)$ (resp.~$\Local(\mAPX)$) is the total search problem that, given a non-uniform query algorithm representing a strategy of the inspector in the Inspector-Adversary game for the theory $\APX$ (resp.~$\mAPX$), outputs a transcript on which the inspector fails (see \Cref{def: IAG,def: IAG APX}). By definition, $\Local(\APX)$ and $\Local(\mAPX)$ are in $\TFNP$ as the standard models of $\APX,\mAPX$ give a winning strategy for the adversary. 

We first recall the definition of some search problems. 

\begin{definition}[$\WeakPigeon$; see, e.g., \cite{jerabek:factoring}]
$\WeakPigeon$ is defined as the following search problem in $\TFNP$: Given a circuit $C:\zo^n\to\zo^{n-1}$, a solution is a pair of distinct strings $x_1,x_2\in\zo^n$ such that $C(x_1) = C(x_2)$. The complexity class $\PWPP$ of search problems is defined as the $\TFNP$ problems that are polynomial-time reducible to $\WeakPigeon$. 
\end{definition}

\begin{definition}[$\Avoid$; see, e.g., \cite{Korten-survey} and references therein]\label{def: avoid}
$\Avoid$ is defined as the following search problem in $\TFSigma{2}$: Given a circuit $C:\zo^{n-1}\to\zo^{n}$, a solution is a string $y\in\zo^n$ such that for every $x\in\zo^{n-1}$, $C(x)\ne y$. 
\end{definition}

\begin{theorem}\label{thm: reduction to Local}
  $\WeakPigeon\le_p\Local(\mAPX)$ via single-query polynomial-time reducibility. 
\end{theorem}

\begin{proof}
This follows directly from the unrelativized version of \Cref{thm: ub WPHP} and \Cref{lmm: witnessing TFNP}. 
\end{proof}

We note that \Cref{thm: UB dwPHP} or \Cref{thm: UB dwPHP stronger} cannot be used to prove $\Avoid\le_p\Local(\mAPX)$ as $\Avoid$ is not a $\TFNP$ problem. In fact, under plausible assumptions, $\Avoid\nleq_{p}\TFNP$ since otherwise there will be an $\NP$ search algorithm for $\Avoid$ (see \cite{CL24,Ilango25,RWZ26,LRZ26}).

\section{Finite Combinatorial Principles, Infinite Models, and the Limits of \texorpdfstring{$\APX$}{APX1}}
\label{sec:riis}

In this section, we establish \Cref{thm:intro-riis} and \Cref{cor:intro-riis}.  The proof of \Cref{thm:intro-riis} is a combination of two steps. First, we introduce a certain game played between an inspector and an adversary. We show in \Cref{sec:adversary_winning} that an infinite model provides a winning strategy for the adversary in this game. In contrast, we show in \Cref{sec:proofs_to_games} that a proof of the relevant sentence in the theory yields an inspector that wins the game. Combining these two results implies the  unprovability in \Cref{thm:intro-riis}.

\RiisThm*

For simplicity, we work with quantifier-free $\mathcal L$-formulas in \Cref{sec:adversary_winning} below. In \Cref{sec:red-herbrand} and \Cref{sec:proofs_to_games}, we describe how standard techniques allow us to reduce the case of quantified formulas to the simpler setting of quantifier-free formulas.

\subsection{Winning Strategies from Infinite Models}\label{sec:adversary_winning}

In this section, we fix some finite first-order language $\cL$ consisting of function and constant symbols, and a \emph{quantifier-free} formula $\phi(v_1\ldots,v_k)$ over $\cL$ with equality and exactly $k$ free variables $v_1,\ldots,v_k$. We assume that $\phi$ exhausts all the symbols in $\cL$, so that $\phi$ determines $\cL$ (this is for notation, so that various parameters with dependence on $\cL, \phi$ can be defined only in terms of $\phi$). We will consider a game $\mathcal{G}_\phi(N)$ played over finite universes of size $N$, where $\phi$ is fixed and $N$ grows to infinity. We start by introducing some terminology necessary to describe the game:

\begin{definition}[Partial Structures, witnesses and decision trees]
A partial $\cL$-structure $\rho$ on $[N]$ assigns a value $c^\rho$ in $[N]$ to some subset of the constant symbols and assigns a partial function $f^\rho: [N]^d\to[N]$ to each $d$-ary function symbol $f$. For two partial structures $\rho, \rho'$ we say $\rho' \supseteq \rho$ (``$\rho'$ extends $\rho$'') if every constant symbol defined in $\rho$ is defined and has the same value in $\rho'$, and every partial function in $\rho'$ extends the corresponding partial function in $\rho$. 
For an assignment $x_1,\ldots,x_k \in [N]$ to the variables $v_1,\ldots,v_k$ of $\phi$, we say that $\rho \models \phi[\bar x]$ if each term in $\phi[\bar x]$ has a well-defined value in $[N]$ under the partial structure $\rho$ and the assignment $\bar x$ (according to the natural interpretation\footnote{A formal definition involves an induction on the depth of terms. If $t = v_i$ its value is $x_i$; if $t = c$ for a constant symbol $c$, its value is $c^\rho$ if $c^\rho$ is defined, otherwise it is undefined. If $t = f(t_1,\ldots,t_d)$ for $f$ a function symbol of arity $d$, and each $t_i$ has a defined value $z_i \in [N]$, and $f^{\rho}(\bar z)$ is defined and equals $y$, then the value of $t $ is $ y$; in all other cases it is undefined.}), and after substituting these values for the terms, the sentence $\phi[\bar x]$ is true.
We say that a partial structure is ``$\phi$-witnessing'' if there exists some $(x_1,\ldots,x_k) \in [N]^k$ such that $\rho \models \phi[x_1,\ldots,x_k]$.

An $(N,\phi)$ decision tree $T$ is a finite rooted tree. Each internal node queries either a constant $c$ or a ``table entry'' $f(\bar x)$, where $\bar x\in[N]^{\operatorname{arity}(f)}$; this node has one child for every possible answer in $[N]$. Every leaf $\ell \in T$ is labeled with a bit value $b_\ell \in \{0,1\}$. For two such trees $T_1, T_2$, we define $T_1 \circ T_2$ to be the tree obtained by attaching a copy of $T_2$ to each leaf of $T_1$ (the root of each copy is identified with its associated leaf). We will not bother to define the bit values $b_\ell$ for $T_1 \circ T_2$ since they have no relevance to us. The height of the tree $T$ is the maximum number of queries on a root-to-leaf path. An $(N,\phi)$ RDT (randomized decision tree) is a probability distribution over $(N,\phi)$ decision trees; its height is the maximum height of a tree in its support. When $N,\phi$ are clear from context, we simply refer to them as decision trees and RDTs respectively.

For a partial structure $\rho$ and a tree $T$, we say that $\rho$ ``determines a path through $T$ to a leaf'' if, starting at the root of $T$ and answering each query according to $\rho$, no query ever yields an undefined value (and hence we trace a path to some leaf).

\end{definition}

\begin{definition}[Game $\mathcal{G}_\phi(N; m,h,\epsilon)$] \label{def:game}
For $\phi$ as above, $N,m,h\in \mathbb{N}$, $\epsilon \in (0,1]$, we define a game $\mathcal{G}_\phi(N;m,h,\epsilon)$ as follows. The players are called the ``inspector'' and the ``adversary.'' The game proceeds for $m$ rounds. In round $i$, the inspector supplies an $(N,\phi)$ RDT $\mathcal{T}_i$ of height at most $h$, and the adversary responds with a value $p_i \in [0,1]$. After all rounds are complete, the adversary supplies a partial $\cL$-structure $\rho$ on $[N]$. The adversary wins when all of the following conditions hold:
\begin{enumerate}
    \item $\rho$ is not $\phi$-witnessing. 
    \item (\emph{Deterministic Consistency}).\; If $\operatorname{supp}(\calt_i)=\{T\}$ for some deterministic tree $T$, then $\rho$ determines a path in $T$ ending at a leaf $\ell \in T$ with $b_\ell = p_i$; in particular, $p_i\in\zo$.
    \item (\emph{Mixture Consistency}).\; For any three round indices $i_1,i_2,i_3$ such that $\calt_{i_1}=\tfrac12\calt_{i_2}+\tfrac12\calt_{i_3}$, we have  $\left|p_{i_1}-\tfrac12p_{i_2}-\tfrac12p_{i_3}\right|\leq\eps$.
\end{enumerate}
\end{definition}
We are now ready to state the main theorem of this section:
\begin{theorem}[Winning strategy]\label{thm:riis:main}
Suppose there is an infinite $\cL$-structure $\cM$ satisfying
\[
    \cM\models\forall v_1\cdots\forall v_k\,\neg\phi(v_1,\ldots,v_k).
\]
Let $a$ be the maximum arity of a function symbol in $\cL$ \emph{(}$a=0$ if $\cL$ only has constant symbols\emph{)}. The adversary has a winning strategy in $\mathcal{G}_\phi(N;m,h,\eps)$ whenever
\begin{equation}\label{eq:riis:main-bound}
    N\ \geq\ 10^5(a+1)^2m^2h^2\eps^{-4}\bigl(\ln(2m)\bigr)^2.
\end{equation}
\end{theorem}

Let $\cU$ be the (infinite) universe of $\cM$. Our argument will center around the construction of a certain partial injection $\pi: [N] \to \cU$. For any such partial injection we can naturally associate a partial $\cL$ structure $\rho_\pi$ on $[N]$: for a constant $c$, if $c^{\mathcal{M}} = X$ and $\pi^{-1}(X)$ is defined, we set $c^{\rho_\pi} = \pi^{-1}(X)$. Similarly, if $f$ is a function symbol of arity $d$, $f^{\cM}(\bar X) = Y$, and $\pi^{-1}(X_1),\ldots,\pi^{-1}(X_d), \pi^{-1}(Y)$ are all defined, then we set $f^{\rho_\pi}(\pi^{-1}(X_1),\ldots,\pi^{-1}(X_d)) = \pi^{-1}(Y)$. In all other cases $\rho_\pi$ is undefined. Observe that if $\pi' \supseteq \pi$ then $\rho_{\pi'} \supseteq \rho_{\pi}$, and for all $\pi$, $\rho_\pi$ is not $\phi$-witnessing. We use $|\pi|$ as shorthand for $|\{ x\in [N] \mid \pi \text{ is defined at } x\}|$. We will describe a randomized strategy for the adversary; before the $i^{th}$ round, the adversary will have generated some partial injection $\pi_{i- 1}$, and after seeing the inspector's next move $\cT_i$, the adversary will choose an extension $\pi_{i} \supseteq \pi_{i-1}$ (using randomness) and some answer $p_i$ for that round. At the end, the adversary will supply $\rho_{\pi_m}$ as the final partial $\cL$-structure. We will argue that for any inspector strategy, with positive probability over the internal randomness of our adversary, the resulting play-sequence will be winning for the adversary. Since $\rho_{\pi_m}$ will always be nonwitnessing, we will only have to confirm the other two win conditions; in fact, once our strategy is defined it will be clear also that condition (2) for singleton RDTs is handled automatically, so the entire crux of the proof will be in demonstrating that the affine constraints in (3) are satisfied.

We will now define one of the two sources of randomness used by the adversary. Fix an arbitrary set $\cS\subseteq\cU$ with $|\cS|=N$.  Define a ``priority oracle'' $\omega$ to consist of a total ordering $\omega_x$ on $\cS$ for each $x \in [N]$, as well as
a total ordering $\omega_X$ on $[N]$ for each $X \in \cU$.
Let $\Omega$ be the set of all priority oracles, equipped with the product of the uniform measures on the component orderings; $\omega\sim\Omega$ denotes a sample from this distribution. For a priority oracle $\omega$, a decision tree $T$ of height $h$, and a partial injection $\pi:[N]\to\cU$ satisfying $|\pi|+(a+1)h\leq N$, we define $T^\omega(\pi)$ to be the extension of $\pi$ obtained by the following traversal. Start at the root of $T$ with $\pi'=\pi$. At each internal node querying $f(x_1,\ldots,x_d)$, perform the following steps, treating a constant query $c$ as a zero-arity query with value $c^\cM$:
\begin{enumerate}
    \item For $i=1,\ldots,d$ in order, if $x_i\notin\operatorname{dom}(\pi')$, extend $\pi'$ by setting $\pi'(x_i)$ to the $\omega_{x_i}$-least element of $\cS\setminus\range(\pi')$.
    \item Set $X=f^\cM(\pi'(x_1),\ldots,\pi'(x_d))$. If $X\notin\range(\pi')$, extend $\pi'$ by setting $\pi'(y)=X$, where $y$ is the $\omega_X$-least element of $[N]\setminus\operatorname{dom}(\pi')$.
    \item Follow the edge in $T$ labeled $(\pi')^{-1}(X)$.
\end{enumerate}
Upon reaching a leaf, set $T^\omega(\pi)=\pi'$. Let $\out^\omega(T,\pi)$ denote the bit labeling the leaf reached in this traversal. 

We are now ready to describe the adversary's strategy. At the start, the adversary samples once and for all $\omega \sim \Omega$, and initializes $\pi_0 = \emptyset$. Fix some parameter $s$ depending on $N,m,h,\epsilon$ which we will set later. Say that round $i-1$ is complete and the current injection is $\pi_{i-1}$. In round $i$, if the inspector supplies $\cT_i$, the adversary uses fresh independent randomness to sample $T_{i,1},\ldots T_{i,s} \sim \cT_i$, sets $\pi_i = (T_{i,1} \circ \cdots \circ T_{i,s})^\omega(\pi_{i-1})$, and sets $p_i = \frac{1}{s} \sum_{j \leq s} \out^\omega(T_{i,j},\pi_i)$. At the end of the last round $m$, the adversary outputs $\rho_{\pi_m}$ as the final partial structure. It is now easy to observe (as promised earlier) that winning condition (2) will always be satisfied for the adversary. Since $\pi_m \supseteq \pi_i$, whenever $\cT_i$ is a singleton supported on a fixed tree $T_i$, we have 
$p_i = \out^\omega(T_i,\pi_i) = \out^\omega(T_i,\pi_m)$. It remains to argue that with nonzero probability, the final state of the game satisfies  condition (3).

Given a transcript of the game (including all the sampled randomness used in the adversary's strategy), we define 3 more sequences of numbers $(p_i^\emp, p_i^\true, p_i^\final \in [0,1])_{i \leq m}$ which will be used in the analysis of the game's outcome. These values are given by:
\[
    p_i^\emp = \frac{1}{s} \sum_{j \leq s} \out^\omega(T_{i,j},\pi_{i-1}),\quad
    p_i^\true = \mathbb{E}_{T_i \sim \cT_i}[\out^\omega(T_i, \pi_{i-1})],\quad
    p_i^\final = \mathbb{E}_{T_i \sim \cT_i}[\out^\omega(T_i, \pi_{m})].
\]
Observe that whenever $i_1,i_2,i_3$ are such that $\calt_{i_1}=\tfrac12\calt_{i_2}+\tfrac12\calt_{i_3}$, we must have $p^\final_{i_1 } = \frac{1}{2} p^\final_{i_2} + \frac{1}{2} p^\final_{i_3}$ exactly. We will argue that with nonzero probability over $\omega$ and the additional subsampling randomness of the adversary, the following holds for all $i$:
\[
p_i = p_i^{\emp}, \quad |p_i^{\emp}- p_i^\true| \leq \frac{\epsilon}{4}, \quad|p_i^\true - p_i^\final| \leq \frac{\epsilon}{4}
\]
This will imply that for all such $i_1,i_2,i_3$, $|p_{i_1} - (\frac{1}{2} p_{i_2} + \frac{1}{2} p_{i_3})| \leq \epsilon $ and hence the game state is winning for the adversary (recall that the other two win conditions are satisfied by design).

We need two technical claims before we can proceed with the proof of Theorem~\ref{thm:riis:main}. We first define a ``fresh state'' which consists of a pair $(\pi, \widetilde{\Omega})$ with $\pi: [N] \to \cU$ a partial injection and $\widetilde{\Omega}$ a distribution of priority oracles satisfying the following: all of the orders $\widetilde{\Omega}_x, \widetilde{\Omega}_X$ with $x \in [N], X \in \cU$ unmatched in $\pi$ are jointly independent and uniformly distributed (``unmatched'' means $\pi(x)$ is undefined, and $X \notin \range(\pi)$). The first technical claim is the simple observation that, if we start in a fresh state, then run a decision tree and condition on its outcome, we end up in another fresh state. In other words, the priority orders of unmatched elements remain independent and uniform, since consulting an element's order immediately matches that element, leaving the orders of all remaining unmatched elements untouched.

\begin{observation}\label{obs:riis:fresh}
Let $(\pi, \widetilde{\Omega})$ be a fresh state and let $T$ be a decision tree with $|\pi|+(a+1) \mathrm{height}(T)\leq N$ (so that $T^\omega(\pi)$ is well-defined). Let $\pi' \supseteq \pi$ be such that $\Pr_{\omega \sim \widetilde{\Omega}}[ T^\omega(\pi)=\pi']>0$, and let  $\Omega'$ be the distribution of $\omega \sim \widetilde{\Omega}$ conditioned on $T^\omega(\pi)=\pi'$. Then $(\pi', \Omega')$ is a fresh state.
\end{observation}

The main technical lemma underlying the proof of Theorem~\ref{thm:riis:main}:
\begin{lemma}[Swapping Orders]\label{lem:riis:swapping-orders}
Fix a fresh state $(\pi, \widetilde{\Omega})$ and two
decision trees $T_1,T_2$ of heights at most $h_1,h_2$, respectively.
If $|\pi|+(a+1)(h_1+h_2)<N$, then
\[
    \Pr_{\omega\sim\widetilde{\Omega}}\!\left[
        (T_1\circ T_2)^\omega(\pi)\ne(T_2\circ T_1)^\omega(\pi)
    \right]
    \leq \frac{2(a+1)^2h_1h_2}{N-|\pi|-(a+1)(h_1+h_2)}.
\]
\end{lemma}
\begin{proof}
We will start by proving the lemma assuming it holds in the special case $h_1=h_2=1$. At the very end we prove it in this special case. Define a ``schedule'' to be a string $\sigma \in \{1,2\}^{h_1 + h_2}$ which has exactly $h_1$ 1's and $h_2$ 2's. We say that $\sigma, \sigma'$ are adjacent schedules if there is some index $i < h_1+h_2$ such that $(\sigma_{i}, \sigma_{i+1}) = (\sigma'_{i+1}, \sigma'_i)$ and $\sigma_j = \sigma'_j$ for all $j \notin \{ i, i+1\}$. For a schedule $\sigma$, we define a decision tree $T_\sigma$ as follows. $T_\sigma$ starts by performing the first query of $T_{\sigma_1}$, and maintains the node reached in $T_{\sigma_1}$ after that query. It then performs the next query of $T_{\sigma_2}$ (if $\sigma_1=\sigma_2$ this is the second step in the same tree, otherwise it is the first step in the other tree) and maintains the node reached in that tree; this continues on for each $i \in [h_1+h_2]$. If a leaf in one tree $T_b$ is reached before the $h_1+h_2$ phases are complete, then on all subsequent steps $i$ with $\sigma_i = b$ we perform no further step in round $i$. We will argue that, whenever $\sigma, \sigma'$ are adjacent schedules, we must have
\begin{gather}
    \Pr_{\omega\sim\widetilde{\Omega}}\!\left[
        T_\sigma^\omega(\pi)\ne T_{\sigma'}^\omega(\pi)
    \right]
    \leq \frac{2(a+1)^2}{N-|\pi|-(a+1)(h_1+h_2)} \label{eq:adj-schedule}
\end{gather}
This will immediately yield the lemma, since $T_1 \circ T_2 = T_{1^{h_1} 2^{h_2}}$, $T_2 \circ T_1 = T_{2^{h_2}1^{h_1}}$, and the schedule $1^{h_1}2^{h_2}$ can be transformed into the schedule $2^{h_2}1^{h_1}$ by passing through a sequence of adjacent schedules of length $h_1h_2+1$. We now claim that (\ref{eq:adj-schedule}) follows from the special case of the current lemma with $h_1=h_2=1$. To see this, say that $\sigma$, $\sigma'$ are adjacent along $i, i+1$. If we can show that after the first $i+1$ steps the partial injections generated by $T_{\sigma}^\omega(\pi)$ and $ T_{\sigma'}^\omega(\pi)$ are identical then it will obviously hold for the remaining steps (corresponding to indices where $\sigma$, $\sigma'$ are equal); we can hence prune the remaining steps, or equivalently, assume $i = h_1 + h_2 - 1$. So assume now $i = h_1+h_2-1$, and let $Q$ be the truncation of $T_{\sigma}$ to the first $h_1 + h_2- 2$ phases of the joint $\sigma$-scheduled simulation of $T_1,T_2$ and let $\Omega'$ be the distribution of $\omega \sim \widetilde{\Omega}$ conditioned on some fixing of the value $\pi' = Q^\omega(\pi)$. We see that there is a pair of trees $Q_1, Q_2$ of height at most one, so that $T_{\sigma}^\omega(\pi) = (Q_1 \circ Q_2)^\omega(\pi')$ and $T_{\sigma'}^\omega(\pi) = (Q_2 \circ Q_1)^\omega(\pi')$. For this conditioning, we have that $(\pi', \Omega')$ is a fresh state, and hence we may apply the height-one case of the current lemma and conclude 
\begin{gather*}
\Pr_{\omega\sim \Omega'}\!\left[
        T_\sigma^\omega(\pi)\ne T_{\sigma'}^\omega(\pi) 
    \right]
    \leq  \Pr_{\omega\sim \Omega'}\!\left[
        (Q_1\circ Q_2)^\omega(\pi')\ne (Q_2\circ Q_1)^\omega(\pi') 
    \right] \\\leq
     \frac{2(a+1)^2}{N-|\pi'|-2(a+1)} \leq \frac{2(a+1)^2}{N-|\pi|-(a+1)(h_1+h_2)}
\end{gather*}
using $|\pi'| \leq |\pi| + (a+1)(h_1 + h_2-2)$.
Averaging this bound over our conditioning establishes (\ref{eq:adj-schedule}) and hence finishes the proof of the general case (assuming the special case).

It remains to prove the special case $h_1=h_2=1$. Say $T_1,T_2$ each have height exactly 1 (the lemma is trivial if one has height zero). Say that $T_1$ makes a query to $f_1(\bar x)$ for a function symbol $f_1$ of arity $d_1$, $\bar x \in [N]^{d_1}$, and $T_2$ makes a query to $f_2(\bar z)$ for a function symbol $f_2$ of arity $d_2$, $\bar z \in [N]^{d_2}$ (querying constant symbols is treated as the special case of an arity zero function). In the execution of $T_1 \circ T_2$, there are $(d_1 + 1) + (d_2 + 1)$ ``basic matching operations'' that must be performed: first we must pair each of $x_1,\ldots,x_{d_1} \in [N]$ with elements in $\cS$, then pair $f_1^{\cM}(\pi(x_1),\ldots,\pi(x_{d_1}))$ with an element of $[N]$, and then repeat the same process for $f_2$. We will apply the same reasoning here as we did in the general case above: if we can analyze the process of swapping two adjacent basic matching operations, and prove that the probability of such a swap starting from a current matching $\pi$ will change the outcome with probability at most $\frac{2}{N-|\pi|}$, then we can consider a sequence of $(d_1+1)(d_2+1) \leq (a+1)^2$ of these swaps taking us from $T_1 \circ T_2$ to $T_2 \circ T_1$ and apply a union bound. Clearly freshness is preserved by these basic matching operations. At each step, the matching $\pi'\supseteq \pi$ we obtain after applying some basic matching operations satisfies $|\pi'| \leq |\pi| + d_1 + d_2 + 2 \leq |\pi| + 2(a+1)$; so overall we get the desired bound \[
    \frac{2(d_1+1)(d_2+1)}{N-|\pi|-2(a+1)}
    \leq \frac{2(a+1)^2}{N-|\pi|-2(a+1)}
\]
on the probability that $(T_1 \circ T_2)^\omega(\pi) \neq (T_2 \circ T_1)^\omega(\pi)$.

To finish the proof we need to analyze a single swap of two basic matching operations. Say that $\pi$ is a current matching, and the next two moves demand that we first extend $\pi$ to pair $\alpha$, and then further extend it to pair $\beta$, for some $\alpha, \beta \in [N] \cup \cU$. Obviously we may assume both $\alpha, \beta$ are currently unpaired in $\pi$. If $\alpha, \beta \in [N]$ or $\alpha, \beta \in \cU$ and $\alpha \neq \beta$ then the swap will only have consequence if, starting from the initial matching $\pi$, both $\alpha$ and $\beta$ would prefer the same partner $\gamma$. This requires that $\omega_{\alpha}, \omega_{\beta}$ have the same least element unpaired in $\pi$; by the freshness of the state $(\pi, \widetilde{\Omega})$ this happens with probability at most $\frac{1}{N - |\pi|}$. If $\alpha=\beta$ then clearly the swap has no effect. On the other hand, say that $\alpha \in [N], \beta \in \cU$; let $Y\in \cU, y\in [N]$ be their preferred mates according to $\omega$ (computed with respect to $\pi$); then a swap will only have an effect when either $\beta = Y$ or $\alpha = y$. Each of these events occurs with probability at most $\frac{1}{N-|\pi|}$, and hence the swap will have an effect with probability at most $\frac{2}{N-|\pi|}$. The case $\alpha \in \cU, \beta \in [N]$ is identical.
\end{proof}

At this point we are ready to prove Theorem~\ref{thm:riis:main}.

\begin{proof}[Proof of Theorem~\ref{thm:riis:main}]
As per the discussion above, it suffices to prove that with nonzero probability, the following inequalities hold for all $i \leq m$:
\[
(1) \; p_i = p_i^{\emp} \qquad (2)\; |p_i^{\emp}- p_i^\true| \leq \frac{\epsilon}{4} \qquad (3) \; |p_i^\true - p_i^\final| \leq \frac{\epsilon}{4}
\]
We will show that for each $i$, each inequality holds with probability $>1- \frac{1}{3m}$ and then take a union bound. For the remainder of the argument, we will fix some particular inspector strategy once and for all. For a single such choice of strategy, we have that $\cT_i$ is a deterministic function of $\omega$ and the subsampling randomness used through the first $i-1$ rounds.

We start with (2) which is easiest; this does not require Lemma~\ref{lem:riis:swapping-orders} and instead follows directly from a concentration bound for an appropriate choice of $s$. Condition on $\omega$ and the choice of subsampling randomness through the first $i-1$ rounds. The summands defining $p_i^\emp$ are independent Bernoulli variables with mean $p_i^\true$; by Hoeffding, we have $\Pr[|p_i^\emp- p_i^\true|> \frac{\epsilon}{4}]\leq 2 e^{- s\epsilon^2/8}$. Setting $s = \lceil 8 \epsilon^{-2} \ln (12m) \rceil$, we have $2 e^{- s\epsilon^2/8} \leq \frac{1}{6m}< \frac{1}{3m}$. We also note that for this setting of $s$ and our assumption on $N$, we have $N \geq (a+1)(ms + 1)h_{\mathrm{max}}$, where $h_{\mathrm{max}} = (ms+1)h$ upper bounds the maximum height of any decision tree $T$ considered in the remainder of this proof; this will guarantee that for all trees $T$ and matchings $\pi$ we consider, $T^\omega(\pi)$ is well-defined.

We now turn to (1). Condition on $\pi_{i-1}$ and all subsampling randomness used through round $i$; we claim that this conditioning amounts to some fresh state $(\pi_{i-1}, \widetilde{\Omega})$. This follows from the fact that the play of the game across the first $i-1$ rounds (once the subsampling randomness has been fixed) amounts precisely to the execution of some decision tree $Q_{<i}$ of depth $\leq (i-1)hs$ on $\omega \sim \Omega$ starting from $\pi_0$\footnote{To define this tree $Q_{<i}$ formally, we start with the first round, in which the inspector supplies some $\cT_1$; since the subsampling randomness has been fixed, this completely determines a sequence $T_{1,1},\ldots,T_{1,s}$ of decision trees. The first part of $Q_{<i}$ will consist of $T_{1,1}\circ\cdots\circ T_{1,s}$; after running these some value $p_1$ is produced, which then determines $\cT_2$, and so on. We halt this process after the first $i-1$ rounds to obtain $Q_{<i}$; the partial matching reached from $\pi_0$ using $\omega$ is precisely $\pi_{i-1} = Q_{<i}^\omega(\pi_0)$.} using Observation~\ref{obs:riis:fresh}. It will suffice to show that for each $j \leq s$, with probability $> 1 - \frac{1}{3ms}$ over $\omega \sim \widetilde{\Omega}$, $\out^\omega(T_{i,j}, \pi_{i-1})= \out^\omega(T_{i,j}, \pi_i)$; taking a union bound over $j \leq s$ will then give (1). We consider the two trees 
\[
T = T_{i,1} \circ \cdots \circ T_{i,s},  \qquad T' = T_{i,j} \circ T_{i, 1} \circ \cdots T_{i, j-1} \circ T_{i, j+1} \circ \cdots T_{i, s}
\]
If we can prove that $\Pr_{\omega \sim \widetilde{\Omega}}[T^\omega(\pi_{i-1}) \neq T'^{\,\omega}(\pi_{i-1})]< \frac{1}{3ms}$ then we are done. This follows directly from Lemma~\ref{lem:riis:swapping-orders}; we are simply swapping the order of $T_{i,1} \circ \cdots \circ T_{i, j-1}$ and $T_{i,j}$, and the remaining suffix $T_{i,j+1} \circ \cdots \circ T_{i,s}$ has no effect. We can apply this lemma provided $N > (a+1)msh + 6(a+1)^2ms(s-1)h^2$ which follows from the Theorem's assumptions and our choice of $s$.

We now turn to (3); the argument here is similar. Fix all subsampling randomness used throughout all $m$ rounds; we will argue that the bound in (3) holds for every conditioning of the subsampling randomness, and hence on average over it. Now, as in (1) we condition on $\pi_{i-1}$; such conditioning again amounts to a fresh state $(\pi_{i-1},\widetilde{\Omega})$. By the same argument used above to define $Q_{<i}$, we can determine that there is a decision tree $Q_{\geq i}$ (depending on our particular choice of $\pi_{i-1}$) of height at most $(m-i+1)sh$ such that $\pi_m = Q_{\geq i}^\omega(\pi_{i-1})$. Observe that $\cT_i$ is completely determined (since $\pi_{i-1}$ has been fixed). Now, our goal is to show that with probability $> 1 - \frac{1}{3m}$ over $\omega \sim \widetilde{\Omega}$, we have:
\[
|\mathbb{E}_{T_i \sim \cT_i}[ \out^\omega(T_i, \pi_{i-1})] - \mathbb{E}_{T_i \sim \cT_i}[ \out^\omega(T_i, \pi_{m})] |  \leq \frac{\epsilon}{4}
\]
By Markov, to prove this it will suffice to show:
\[
\Pr_{\omega \sim \widetilde{\Omega}, T_i \sim \cT_i}[\out^\omega(T_i, \pi_{i-1}) \neq \out^\omega(T_i, \pi_{m})] < \frac{\epsilon}{12m}
\]
where $\pi_m$ implicitly depends on $\omega$ in the above expression. We show the above for every fixed choice of $T_i$. Applying Lemma~\ref{lem:riis:swapping-orders}, we know that $(T_i \circ Q_{ \geq i})^{\omega}(\pi_{i-1})= (Q_{\geq i} \circ T_i)^\omega(\pi_{i-1}) $ holds with probability $>1- \frac{\epsilon}{12m}$ over $\omega \sim \widetilde{\Omega}$, provided $N > (a+1)(ms+1)h + 24(a+1)^2m^2sh^2\epsilon^{-1}$ which follows from the Theorem's assumptions and our choice of $s$. This gives the desired result, since if $(T_i \circ Q_{ \geq i})^{\omega}(\pi_{i-1}) =(Q_{\geq i} \circ T_i)^\omega(\pi_{i-1}) = \pi$, then $\pi$ determines the path traversed in $T_i^\omega$ during both executions, and hence this path must agree in both. In the first execution, we run $T_i$ directly on $\pi_{i-1}$ using $\omega$ and hence trace a path to a leaf in $T_i$ labeled $\out^\omega(T_i, \pi_{i-1})$; in the second, we first run $Q_{\geq i}$ on $\pi_{i-1}$ using $\omega$, which yields $\pi_m$, and then trace a path through $T_i$ starting at $\pi_m$ using $\omega$, which reaches a leaf labeled $\out^\omega(T_i, \pi_m)$. Hence we must have $\out^\omega(T_i, \pi_{i-1}) = \out^\omega(T_i, \pi_{m})$.
\end{proof}

\subsection{Auxiliary Results for the Reduction to Quantifier-Free Formulas}
\label{sec:red-herbrand}

\newcommand{\calL}{\mathcal{L}}
\newcommand{\FO}{\mathrm{FO}}
\newcommand{\HNF}{\mathsf{HNF}}
\newcommand{\calM}{\mathcal{M}}

We first recall some basic definitions. Let $\phi(x_1,\dots,x_k)$ be a formula over the language $\calL$. The \emph{Herbrand normal form} of $\phi$ is defined as follows. We first convert $\phi$ into prenex normal form, say 
\[ \vdash_{\FO}\phi(\overline x)\leftrightarrow \forall y_1~\exists y_2~\dots~\forall y_\ell~\psi(\overline x,y_1,\dots,y_\ell),\]
where $\psi$ is a quantifier-free $\calL$-formula. Let $f_1,f_3,\dots, f_\ell$ be new function symbols. The Herbrand normal form of $\phi$, denoted $\HNF(\phi)$, is defined as the formula
\[
\exists y_2~\exists y_4~\dots~\exists y_{\ell-1}~\psi(\overline x,f_1(\overline x),y_2,f_3(\overline x,y_2),\dots,f_\ell(\overline x,y_2,y_4,\dots,y_{\ell-1}))
\]
in the language $\calL_\HNF\eqdef \calL\cup\{f_1,f_3,\dots,f_\ell\}$. We will need the following  basic properties of first-order logic.

\begin{proposition}\label{prop:Herbrandization-model}
    $\forall \overline x~\lnot \phi(\overline x)$ has a model over the universe $\disU$ iff\;$\forall\overline x~\lnot \HNF(\phi)$ has a model over the universe $\disU$. In particular, $\forall \overline x~\lnot \phi(\overline x)$ has an infinite model iff\; $\forall\overline x~\lnot \HNF(\phi)$ has an infinite model. 
\end{proposition}

\begin{proof}
    We may assume without loss of generality that $\phi$ is in prenex normal form, as logically equivalent formulas have the same truth value over any model. 
    
    The (\emph{if}) direction is straightforward: the model $\calM$ for $\forall\overline x~\lnot\HNF(\phi)$ is also a model for $\forall \overline x~\lnot \phi(\overline x)$ by removing the interpretations for $f_1,f_3,\dots,f_\ell$ on the same universe. For the (\emph{only if}) direction, we consider the simplified case where $\phi(\overline x)\eqdef \forall y_1~\exists y_2~\psi(\overline x,y_1,y_2)$, and the general case can be proved analogously. Assume that $\calM$ is a model for $\forall\overline x~\lnot \phi(\overline x)$. Then,  by definition, for every $\overline x^*$ in the universe $\disU$, there is an $y_1^*\in\disU$ such that for every $y_2^*\in\disU$, $\psi^\calM(\overline x^*, y_1^*,y_2^*)$ does not hold. Let $f^*_1$ be any function that fixes some $y_1^*\in \disU$ satisfying the above given $\overline x^*$. Then, by definition, $\calM$ augmented with the interpretation $f_1\mapsto f_1^*$ is a model for $\forall \overline x~\lnot\HNF(\phi)$. This completes the proof. 
\end{proof}

\begin{proposition}[{\cite[Remark 2.15]{applied-proof-theory}}]\label{prop:herbrand-imp}
    $\vdash_\FO\phi\to\HNF(\phi)$. 
\end{proposition}

Let $\phi(x_1,\dots,x_k)$ be any formula in the language $\calL$ with equality. Following the notation from \Cref{sec:intro_results_riis}, we define 
$$
\Theta_\phi \eqdef
\forall N\geq1\;
\left(
\operatorname{Closed}_{\widehat{\mathcal L}}(N)
\;\rightarrow\;
\exists x_1,\ldots,x_k<N\;
\phi^{[N]}(x_1,\ldots,x_k)
\right).
$$

\begin{proposition}\label{prop:phi-to-HNF}
    $\vdash_\FO \Theta_\phi\to\Theta_{\HNF(\phi)}$. 
\end{proposition}

\begin{proof}
    Let $\calL_\HNF\eqdef \calL\cup \{f_1,f_3,\dots,f_\ell\}$ be the language of $\HNF(\phi)$. Assume, towards a contradiction, that there is a model $\cM=(\cU,\cI)$ such that $\cM\vDash\Theta_\phi$ but $\cM\nvDash \Theta_{\HNF(\phi)}$.  Consider any $N^*\in\cM$ such that \(\cM\vDash N^*\geq1\) and
    \[
    \cM\vDash \operatorname{Closed}_{\widehat{\mathcal L_\HNF}}(N^*)
\;\land\;
\forall x_1,\ldots,x_k<N^*\;
\lnot\HNF(\phi)^{[N^*]}(x_1,\ldots,x_k).
    \]
    Since \(\operatorname{Closed}_{\widehat{\mathcal L_\HNF}}(N^*)\) implies \(\operatorname{Closed}_{\widehat{\mathcal L}}(N^*)\), the assumption \(\cM\vDash\Theta_\phi\) gives witnesses \(x_1^*,\dots,x_k^*<^\cI N^*\) such that \(\cM\vDash\phi^{[N^*]}(x_1^*,\dots,x_k^*)\). 
    
    Let \(\cU_{N^*}\eqdef\{x\in\cU\mid x<^\cI N^*\}\). It can be verified that this set is nonempty: if \(k\geq1\), it contains \(x_1^*\); if \(k=0\), the first Herbrand symbol \(f_1\) is a constant, and expanded closure gives \(\widehat f_1^{\cM}(N^*)<^\cI N^*\). Let $\cI_{N^*}:f(\cdot)\mapsto \widehat f^{\cI}(N^*,\cdot)$, where constants are treated with functions with no input. By the definition of finitary translation, \(\cM_{N^*}\eqdef(\cU_{N^*},\cI_{N^*})\) satisfies both \(\forall\overline x\,\lnot\HNF(\phi)(\overline x)\) and \(\phi(x_1^*,\dots,x_k^*)\). This leads to a contradiction with \Cref{prop:herbrand-imp}.
\end{proof}

\begin{corollary}\label{cor:Herbrandization-theory}
Let $\calL_\HNF\eqdef \calL\cup \{f_1,f_3,\dots,f_\ell\}$ be the language of $\HNF(\phi)$, and let $\cT\in\{\PV,\APX\}$. If $\cT(\widehat\calL)\vdash \Theta_\phi$, then $\cT(\widehat{\calL_\HNF})\vdash \Theta_{\HNF(\phi)}$. 
\end{corollary}

\subsection{Unprovability in \texorpdfstring{$\APX(\mathcal L)$}{APX1(L)}}\label{sec:proofs_to_games}

In this section, we complete the proof of \Cref{thm:intro-riis}. 
Let $\phi(x_1,\ldots,x_k)$ be a formula in the
language $\mathcal L$ (with equality) of arbitrary quantifier complexity, and write
$$
\Theta_\phi \eqdef
\forall N\geq1\;
\left(
\operatorname{Closed}_{\widehat{\mathcal L}}(N)
\;\rightarrow\;
\exists x_1,\ldots,x_k<N\;
\phi^{[N]}(x_1,\ldots,x_k)
\right).
$$
Assume that
$\forall x_1\cdots\forall x_k\;\neg\phi(x_1,\ldots,x_k)$
has an infinite model. 

\begin{lemma}\label{lmm:quantifier-free-riis}
Assume that \Cref{thm:intro-riis} holds for all quantifier-free formulas $\phi$. Then \Cref{thm:intro-riis} holds. 
\end{lemma}

\begin{proof}
    Let $\HNF(\phi)\eqdef \exists \overline y~\beta(\overline x,\overline y)$ be the Herbrand normal form of $\phi$. Assume that $\APX(\widehat\calL)\vdash\Theta_\phi$. We know by \Cref{cor:Herbrandization-theory} that $\APX(\widehat{\calL_\HNF})\vdash\Theta_{\HNF(\phi)}$. By the definition of the finitary translation, we can see that $\Theta_{\HNF(\phi)}$ and $\Theta_{\beta}$ are logically equivalent, and thus $\APX(\widehat{\calL_\HNF})\vdash\Theta_{\beta}$.
    
    Since by assumption \Cref{thm:intro-riis} holds for any quantifier-free formula, applying it to the formula $\beta$ shows that $\forall\overline x~\forall\overline y~\lnot \beta(\overline x,\overline y)$ has no infinite model, or equivalently, $\forall\overline x~\lnot \HNF(\phi)$ has no infinite model. By \Cref{prop:Herbrandization-model}, $\forall\overline x~\lnot\phi$ also has no infinite model. 
\end{proof}

In the rest of the proof, we consider the case where $\phi$ is a quantifier-free formula. We prove \Cref{thm:intro-riis} (for quantifier-free $\phi$) by showing
that a proof of $\Theta_\phi$ in $\APX(\widehat{\mathcal L})$
yields a \emph{winning inspector}. In more detail, we call a play
of the game $\mathcal{G}_\phi$ (\Cref{def:game}) \emph{consistent}
if it satisfies the adversary's conditions~(2) and~(3), i.e.,
\emph{deterministic consistency} and \emph{mixture consistency},
respectively. Our inspector forces every consistent play to end
with a partial $\mathcal L$-structure witnessing $\phi$, thereby
violating condition~(1). Since this is in contradiction with
\Cref{thm:riis:main}, it follows that
$\APX(\widehat{\mathcal L})\nvdash\Theta_\phi$, as desired.

Formally, we will prove the following result.

\begin{lemma}[Proofs yield winning inspectors]
\label{lem:riis:winning_inspector}
Let $\phi$ be a quantifier-free formula. If $\APX(\widehat{\mathcal L})\vdash\Theta_\phi$, there is a polynomial $Q(n)$ depending only on the proof such that the inspector wins
$$
\mathcal G_\phi\bigl(2^n;Q(n),Q(n),1/Q(n)\bigr)
$$
for every $n\geq1$.
\end{lemma}

First, we derive \Cref{thm:intro-riis} from the lemma.

\begin{proof}[Proof of \Cref{thm:intro-riis}]
By \Cref{lmm:quantifier-free-riis}, we may assume that $\phi$ is quantifier-free. Suppose $\APX(\widehat{\mathcal L})\vdash\Theta_\phi$, and take
$Q$ from \Cref{lem:riis:winning_inspector}. Let $a$ be the maximum
arity in the original language $\mathcal L$, with $a=0$ if it
contains only constants. Choose $n$ sufficiently large such that
$$
2^n\geq10^5(a+1)^2Q(n)^8\bigl(\ln(2Q(n))\bigr)^2.
$$
The infinite model and \Cref{thm:riis:main} give a winning adversary
in $\mathcal G_\phi(2^n;Q(n),Q(n),1/Q(n))$, contradicting the lemma. This shows that
$\APX(\widehat{\mathcal L})\nvdash\Theta_\phi$.
\end{proof}

The remainder of this section is dedicated to the proof of the lemma.

\begin{proof}[Proof of \Cref{lem:riis:winning_inspector}]
    Roughly speaking, we use Herbrand's theorem to construct an inspector from a proof of $\Theta_\phi$ in $\APX(\widehat {\mathcal L})$ and prove that the inspector wins the game. More details follow:

    \paragraph{Step I: Applying Herbrand's theorem.} Consider the theory $\APX(\widehat{\mathcal L})$ augmented with a fresh constant $\nu$ and the following axioms:
    \[\nu\ge 1\qquad\text{and}\qquad \operatorname{Closed}_{\widehat{\mathcal L}}(\nu).\]
    Intuitively, the constant $\nu$ denotes the model size $N = 2^n$ we are working with. Since $\APX(\widehat{\mathcal L})$ proves $\Theta_\phi$, the augmented theory proves $\exists \bar{x}\,W(\nu, \bar{x})$ where
    \[W(N, \bar{x})\eqdef \mleft(\bigwedge_{\ell = 1}^k x_\ell < N\mright)\land \phi^{[N]}(\bar{x}).\]
    Since the augmented theory is a universal theory, by Herbrand's theorem (\autoref{thm:finite_herbrand}), the following \emph{finite} implication holds:
    \begin{equation}\label{eq:Herbrand_finite_implication}
        \vdash_{\mathrm{FO}}\;\bigwedge_{j=1}^m \theta_j(\bar{s}_j(\nu)) \;\longrightarrow\; \bigvee_{i=1}^{r}W(\nu,\bar t_i(\nu)).
    \end{equation}
    where each $\theta_j$ is a ground instance of a universal axiom $\forall \bar{u}_j\,\theta_j(\bar{u}_j)$ of the augmented theory (with $\theta_j(\bar{u}_j)$ quantifier-free), and $\bar{s}_j$ and $\bar{t}_i$ are tuples of ground terms in its language.

    \paragraph{Step II: From the finite implication to an inspector.} We define an inspector that evaluates every antecedent $\theta_j(\bar{s}_j(\nu))$ and every succedent $W(\nu, \bar{t}_i(\nu))$ in \eqref{eq:Herbrand_finite_implication}. Given a total $\mathcal{L}$-structure $\rho$ on $[N]$, define its interpretation $\widehat{\rho}$ by
    \begin{equation}\label{eq:riis:slice_extension1}
    \widehat f^{\widehat{\rho}}(s,\bar u)=
    \begin{cases}
    f^{\rho}(\bar u),&
    s=N\text{ and }\bar u\in[N]^a,\\
    0,&\text{otherwise},
    \end{cases}
    \end{equation}
    for each function $f$ with arity $a$ (constants have arity $a=0$). Then the inspector evaluates every term and atomic subformula occurring in the antecedents $\theta_j(\bar{s}_j(\nu))$ and succedents $W(\nu, \bar t_i(\nu))$ in \eqref{eq:Herbrand_finite_implication} on the interpretation $\widehat{\rho}$. During this procedure, the inspector needs to evaluate $\mathcal L$-functions and $\P$-oracles. Whenever this occurs, we translate them into (randomized) decision trees and submit them as queries to the adversary.
    
    In more detail, fix a large enough polynomial $P(n)$ that upper bounds the running time of each $\theta_j(\bar{s}_j(\nu))$ and $W(\nu, \bar{t}_i(\nu))$. Note that $P(n)$ also upper bounds their query complexity, the precision parameters $\delta^{-1} = \|\Delta\|$ of each invocation of $\P$, and the inverse of slack parameters $\beta^{-1}$ in each approximate counting axioms mentioned in \eqref{eq:Herbrand_finite_implication}. We set $Q(n) := (m+r)\cdot \lceil\log N\rceil\cdot P(n) \le \poly(n)$ and consider the game $\mathcal{G}_\phi(N; Q(n), Q(n), 1/Q(n))$.
    \begin{itemize}
        \item {\bf Case I: evaluating a function $f$.} A query to $f(\bar{u})$ is translated into $\lceil\log N\rceil$ queries to the adversary, each corresponding to one output bit of $f(\bar{u})$. That is, for each output bit $i\in [\lceil \log N\rceil]$, let $B_i$ denote the decision tree that outputs the $i$-th bit of $f(\bar{u})$; the inspector submits the singleton distribution concentrated at $B_i$. Denote the adversary's response by $p_i$. If one of the numbers $p_i$ is non-Boolean, then the {\it (Deterministic Consistency)} of the adversary already fails and the inspector wins the game. Otherwise, we have $f(\bar{u}) = \sum_i 2^ip_i$.

        \item {\bf Case II: evaluating a $\P$ oracle.} A query to $\P(C, \Delta)$, where $C$ is a valid circuit and $1\le \|\Delta\|\le P(n)$, is translated into a randomized decision tree query to the adversary. For each input $z$ of $C$ (recall that $z$ is treated as ``randomness''), the computation of $C(z)$ can be expressed as a decision tree that queries $\mathcal L$-functions; we denote this decision tree as $T_{C, z}$. Let $\mathcal{T}_C$ denote the randomized decision tree that is the uniform distribution of $T_{C, z}$ (over $z$). The inspector submits $\mathcal{T}_C$ as a randomized decision tree to the adversary and obtains an answer $p_C \in [0, 1]$. We round $p_C$ to the closest integer multiple of $2^{-\lceil\log \|\Delta\|\rceil}$ as the answer to this query.
        
        Note that if $C$ does not read its input (hence $\mathcal{T}_C$ can be written as a singleton distribution) and $p_C$ is non-Boolean, then the {\it (Deterministic Consistency)} of the adversary fails and the inspector wins the game. Also, the inspector stores a cache of already returned answers $p_C$ so that if the same circuit $C$ occurs later, the inspector does not issue any query and instead returns (the rounding of) $p_C$ as stored.
    \end{itemize}

    The game proceeds in at most $(m+r)\cdot \log N\cdot P(n) \le Q(n)$ rounds and each randomized decision tree submitted by the inspector has height at most $P(n) \le Q(n)$.

    \paragraph{Step III: Proving the inspector wins.} We now show that the inspector wins the game; this is an immediate corollary of \eqref{eq:Herbrand_finite_implication}.
    
    Start with the standard arithmetic structure on binary strings. Recall that at the end of the game, the adversary returns a partial $\mathcal L$-structure $\rho$ on $[N]$. We extend $\rho$ to a total structure $\rho'$ by arbitrarily assigning unassigned constants and function values in $\rho$. Every $\widehat{\mathcal L}$-function symbol $\widehat f$ is interpreted as the function $\widehat{f}^{\rho'}$ as in \eqref{eq:riis:slice_extension1}. The symbol $\nu$ is interpreted as $N$. We also interpret $\P(C, \Delta)$ as $p_C$, rounded to the nearest integer multiple of $2^{-\lceil\log\|\Delta\|\rceil}$, whenever the inspector queried $P(C, \Delta)$ during the game, and $0$ otherwise. Finally, we interpret every $\PV(\widehat{\mathcal L}, \P)$-function by its defining oracle program. Since \eqref{eq:Herbrand_finite_implication} holds in this structure, one of the following is true:
    \begin{itemize}

        \item One of the succedents $W(\nu, \bar{t}_i(\nu))$ is true. We show that this implies $\rho$ is $\phi$-witnessing. Indeed, let $\bar{x}$ be the value of $\bar{t}_i(N)$ in the completed arithmetic structure, then $\bar{x} < N$ and $\phi^{[N]}(\bar{x})$ holds there. Every term of $\phi^{[N]}(\bar x)$ was evaluated, and one can use induction to show that $\rho\models\phi^{[N]}[\bar{x}]$. It follows that $\rho$ is $\phi$-witnessing and the inspector wins the game.
        
        \item One of the antecedents $\theta_j(\bar{s}_j(\nu))$ is false. Among all axioms of $\APX(\widehat{L})$, the axioms $\nu\ge 1$, $\operatorname{Closed}_{\widehat{\mathcal L}}(\nu)$, and axioms of $\PV(\widehat{L}, \P)$ are clearly true under our interpretation. Thus it suffices to consider the case that $\theta_j(\bar{s}_j(\nu))$ is an approximate counting axiom.
        \begin{enumerate}
            \item {\it Basic Axiom Instances.} Since the adversary always returns a number in $[0, 1]$, the basic axiom instance $\P_\delta(C) \in [0, 1]\cap \mathbb{Q}$ always holds.
            \item {\it Boundary Axiom Instances.} It is easy to see that either the adversary violates the {\it (Deterministic Consistency)} or any Boundary Axiom instance appearing as some $\theta_j(\bar{s}_j(\nu))$ is satisfied.
            \item {\it Precision Consistency Axiom Instances.} Let $C$ be a circuit and $\delta_1^{-1}, \delta_2^{-1}, \beta^{-1}\in \Log$ such that the axiom $|\P_{\delta_1}(C) - \P_{\delta_2}(C)| \le \delta_1 + \delta_2 + \beta$ appears as some $\theta_j(\bar{s}_j(\nu))$. Since both $\P_{\delta_1}(C)$ and $\P_{\delta_2}(C)$ are the rounding of the same stored response $p_C$, and the roundings have errors at most $\delta_1$ and $\delta_2$ respectively, the axiom is satisfied.
            \item {\it Local Consistency Axiom Instances.} Finally, it suffices to show that if $C$ is a circuit, $\delta^{-1}, \beta^{-1}\in\Log$, and the Local Consistency Axiom
            \begin{equation}\label{eq:falsified-local-consistency}
                \mleft|\P_\delta(C) - \frac{\P_\delta(\Fix_0(C)) + \P_\delta(\Fix_1(C))}{2}\mright| \le 2\delta + \beta
            \end{equation}
            appeared as some $\theta_j(\bar{s}_j(\nu))$ and is falsified, then the adversary violates {\it (Mixture Consistency)} and loses the game. Note that $\mathcal{T}_C = \frac{1}{2}\mathcal{T}_{\Fix_0(C)} + \frac{1}{2}\mathcal{T}_{\Fix_1(C)}$. However, if \eqref{eq:falsified-local-consistency} is false, then
            \[\mleft|p_C - \frac{p_{\Fix_0(C)} + p_{\Fix_1(C)}}{2}\mright| > \beta \ge 1/Q(n),\]
            meaning the adversary violates {\it (Mixture Consistency)}.\qedhere
        \end{enumerate}
    \end{itemize}
\end{proof}

\subsection{Finite Combinatorial Principles Provable in \texorpdfstring{$\APX$}{APX1}}\label{sec:proof_corollary_riis}

In this section, we provide the proof of \Cref{cor:intro-riis}. We will use the following result.

\begin{proposition}\label{prop:phi-imply-Theta}
    If $\vdash_\FO\exists \overline x~\phi(\overline x)$, then $\vdash_\FO (\forall N\ge 1~\exists x~x<N)\to \Theta_\phi$. 
\end{proposition}

\begin{proof}
    Suppose for a contradiction that there is a model $\cM=(\cU,\cI)$ such that $\cM\vDash \forall N\ge 1~\exists x~x<N$ but  $\cM\nvDash\Theta_\phi$. Let $N^*\in\cM$ be an element such that $\cM\vDash N^*\ge 1$, and 
    \[
    \cM\vDash \left(
\operatorname{Closed}_{\widehat{\mathcal L}}(N^*)
\;\land\;
\forall x_1,\ldots,x_k<N^*\;
\lnot \phi^{[N^*]}(x_1,\ldots,x_k)
\right).
    \]
    Let $\cU_{N^*}\eqdef \{x\in\cM\mid x < N^*\}$, and let $\cI_{N^*}\colon f(\cdot)\mapsto \cI(\widehat f(N^*,\cdot))$. Since $\cM\vDash \forall N\ge 1~\exists x~x<N$, we know that $\cU_{N^*}$ is non-empty. By the definition of the finitary translation, we know that $\cM_{N^*}\eqdef (\cU_{N^*},\cI_{N^*})$ is a model of $\forall x_1,\dots,x_k~\lnot \phi$. This leads to a contradiction with the assumption that $\vdash_\FO\exists \overline x~\phi(\overline x)$. 
\end{proof}

\CorFinite*

\begin{proof}
The implication $(2)\Rightarrow(1)$ is immediate, since
$\APX(\widehat{\mathcal L})$ extends $\PV(\widehat{\mathcal L})$.
For $(1)\Rightarrow(3)$, if $(3)$ fails, then
$\forall\bar x\,\neg\phi(\bar x)$ has an infinite model.
\Cref{thm:intro-riis} therefore gives
$\APX(\widehat{\mathcal L})\nvdash\Theta_\phi$, contradicting $(1)$.

It remains to prove $(3)\Rightarrow(2)$.
Assume $(3)$. Together with the hypothesis concerning finite
structures, this implies that $\vdash_\FO \exists\bar x\,\phi(\bar x)$. By \Cref{prop:phi-imply-Theta}, $\vdash_\FO (\forall N\ge 1~\exists x~x<N)\to \Theta_\phi$. Then the provability in $\PV(\widehat{\cL})$ follows from the simple fact that $\PV\vdash \forall N\ge 1~\exists x~x<N$.  
\end{proof}

\small

\bibliographystyle{alpha}
\bibliography{ref}

\normalsize

\appendix

\section{Auxiliary Results and Missing Formalizations}
\label{sec: missing formalizations}

\subsection{Standard Models and Admissible Models}
\label{sec: missing standard admissible}

\propStandardAdmissible*

\AI{Drafted by Claude Fable 5 with proof idea given as prompt, verified and edited by authors.}

\begin{proof}
    ($\Rightarrow$): it can be verified by straightforward calculation. ($\Leftarrow$): Suppose that $\bbM(\hat\P)$ is admissible. The \nameref{axiom: basic mAPX} and the length bound imposed on $\P$ in the language $\PV(\P)$ ensure condition 1 of \Cref{def: standard models mAPX} except for the requirement $q\in(1\pm1/\|\Delta\|)\cdot p(C)$, and the \nameref{axiom: boundary mAPX} ensures condition 2. Suppose, towards a contradiction, that there are a circuit $C:\zo^n\to\zo$ and a string $\Delta$ such that $q\eqdef\hat\P(C,\Delta)\notin(1\pm 1/\|\Delta\|)\cdot p(C)$, where $p(C)$ denotes the acceptance probability of $C$. Note that both $C$ and $\Delta$ are fixed finite strings and $n\in\N$ is a standard integer. Then there are two cases:

    \smallskip\noindent(Case 1: \emph{$q>(1+1/\|\Delta\|)\cdot p(C)$}). Let $\eps\in(0,1)$ such that $q>(1+\eps)(1+1/\|\Delta\|)\cdot p(C)$. Let $\Xi\eqdef 1^{64(n+1)/\eps}$ and $\xi\eqdef 1/\|\Xi\| = \eps / (64(n+1))$, and instantiate the notation $\mlsim$ in every axiom below at the slack $\beta\eqdef\xi$. By the \nameref{axiom: precision consistency mAPX} (with $\delta_1\eqdef\xi$ and $\delta_2\eqdef 1/\|\Delta\|$),
    \[
    \hat\P(C,\Xi) \;\ge\; \frac{(1-\xi)\,q}{(1+\xi)(1+1/\|\Delta\|)} \;>\; (1+\eps)\cdot\frac{1-\xi}{1+\xi}\cdot p(C) \;\ge\; (1+\eps)(1-2\xi)\cdot p(C).
    \]
    We construct a sequence of circuits $C^{(0)}\eqdef C$ and $C^{(i+1)}\eqdef\Fix_{\sigma_i}(C^{(i)})$ maintaining the invariant
    \[
    \hat\P(C^{(i)},\Xi) \;>\; t_i\cdot p(C^{(i)}),
    \qquad t_0\eqdef(1+\eps)(1-2\xi),\quad t_{i+1}\eqdef(1-3\xi)\cdot t_i;
    \]
    note that when $p(C^{(i)})=0$ the invariant asserts $\hat\P(C^{(i)},\Xi)>0$. For the inductive step, the \nameref{axiom: local consistency mAPX} at precision $\xi$ gives
    \[
    \frac{\hat\P(\Fix_0(C^{(i)}),\Xi)+\hat\P(\Fix_1(C^{(i)}),\Xi)}{2} \;\ge\; \frac{1-\xi}{(1+\xi)^2}\cdot \hat\P(C^{(i)},\Xi) \;>\; t_{i+1}\cdot\frac{p(\Fix_0(C^{(i)}))+p(\Fix_1(C^{(i)}))}{2},
    \]
    where the last inequality follows from $(1-3\xi)(1+\xi)^2\le 1-\xi$ and $p(C^{(i)})=(p(\Fix_0(C^{(i)}))+p(\Fix_1(C^{(i)})))/2$. By averaging, there is $\sigma_i\in\zo$ with $\hat\P(\Fix_{\sigma_i}(C^{(i)}),\Xi)>t_{i+1}\cdot p(\Fix_{\sigma_i}(C^{(i)}))$. After $n$ steps, the circuit $C^{(n)}$ does not read any input bit, and
    \[
    t_n \;\ge\; (1+\eps)(1-3\xi)^{n+1} \;\ge\; (1+\eps)\bigl(1-3(n+1)\xi\bigr) \;=\; (1+\eps)(1-3\eps/64) \;>\; 1.
    \]
    By the \nameref{axiom: boundary mAPX}, $\hat\P(C^{(n)},\Xi)=p(C^{(n)})\in\zo$, while both cases contradict $\hat\P(C^{(n)},\Xi)>p(C^{(n)})$.

    \smallskip\noindent(\emph{Case 2: $q<(1-1/\|\Delta\|)\cdot p(C)$}). This case is symmetric and essentially similar to Case 1. Specifically, we construct $C^{(i)}$ for $i=0,1,\dots,n$ such that $\hat\P(C^{(i)},\Xi)< s_i\cdot p(C^{(i)})$; we set $\Xi$ to be sufficiently large so $s_n<1$ and it leads to a contradiction. 
\end{proof}

\subsection{Basic Counting Functionalities in \texorpdfstring{$\mAPX$}{APX*}}
\label{sec: missing counting mAPX}

In this subsection, we prove the lemmas deferred from \Cref{sec: counting in mAPX}.

\lmmGlobalConsistency*

\AI{The first version was written by the authors. Claude Opus 5.5 then fixed gaps in it (the choice of $\eta$ and the case of zero probabilities), and the fix was checked by Claude Fable 5.1. The authors further edited it interactively.}
\begin{proof}
    We follow the bit-by-bit fixing trick as in \cite{CLOW26}, with changes to take care of the multiplicative error in approximate counting. We argue in $\mAPX$. Assume, towards a contradiction, that there exist $C_1,C_2:\zo^n\to\zo$ such that $C_1(x)\le C_2(x)$ for every $x\in\zo^n$, and for some $\beta^{-1}\in\Log$, 
    \begin{equation} \label{equ: inequality C1 C2}
      (1-\delta)\cdot \P_\delta(C_1) > (1+\beta)\cdot (1+\delta)\cdot \P_{\delta}(C_2). 
    \end{equation} 
    We may assume that $\delta<1$ as the other case is trivial. Let $\eta^{-1}\in\Log$ be determined later (satisfying $\eta\le 1/16$) and let $\kappa\eqdef(1+\eta)^2/(1-\eta)$. (Note that every occurrence of $\mlsim$ in the axioms below is instantiated at the slack $\eta$.) 
    
    By the \nameref{axiom: precision consistency mAPX}, we have 
    \begin{align*}
    & \P_{\delta}(C_1)\le \frac{(1+\eta)(1+\delta)}{1-\eta}\cdot \P_{\eta}(C_1) \le \kappa\cdot(1+\delta)\cdot \P_{\eta}(C_1), \\ 
    & \P_{\delta}(C_2) \ge \frac{1-\delta}{(1+\eta)^2}\cdot\P_{\eta}(C_2)\ge \kappa^{-1}\cdot(1-\delta)\cdot \P_\eta(C_2), 
    \end{align*}
    where the last inequality in each line follows from $(1+\eta)/(1-\eta)\le\kappa$ and $(1+\eta)^2\le\kappa$. By combining these two inequalities with \eqref{equ: inequality C1 C2}, we have 
    \begin{equation}
    \kappa^2\cdot \P_\eta(C_1) > (1+\beta)\cdot \P_\eta(C_2). \label{equ: ratio lb}
    \end{equation}

    Let $\Fix_b(C)$ be the circuit obtained by fixing the rightmost input bit of $C$ to $b$. Similarly, we define $\Fix_{z}(C)$ as the circuit that fixes the last $\|z\|$ input bits of $C$ to $z$. By the \nameref{axiom: local consistency mAPX}, for every circuit $D$ with at least one input bit, we have  
    \begin{equation}
    \kappa^{-1}\cdot\frac{\P_\eta(\Fix_0(D))+\P_\eta(\Fix_1(D))}{2}\le\P_\eta(D)\le\kappa\cdot\frac{\P_\eta(\Fix_0(D))+\P_\eta(\Fix_1(D))}{2}.\label{equ: approx PD}
    \end{equation}%
    
    Consider the algorithm that iteratively fixes the rightmost input bit of both circuits: starting from $D^{(0)}_1\eqdef C_1$ and $D^{(0)}_2\eqdef C_2$, in round $j=0,1,\dots,n-1$ it computes a bit $b_j$ (specified below) and sets $D^{(j+1)}_i\eqdef\Fix_{b_j}(D^{(j)}_i)$ for $i\in\{1,2\}$. The algorithm maintains the invariant
    \begin{equation}\label{equ: GC invariant}
    \kappa^{2(j+1)}\cdot\P_\eta(D^{(j)}_1)>(1+\beta)\cdot\P_\eta(D^{(j)}_2),
    \end{equation}
    which holds for $j=0$ by \eqref{equ: ratio lb}. Suppose that it holds for some $j<n$, and let $S_i\eqdef\P_\eta(\Fix_0(D^{(j)}_i))+\P_\eta(\Fix_1(D^{(j)}_i))$ for $i\in\{1,2\}$. Applying \eqref{equ: approx PD} to $D^{(j)}_1$ and $D^{(j)}_2$, we obtain
    \[
    \kappa^{2(j+1)}\cdot\kappa\cdot\frac{S_1}{2}\ge\kappa^{2(j+1)}\cdot\P_\eta(D^{(j)}_1)>(1+\beta)\cdot\P_\eta(D^{(j)}_2)\ge(1+\beta)\cdot\kappa^{-1}\cdot\frac{S_2}{2}.
    \]
    Multiplying by $2\kappa>0$ gives $\kappa^{2(j+2)}\cdot S_1>(1+\beta)\cdot S_2$. Subsequently, there exists $b\in\zo$ such that $\kappa^{2(j+2)}\cdot\P_\eta(\Fix_{b}(D^{(j)}_1))>(1+\beta)\cdot\P_\eta(\Fix_{b}(D^{(j)}_2))$. The algorithm sets $b_j\eqdef b$ so the invariant \eqref{equ: GC invariant} holds for $j+1$. By induction on $j\le n$ on a quantifier-free formula, \eqref{equ: GC invariant} also holds for $j=n$.

    Let $z\eqdef b_{n-1}\cdots b_1b_0\in\zo^n$, so that $D^{(n)}_i=\Fix_z(C_i)$ for $i\in\{1,2\}$. Since $\Fix_z(C_i)$ does not read its input and outputs $C_i(z)$, the \nameref{axiom: boundary mAPX} gives $\P_\eta(D^{(n)}_i)=C_i(z)$, so \eqref{equ: GC invariant} for $j=n$ means that $\kappa^{2(n+1)}\cdot C_1(z)>(1+\beta)\cdot C_2(z)$. Since the RHS is non-negative, $C_1(z)=1$, we know that $C_2(z)=1$ (as $C_1(z)\le C_2(z)$). Therefore, $\kappa^{2(n+1)}>1+\beta$. 
    
    Finally, we set $\eta\eqdef\beta/(16(n+1))\le 1/16$ to derive a contradiction. Since $\eta\le 1/5$, we have $(1+\eta)^2\le(1+4\eta)(1-\eta)$, i.e., $\kappa\le 1+4\eta$. We use the inequality $(1+s)^k\le 1+2ks$ for $s\ge 0$ and $ks\le 1/2$, which can be proved in $\PV$. Applying it with $s=4\eta$ and $k=2(n+1)$ (where $ks=\beta/2$), we obtain $\kappa^{2(n+1)}\le(1+4\eta)^{2(n+1)}\le 1+16(n+1)\eta=1+\beta$, which leads to a contradiction.
\end{proof}

\lmmNonExistence*

\AI{Drafted by Claude Opus 5.5 following the proofs of \Cref{lmm: global consistency mAPX} and \cite[Lemma 3.7]{CLOW26}, and checked by Claude Fable 5.1. The authors further edited it interactively.}
\begin{proof}
  We use the bit-by-bit fixing argument, as in the proofs of \Cref{lmm: global consistency mAPX} and \cite[Lemma 3.7]{CLOW26}. We argue in $\mAPX$. Fix $n,\delta^{-1}\in\Log$, a circuit $C:\zo^n\to\zo$, an input $x=x_1\cdots x_n\in\zo^n$ such that $C(x)=1$, and $\beta^{-1}\in\Log$; our goal is to show that $(1-\delta)\cdot 2^{-n}\le(1+\beta)\cdot\P_\delta(C)$. Let $\eta^{-1}\in\Log$ be a parameter to be determined (satisfying $\eta\le 1/8$) and $\kappa\eqdef(1+\eta)^2/(1-\eta)$. (Note that every occurrence of $\mlsim$ in the axioms below is instantiated at the slack $\eta$.)

  Let $D_0\eqdef C$ and $D_{j+1}\eqdef\Fix_{x_{n-j}}(D_j)$ for $j=0,1,\dots,n-1$, so that $D_j$ is obtained from $C$ by fixing its last $j$ input bits to $x_{n-j+1}\cdots x_n$. In particular, $D_n$ does not read its input and outputs $C(x)=1$, so the \nameref{axiom: boundary mAPX} gives $\P_\eta(D_n)=1$. For every $j<n$, the \nameref{axiom: local consistency mAPX} at precision $\eta$ gives
  \[
  (1-\eta)\cdot\frac{\P_\eta(\Fix_0(D_j))+\P_\eta(\Fix_1(D_j))}{2}\le(1+\eta)^2\cdot\P_\eta(D_j).
  \]
  Since $\P_\eta(\Fix_{1-x_{n-j}}(D_j))\ge 0$ by the \nameref{axiom: basic mAPX} and $\Fix_{x_{n-j}}(D_j)=D_{j+1}$, this implies that
  \[
  \P_\eta(D_j)\ge\frac{1-\eta}{2(1+\eta)^2}\cdot\P_\eta(D_{j+1})=\frac{1}{2\kappa}\cdot\P_\eta(D_{j+1}).
  \]
  By induction on $i\le n$ on a quantifier-free formula, $\P_\eta(D_{n-i})\ge(2\kappa)^{-i}$ for every $i\in\{0,1,\dots,n\}$; in particular, $\P_\eta(C)=\P_\eta(D_0)\ge(2\kappa)^{-n}$.

  It remains to switch the counting precision back to $\delta$. By the \nameref{axiom: precision consistency mAPX} instantiated with $(\delta_1,\delta_2)\eqdef(\delta,\eta)$, we have $(1-\delta)\cdot\P_\eta(C)\le(1+\eta)^2\cdot\P_\delta(C)\le\kappa\cdot\P_\delta(C)$, and therefore, as $\delta\le 1$,
  \[
  (1-\delta)\cdot 2^{-n}=\kappa^n\cdot(1-\delta)\cdot(2\kappa)^{-n}\le\kappa^n\cdot(1-\delta)\cdot\P_\eta(C)\le\kappa^{n+1}\cdot\P_\delta(C).
  \]
  Finally we set $\eta\eqdef\beta/(8(n+1))\le 1/8$. Since $\eta\le 1/5$, we have $(1+\eta)^2\le(1+4\eta)(1-\eta)$, i.e., $\kappa\le 1+4\eta$. We use the inequality $(1+s)^k\le 1+2ks$ for $s\ge 0$ and $ks\le 1/2$, which can be proved in $\PV$. Applying it with $s=4\eta$ and $k=n+1$ (where $ks=\beta/2$), we obtain $\kappa^{n+1}\le(1+4\eta)^{n+1}\le 1+8(n+1)\eta=1+\beta$. Hence $(1-\delta)\cdot 2^{-n}\le(1+\beta)\cdot\P_\delta(C)$, as desired.
\end{proof}

\lmmAveraging*

\AI{The first version was drafted by Claude Opus 5.5 following the proof of \Cref{lmm: global consistency mAPX}, and checked by Claude Fable 5.1. The authors further edited it interactively.}
\begin{proof}
  We follow the bit-by-bit fixing trick in the proof of \Cref{lmm: global consistency mAPX}. We argue in $\mAPX$. Fix $\beta^{-1}\in\Log$, and for $z\in\zo^n$ let $\Lambda^+(z)\eqdef\sum_{i\in[m]}\lambda_i^+\cdot C_i(z)$ and $\Lambda^-(z)\eqdef\sum_{i\in[m]}\lambda_i^-\cdot C_i(z)$. Our goal is to find $z\in\zo^n$ such that
  \begin{equation}\label{equ: averaging goal}
  \alpha^+\cdot\Lambda^+(z)\ge\alpha^-\cdot\Lambda^-(z),\qquad\text{where}\quad\alpha^+\eqdef(1+\beta)(1+\delta)\cdot\mu^-\quad\text{and}\quad\alpha^-\eqdef(1-\delta)\cdot\mu^+.
  \end{equation}
  Note that $\alpha^+,\alpha^-\ge 0$, as $\lambda_i^+,\lambda_i^-\ge 0$, $\P_\delta(C_i)\ge 0$ by the \nameref{axiom: basic mAPX}, and $\delta\le 1$. Let
 $\eta^{-1}\in\Log$ be a parameter to be determined and $\kappa\eqdef(1+\eta)^2/(1-\eta)$. (Note that every occurrence of $\mlsim$ in the axioms below is instantiated at the slack $\eta$.)

  For a tuple $\vec D=(D_1,\dots,D_m)$ of circuits on a common input length and $b\in\zo$, let $\Fix_b(\vec D)$ denote the tuple $(\Fix_b(D_1),\dots,\Fix_b(D_m))$, and let
  \[
  A(\vec D)\eqdef\sum_{i\in[m]}\lambda_i^+\cdot\P_\eta(D_i)\qquad\text{and}\qquad B(\vec D)\eqdef\sum_{i\in[m]}\lambda_i^-\cdot\P_\eta(D_i).
  \]
  For every circuit $D$ with at least one input bit, the \nameref{axiom: local consistency mAPX} at precision $\eta$ gives $(1-\eta)\P_\eta(D)\le(1+\eta)^2\cdot p_M$ and $(1-\eta)\cdot p_M\le(1+\eta)^2\cdot\P_\eta(D)$, where $p_M=\frac12(\P_\eta(\Fix_0(D))+\P_\eta(\Fix_1(D)))$; that is, $\kappa^{-1}\cdot p_M\le\P_\eta(D)\le\kappa\cdot p_M$. Multiplying by the non-negative coefficients $\lambda_i^+$ (resp.~$\lambda_i^-$) and summing over $i\in[m]$, we obtain
  \begin{equation}\label{equ: averaging one step}
  \begin{aligned}
  A(\vec D)&\le\frac{\kappa}{2}\cdot\big(A(\Fix_0(\vec D))+A(\Fix_1(\vec D))\big)\qquad\text{and}\\
  B(\vec D)&\ge\frac{1}{2\kappa}\cdot\big(B(\Fix_0(\vec D))+B(\Fix_1(\vec D))\big).
  \end{aligned}
  \end{equation}

  For $j\in\{0,1,\dots,n\}$ and a tuple $\vec D$ of circuits on input length $n-j$, we define the potential function 
  \[
  \Phi_j(\vec D)\eqdef\kappa^{j-n}\cdot\alpha^+\cdot A(\vec D)-\kappa^{n-j}\cdot\alpha^-\cdot B(\vec D).
  \]
  As $\alpha^+,\alpha^-\ge 0$, \eqref{equ: averaging one step} implies that for every $j<n$,
  \[
  \Phi_j(\vec D)\le\frac{1}{2}\cdot\big(\Phi_{j+1}(\Fix_0(\vec D))+\Phi_{j+1}(\Fix_1(\vec D))\big),
  \]
  and hence there is $b\in\zo$ such that $\Phi_{j+1}(\Fix_b(\vec D))\ge\Phi_j(\vec D)$. Consider the algorithm that starts from $\vec C^{(0)}\eqdef(C_1,\dots,C_m)$ and, for $j=0,1,\dots,n-1$, computes such a bit $b_j$ (by querying $\P_\eta$ on the $2m$ circuits in $\Fix_0(\vec C^{(j)})$ and $\Fix_1(\vec C^{(j)})$) and sets $\vec C^{(j+1)}\eqdef\Fix_{b_j}(\vec C^{(j)})$. By induction on $j\le n$ on a quantifier-free formula, $\Phi_n(\vec C^{(n)})\ge\Phi_0(\vec C^{(0)})$. 
  
  Let $z\eqdef b_{n-1}\cdots b_1b_0\in\zo^n$, so that $\vec C^{(n)}=(\Fix_z(C_1),\dots,\Fix_z(C_m))$, where $\Fix_z(C)$ fixes the last $\|z\|$ input bits of $C$ to $z$. Each circuit $\Fix_z(C_i)$ does not read its input and outputs $C_i(z)$, so the \nameref{axiom: boundary mAPX} gives $\P_\eta(\Fix_z(C_i))=C_i(z)$, and therefore
  \begin{equation}\label{equ: averaging final}
  \alpha^+\cdot\Lambda^+(z)-\alpha^-\cdot\Lambda^-(z)=\Phi_n(\vec C^{(n)})\ge\Phi_0(\vec C^{(0)}).
  \end{equation}

  It remains to show that $\Phi_0(\vec C^{(0)})\ge 0$, for which we switch the precision from $\eta$ back to $\delta$. By the \nameref{axiom: precision consistency mAPX} instantiated with $(\delta_1,\delta_2)\eqdef (\eta,\delta)$ and with $(\delta_1,\delta_2)\eqdef (\delta,\eta)$, for every $i\in[m]$,
  \begin{align*}
  & (1+\delta)\cdot\P_\eta(C_i)\ge\frac{1-\eta}{1+\eta}\cdot\P_\delta(C_i)\ge\frac{1}{\kappa}\cdot\P_\delta(C_i)\qquad\text{and}\qquad \\ & (1-\delta)\cdot\P_\eta(C_i)\le(1+\eta)^2\cdot\P_\delta(C_i)\le\kappa\cdot\P_\delta(C_i).
  \end{align*}
  Multiplying by $\lambda_i^+$ (resp.~$\lambda_i^-$) and summing over $i\in[m]$, we obtain $(1+\delta)\cdot A(\vec C^{(0)})\ge\kappa^{-1}\cdot\mu^+$ and $(1-\delta)\cdot B(\vec C^{(0)})\le\kappa\cdot\mu^-$. Plugging in $\alpha^+$ and $\alpha^-$, and using $\mu^+,\mu^-\ge 0$, we have 
  \begin{align*}
  \Phi_0(\vec C^{(0)})& =\kappa^{-n}(1+\beta)\mu^-\cdot(1+\delta)A(\vec C^{(0)})-\kappa^{n}\mu^+\cdot(1-\delta)B(\vec C^{(0)})\\ 
  & \ge\kappa^{-(n+1)}\cdot\mu^+\mu^-\cdot\big((1+\beta)-\kappa^{2(n+1)}\big).
  \end{align*}
  Finally we set the parameters. Let $\eta\eqdef \beta/(32(n+1))\le 1/32$. We have $\kappa = (1+\eta)^2/(1-\eta)\le 1+4\eta$. Recall that $(1+s)^k\le 1+2ks$ if $s\ge 0$ and $ks\le 1/2$. Applying it with $s=4\eta$ and $k=2(n+1)$, we have
  \[
  \kappa^{2(n+1)}\le(1+4\eta)^{2(n+1)}\le 1+16(n+1)\eta=1+\beta/2.
  \]
  Hence $(1+\beta)-\kappa^{2(n+1)}\ge \beta/2$, so $\Phi_0(\vec C^{(0)})\ge\frac{\beta}{2}\kappa^{-(n+1)}\mu^+\mu^-\ge 0$, which together with \eqref{equ: averaging final} gives \eqref{equ: averaging goal}. If moreover $\mu^+>0$ and $\mu^->0$, then $\Phi_0(\vec C^{(0)})>0$, so the inequality in \eqref{equ: averaging goal} is strict.
\end{proof}

\end{document}